\documentclass[longauth]{aa} 

\makeatletter
\renewcommand*\maketitle{%
  \thispagestyle{firstpage}
\begingroup
    \if@wideboxfn
    \setlength\bibindent{1.4\parindent}
    \else
    \setlength\bibindent{\parindent}
    \fi
    \renewcommand*\thefootnote{\@fnsymbol\c@footnote}%
    \renewcommand\@makefntext[1]{%
    \ifaa@longfn\hsize\textwidth\fi
    \noindent
    \hb@xt@\bibindent{\hss\@makefnmark\enspace}##1}
  \ifaa@twocolumn
  \begingroup
    \begin{aa@strip}
          \aa@maketitle
    \end{aa@strip}
    \@thanks            
  \endgroup
  \else
    \begingroup
      \let\thanks\footnote
      \aa@maketitle
    \endgroup
  \fi
\endgroup
  \setcounter{footnote}{0}%
}
\makeatother

\usepackage{graphicx, array}
\usepackage[varg]{txfonts}
\usepackage{natbib,twoopt}
\usepackage{xcolor}
\usepackage{listings}
\usepackage[breaklinks=true]{hyperref} 
\bibpunct{(}{)}{;}{a}{}{,}             
\makeatletter
  \newcommandtwoopt{\citeads}[3][][]{\href{http://adsabs.harvard.edu/abs/#3}%
    {\def\hyper@linkstart##1##2{}%
     \let\hyper@linkend\@empty\citealp[#1][#2]{#3}}}
  \newcommandtwoopt{\citepads}[3][][]{\href{http://adsabs.harvard.edu/abs/#3}%
    {\def\hyper@linkstart##1##2{}%
     \let\hyper@linkend\@empty\citep[#1][#2]{#3}}}
  \newcommandtwoopt{\citetads}[3][][]{\href{http://adsabs.harvard.edu/abs/#3}%
    {\def\hyper@linkstart##1##2{}%
     \let\hyper@linkend\@empty\citet[#1][#2]{#3}}}
  \newcommandtwoopt{\citeyearads}[3][][]%
    {\href{http://adsabs.harvard.edu/abs/#3}
    {\def\hyper@linkstart##1##2{}%
     \let\hyper@linkend\@empty\citeyear[#1][#2]{#3}}}
\makeatother
\hypersetup{colorlinks=true,linkcolor=blue,citecolor=blue,urlcolor=blue}

\definecolor{codegreen}{rgb}{0,0.6,0}
\definecolor{codegray}{rgb}{0.5,0.5,0.5}
\definecolor{codepurple}{rgb}{0.58,0,0.82}
\definecolor{backcolour}{rgb}{0.95,0.95,0.92}

\lstdefinestyle{mystyle}{
    commentstyle=\color{codegreen},
    keywordstyle=\color{magenta},
    numberstyle=\tiny\color{codegray},
    stringstyle=\color{codepurple},
    basicstyle=\ttfamily,
    breakatwhitespace=false,         
    breaklines=true,                 
    captionpos=b,                    
    keepspaces=true,                 
    numbers=none,                    
    showspaces=false,                
    showstringspaces=false,
    showtabs=false,                  
    tabsize=2
}
\newcommand{\orcit}[1]{\protect\href{https://orcid.org/#1}{\protect\includegraphics[width=8pt]{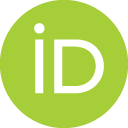}}}

\newcommand\gaia{\textit{Gaia}}
\newcommand\gdr[1]{\gaia~DR#1}
\newcommand\edr[1]{\gaia~EDR#1}
\newcommand\hip{\textsc{Hipparcos}}
\newcommand\tyc{\textit{Tycho}}

\newcommand\gcrf[1]{\gaia-CRF#1}

\newcommand\secref[1]{Sect.~\ref{#1}}

\newcommand\figref[1]{Fig.~\ref{#1}}
\newcommand\figsref[1]{Figs.~\ref{#1}}

\newcommand\tabref[1]{Table~\ref{#1}}
\newcommand\tabsref[1]{Tables~\ref{#1}}

\newcommand\gdrtwovradnum{\ensuremath{7\,224\,631}}

\newcommand\edrtotal{\ensuremath{1\,811\,709\,771}}

\newcommand\edrwithgtot{\ensuremath{1\,806\,254\,432}}
\newcommand\edrwithbptot{\ensuremath{1\,542\,033\,472}}
\newcommand\edrwithrptot{\ensuremath{1\,554\,997\,939}}
\newcommand\edrvradtot{\ensuremath{7\,209\,831}}
\newcommand\edrfiveptot{\ensuremath{585\,416\,709}}
\newcommand\edrsixptot{\ensuremath{882\,328\,109}}
\newcommand\edrtwoptot{\ensuremath{343\,964\,953}}
\newcommand\edrgcrftot{\ensuremath{1\,614\,173}}

\newcommand\edricrfconsidered{\ensuremath{2269}}
\newcommand\edrspinconsidered{\ensuremath{429\,249}}

\newcommand\muas{\ensuremath{\mu\text{as}}}
\newcommand\masyr{\ensuremath{\text{mas~yr}^{-1}}}
\newcommand\muasyr{\ensuremath{\mu\text{as~yr}^{-1}}}
\newcommand\kms{\ensuremath{\text{km~s}^{-1}}}

\newcommand\gbp{\ensuremath{G_\mathrm{BP}}}
\newcommand\grp{\ensuremath{G_\mathrm{RP}}}
\newcommand\bpminrp{\ensuremath{(\gbp-\grp)}}
\newcommand\grvs{\ensuremath{G_\mathrm{RVS}}}

\newcommand\vrad{\ensuremath{v_\mathrm{rad}}}
\newcommand\nueff{\ensuremath{\nu_\mathrm{eff}}}
\newcommand\pmra{\ensuremath{\mu_{\alpha*}}}
\newcommand\pmdec{\ensuremath{\mu_{\delta}}}

\begin{document} 

\title{{\gaia} Early Data Release 3}

\subtitle{Summary of the contents and survey properties}

\author{
{\it Gaia} Collaboration
\and A.G.A.    ~Brown                         \orcit{0000-0002-7419-9679}\inst{\ref{inst:0001}}
\and A.        ~Vallenari                     \orcit{0000-0003-0014-519X}\inst{\ref{inst:0002}}
\and T.        ~Prusti                        \orcit{0000-0003-3120-7867}\inst{\ref{inst:0003}}
\and J.H.J.    ~de Bruijne                    \orcit{0000-0001-6459-8599}\inst{\ref{inst:0003}}
\and C.        ~Babusiaux                     \orcit{0000-0002-7631-348X}\inst{\ref{inst:0005},\ref{inst:0006}}
\and M.        ~Biermann                      \inst{\ref{inst:0007}}
\and O.L.      ~Creevey                       \orcit{0000-0003-1853-6631}\inst{\ref{inst:0008}}
\and D.W.      ~Evans                         \orcit{0000-0002-6685-5998}\inst{\ref{inst:0009}}
\and L.        ~Eyer                          \orcit{0000-0002-0182-8040}\inst{\ref{inst:0010}}
\and A.        ~Hutton                        \inst{\ref{inst:0011}}
\and F.        ~Jansen                        \inst{\ref{inst:0003}}
\and C.        ~Jordi                         \orcit{0000-0001-5495-9602}\inst{\ref{inst:0013}}
\and S.A.      ~Klioner                       \orcit{0000-0003-4682-7831}\inst{\ref{inst:0014}}
\and U.        ~Lammers                       \orcit{0000-0001-8309-3801}\inst{\ref{inst:0015}}
\and L.        ~Lindegren                     \orcit{0000-0002-5443-3026}\inst{\ref{inst:0016}}
\and X.        ~Luri                          \orcit{0000-0001-5428-9397}\inst{\ref{inst:0013}}
\and F.        ~Mignard                       \inst{\ref{inst:0008}}
\and C.        ~Panem                         \inst{\ref{inst:0019}}
\and D.        ~Pourbaix                      \orcit{0000-0002-3020-1837}\inst{\ref{inst:0020},\ref{inst:0021}}
\and S.        ~Randich                       \orcit{0000-0003-2438-0899}\inst{\ref{inst:0022}}
\and P.        ~Sartoretti                    \inst{\ref{inst:0006}}
\and C.        ~Soubiran                      \orcit{0000-0003-3304-8134}\inst{\ref{inst:0024}}
\and N.A.      ~Walton                        \orcit{0000-0003-3983-8778}\inst{\ref{inst:0009}}
\and F.        ~Arenou                        \orcit{0000-0003-2837-3899}\inst{\ref{inst:0006}}
\and C.A.L.    ~Bailer-Jones                  \inst{\ref{inst:0027}}
\and U.        ~Bastian                       \orcit{0000-0002-8667-1715}\inst{\ref{inst:0007}}
\and M.        ~Cropper                       \orcit{0000-0003-4571-9468}\inst{\ref{inst:0029}}
\and R.        ~Drimmel                       \orcit{0000-0002-1777-5502}\inst{\ref{inst:0030}}
\and D.        ~Katz                          \orcit{0000-0001-7986-3164}\inst{\ref{inst:0006}}
\and M.G.      ~Lattanzi                      \orcit{0000-0003-0429-7748}\inst{\ref{inst:0030},\ref{inst:0033}}
\and F.        ~van Leeuwen                   \inst{\ref{inst:0009}}
\and J.        ~Bakker                        \inst{\ref{inst:0015}}
\and C.        ~Cacciari                      \orcit{0000-0001-5174-3179}\inst{\ref{inst:0036}}
\and J.        ~Casta\~{n}eda                 \orcit{0000-0001-7820-946X}\inst{\ref{inst:0037}}
\and F.        ~De Angeli                     \inst{\ref{inst:0009}}
\and C.        ~Ducourant                     \orcit{0000-0003-4843-8979}\inst{\ref{inst:0024}}
\and C.        ~Fabricius                     \orcit{0000-0003-2639-1372}\inst{\ref{inst:0013}}
\and M.        ~Fouesneau                     \orcit{0000-0001-9256-5516}\inst{\ref{inst:0027}}
\and Y.        ~Fr\'{e}mat                    \orcit{0000-0002-4645-6017}\inst{\ref{inst:0042}}
\and R.        ~Guerra                        \orcit{0000-0002-9850-8982}\inst{\ref{inst:0015}}
\and A.        ~Guerrier                      \inst{\ref{inst:0019}}
\and J.        ~Guiraud                       \inst{\ref{inst:0019}}
\and A.        ~Jean-Antoine Piccolo          \inst{\ref{inst:0019}}
\and E.        ~Masana                        \orcit{0000-0002-4819-329X}\inst{\ref{inst:0013}}
\and R.        ~Messineo                      \inst{\ref{inst:0048}}
\and N.        ~Mowlavi                       \inst{\ref{inst:0010}}
\and C.        ~Nicolas                       \inst{\ref{inst:0019}}
\and K.        ~Nienartowicz                  \orcit{0000-0001-5415-0547}\inst{\ref{inst:0051},\ref{inst:0052}}
\and F.        ~Pailler                       \inst{\ref{inst:0019}}
\and P.        ~Panuzzo                       \orcit{0000-0002-0016-8271}\inst{\ref{inst:0006}}
\and F.        ~Riclet                        \inst{\ref{inst:0019}}
\and W.        ~Roux                          \inst{\ref{inst:0019}}
\and G.M.      ~Seabroke                      \inst{\ref{inst:0029}}
\and R.        ~Sordo                         \orcit{0000-0003-4979-0659}\inst{\ref{inst:0002}}
\and P.        ~Tanga                         \orcit{0000-0002-2718-997X}\inst{\ref{inst:0008}}
\and F.        ~Th\'{e}venin                  \inst{\ref{inst:0008}}
\and G.        ~Gracia-Abril                  \inst{\ref{inst:0061},\ref{inst:0007}}
\and J.        ~Portell                       \orcit{0000-0002-8886-8925}\inst{\ref{inst:0013}}
\and D.        ~Teyssier                      \orcit{0000-0002-6261-5292}\inst{\ref{inst:0064}}
\and M.        ~Altmann                       \orcit{0000-0002-0530-0913}\inst{\ref{inst:0007},\ref{inst:0066}}
\and R.        ~Andrae                        \inst{\ref{inst:0027}}
\and I.        ~Bellas-Velidis                \inst{\ref{inst:0068}}
\and K.        ~Benson                        \inst{\ref{inst:0029}}
\and J.        ~Berthier                      \orcit{0000-0003-1846-6485}\inst{\ref{inst:0070}}
\and R.        ~Blomme                        \orcit{0000-0002-2526-346X}\inst{\ref{inst:0042}}
\and E.        ~Brugaletta                    \orcit{0000-0003-2598-6737}\inst{\ref{inst:0072}}
\and P.W.      ~Burgess                       \inst{\ref{inst:0009}}
\and G.        ~Busso                         \orcit{0000-0003-0937-9849}\inst{\ref{inst:0009}}
\and B.        ~Carry                         \orcit{0000-0001-5242-3089}\inst{\ref{inst:0008}}
\and A.        ~Cellino                       \orcit{0000-0002-6645-334X}\inst{\ref{inst:0030}}
\and N.        ~Cheek                         \inst{\ref{inst:0077}}
\and G.        ~Clementini                    \orcit{0000-0001-9206-9723}\inst{\ref{inst:0036}}
\and Y.        ~Damerdji                      \inst{\ref{inst:0079},\ref{inst:0080}}
\and M.        ~Davidson                      \inst{\ref{inst:0081}}
\and L.        ~Delchambre                    \inst{\ref{inst:0079}}
\and A.        ~Dell'Oro                      \orcit{0000-0003-1561-9685}\inst{\ref{inst:0022}}
\and J.        ~Fern\'{a}ndez-Hern\'{a}ndez   \inst{\ref{inst:0084}}
\and L.        ~Galluccio                     \orcit{0000-0002-8541-0476}\inst{\ref{inst:0008}}
\and P.        ~Garc\'{i}a-Lario              \inst{\ref{inst:0015}}
\and M.        ~Garcia-Reinaldos              \inst{\ref{inst:0015}}
\and J.        ~Gonz\'{a}lez-N\'{u}\~{n}ez    \orcit{0000-0001-5311-5555}\inst{\ref{inst:0077},\ref{inst:0089}}
\and E.        ~Gosset                        \inst{\ref{inst:0079},\ref{inst:0021}}
\and R.        ~Haigron                       \inst{\ref{inst:0006}}
\and J.-L.     ~Halbwachs                     \orcit{0000-0003-2968-6395}\inst{\ref{inst:0093}}
\and N.C.      ~Hambly                        \orcit{0000-0002-9901-9064}\inst{\ref{inst:0081}}
\and D.L.      ~Harrison                      \orcit{0000-0001-8687-6588}\inst{\ref{inst:0009},\ref{inst:0096}}
\and D.        ~Hatzidimitriou                \orcit{0000-0002-5415-0464}\inst{\ref{inst:0097}}
\and U.        ~Heiter                        \orcit{0000-0001-6825-1066}\inst{\ref{inst:0098}}
\and J.        ~Hern\'{a}ndez                 \inst{\ref{inst:0015}}
\and D.        ~Hestroffer                    \orcit{0000-0003-0472-9459}\inst{\ref{inst:0070}}
\and S.T.      ~Hodgkin                       \inst{\ref{inst:0009}}
\and B.        ~Holl                          \orcit{0000-0001-6220-3266}\inst{\ref{inst:0010},\ref{inst:0051}}
\and K.        ~Jan{\ss}en                    \inst{\ref{inst:0104}}
\and G.        ~Jevardat de Fombelle          \inst{\ref{inst:0010}}
\and S.        ~Jordan                        \orcit{0000-0001-6316-6831}\inst{\ref{inst:0007}}
\and A.        ~Krone-Martins                 \orcit{0000-0002-2308-6623}\inst{\ref{inst:0107},\ref{inst:0108}}
\and A.C.      ~Lanzafame                     \orcit{0000-0002-2697-3607}\inst{\ref{inst:0072},\ref{inst:0110}}
\and W.        ~L\"{ o}ffler                  \inst{\ref{inst:0007}}
\and A.        ~Lorca                         \inst{\ref{inst:0011}}
\and M.        ~Manteiga                      \orcit{0000-0002-7711-5581}\inst{\ref{inst:0113}}
\and O.        ~Marchal                       \inst{\ref{inst:0093}}
\and P.M.      ~Marrese                       \inst{\ref{inst:0115},\ref{inst:0116}}
\and A.        ~Moitinho                      \orcit{0000-0003-0822-5995}\inst{\ref{inst:0107}}
\and A.        ~Mora                          \inst{\ref{inst:0011}}
\and K.        ~Muinonen                      \orcit{0000-0001-8058-2642}\inst{\ref{inst:0119},\ref{inst:0120}}
\and P.        ~Osborne                       \inst{\ref{inst:0009}}
\and E.        ~Pancino                       \orcit{0000-0003-0788-5879}\inst{\ref{inst:0022},\ref{inst:0116}}
\and T.        ~Pauwels                       \inst{\ref{inst:0042}}
\and J.-M.     ~Petit                         \orcit{0000-0003-0407-2266}\inst{\ref{inst:0125}}
\and A.        ~Recio-Blanco                  \inst{\ref{inst:0008}}
\and P.J.      ~Richards                      \inst{\ref{inst:0127}}
\and M.        ~Riello                        \orcit{0000-0002-3134-0935}\inst{\ref{inst:0009}}
\and L.        ~Rimoldini                     \orcit{0000-0002-0306-585X}\inst{\ref{inst:0051}}
\and A.C.      ~Robin                         \orcit{0000-0001-8654-9499}\inst{\ref{inst:0125}}
\and T.        ~Roegiers                      \inst{\ref{inst:0131}}
\and J.        ~Rybizki                       \orcit{0000-0002-0993-6089}\inst{\ref{inst:0027}}
\and L.M.      ~Sarro                         \orcit{0000-0002-5622-5191}\inst{\ref{inst:0133}}
\and C.        ~Siopis                        \inst{\ref{inst:0020}}
\and M.        ~Smith                         \inst{\ref{inst:0029}}
\and A.        ~Sozzetti                      \orcit{0000-0002-7504-365X}\inst{\ref{inst:0030}}
\and A.        ~Ulla                          \inst{\ref{inst:0137}}
\and E.        ~Utrilla                       \inst{\ref{inst:0011}}
\and M.        ~van Leeuwen                   \inst{\ref{inst:0009}}
\and W.        ~van Reeven                    \inst{\ref{inst:0011}}
\and U.        ~Abbas                         \orcit{0000-0002-5076-766X}\inst{\ref{inst:0030}}
\and A.        ~Abreu Aramburu                \inst{\ref{inst:0084}}
\and S.        ~Accart                        \inst{\ref{inst:0143}}
\and C.        ~Aerts                         \orcit{0000-0003-1822-7126}\inst{\ref{inst:0144},\ref{inst:0145},\ref{inst:0027}}
\and J.J.      ~Aguado                        \inst{\ref{inst:0133}}
\and M.        ~Ajaj                          \inst{\ref{inst:0006}}
\and G.        ~Altavilla                     \orcit{0000-0002-9934-1352}\inst{\ref{inst:0115},\ref{inst:0116}}
\and M.A.      ~\'{A}lvarez                   \orcit{0000-0002-6786-2620}\inst{\ref{inst:0151}}
\and J.        ~\'{A}lvarez Cid-Fuentes       \orcit{0000-0001-7153-4649}\inst{\ref{inst:0152}}
\and J.        ~Alves                         \orcit{0000-0002-4355-0921}\inst{\ref{inst:0153}}
\and R.I.      ~Anderson                      \orcit{0000-0001-8089-4419}\inst{\ref{inst:0154}}
\and E.        ~Anglada Varela                \orcit{0000-0001-7563-0689}\inst{\ref{inst:0084}}
\and T.        ~Antoja                        \orcit{0000-0003-2595-5148}\inst{\ref{inst:0013}}
\and M.        ~Audard                        \orcit{0000-0003-4721-034X}\inst{\ref{inst:0051}}
\and D.        ~Baines                        \orcit{0000-0002-6923-3756}\inst{\ref{inst:0064}}
\and S.G.      ~Baker                         \orcit{0000-0002-6436-1257}\inst{\ref{inst:0029}}
\and L.        ~Balaguer-N\'{u}\~{n}ez        \orcit{0000-0001-9789-7069}\inst{\ref{inst:0013}}
\and E.        ~Balbinot                      \orcit{0000-0002-1322-3153}\inst{\ref{inst:0161}}
\and Z.        ~Balog                         \orcit{0000-0003-1748-2926}\inst{\ref{inst:0007},\ref{inst:0027}}
\and C.        ~Barache                       \inst{\ref{inst:0066}}
\and D.        ~Barbato                       \inst{\ref{inst:0010},\ref{inst:0030}}
\and M.        ~Barros                        \orcit{0000-0002-9728-9618}\inst{\ref{inst:0107}}
\and M.A.      ~Barstow                       \orcit{0000-0002-7116-3259}\inst{\ref{inst:0168}}
\and S.        ~Bartolom\'{e}                 \orcit{0000-0002-6290-6030}\inst{\ref{inst:0013}}
\and J.-L.     ~Bassilana                     \inst{\ref{inst:0143}}
\and N.        ~Bauchet                       \inst{\ref{inst:0070}}
\and A.        ~Baudesson-Stella              \inst{\ref{inst:0143}}
\and U.        ~Becciani                      \orcit{0000-0002-4389-8688}\inst{\ref{inst:0072}}
\and M.        ~Bellazzini                    \orcit{0000-0001-8200-810X}\inst{\ref{inst:0036}}
\and M.        ~Bernet                        \inst{\ref{inst:0013}}
\and S.        ~Bertone                       \orcit{0000-0001-9885-8440}\inst{\ref{inst:0176},\ref{inst:0177},\ref{inst:0030}}
\and L.        ~Bianchi                       \inst{\ref{inst:0179}}
\and S.        ~Blanco-Cuaresma               \orcit{0000-0002-1584-0171}\inst{\ref{inst:0180}}
\and T.        ~Boch                          \orcit{0000-0001-5818-2781}\inst{\ref{inst:0093}}
\and A.        ~Bombrun                       \inst{\ref{inst:0182}}
\and D.        ~Bossini                       \orcit{0000-0002-9480-8400}\inst{\ref{inst:0183}}
\and S.        ~Bouquillon                    \inst{\ref{inst:0066}}
\and A.        ~Bragaglia                     \orcit{0000-0002-0338-7883}\inst{\ref{inst:0036}}
\and L.        ~Bramante                      \inst{\ref{inst:0048}}
\and E.        ~Breedt                        \orcit{0000-0001-6180-3438}\inst{\ref{inst:0009}}
\and A.        ~Bressan                       \orcit{0000-0002-7922-8440}\inst{\ref{inst:0188}}
\and N.        ~Brouillet                     \inst{\ref{inst:0024}}
\and B.        ~Bucciarelli                   \orcit{0000-0002-5303-0268}\inst{\ref{inst:0030}}
\and A.        ~Burlacu                       \inst{\ref{inst:0191}}
\and D.        ~Busonero                      \orcit{0000-0002-3903-7076}\inst{\ref{inst:0030}}
\and A.G.      ~Butkevich                     \inst{\ref{inst:0030}}
\and R.        ~Buzzi                         \orcit{0000-0001-9389-5701}\inst{\ref{inst:0030}}
\and E.        ~Caffau                        \orcit{0000-0001-6011-6134}\inst{\ref{inst:0006}}
\and R.        ~Cancelliere                   \orcit{0000-0002-9120-3799}\inst{\ref{inst:0196}}
\and H.        ~C\'{a}novas                   \orcit{0000-0001-7668-8022}\inst{\ref{inst:0011}}
\and T.        ~Cantat-Gaudin                 \orcit{0000-0001-8726-2588}\inst{\ref{inst:0013}}
\and R.        ~Carballo                      \inst{\ref{inst:0199}}
\and T.        ~Carlucci                      \inst{\ref{inst:0066}}
\and M.I       ~Carnerero                     \orcit{0000-0001-5843-5515}\inst{\ref{inst:0030}}
\and J.M.      ~Carrasco                      \orcit{0000-0002-3029-5853}\inst{\ref{inst:0013}}
\and L.        ~Casamiquela                   \orcit{0000-0001-5238-8674}\inst{\ref{inst:0024}}
\and M.        ~Castellani                    \orcit{0000-0002-7650-7428}\inst{\ref{inst:0115}}
\and A.        ~Castro-Ginard                 \orcit{0000-0002-9419-3725}\inst{\ref{inst:0013}}
\and P.        ~Castro Sampol                 \inst{\ref{inst:0013}}
\and L.        ~Chaoul                        \inst{\ref{inst:0019}}
\and P.        ~Charlot                       \inst{\ref{inst:0024}}
\and L.        ~Chemin                        \orcit{0000-0002-3834-7937}\inst{\ref{inst:0209}}
\and A.        ~Chiavassa                     \orcit{0000-0003-3891-7554}\inst{\ref{inst:0008}}
\and M.-R. L.  ~Cioni                         \orcit{0000-0002-6797-696x}\inst{\ref{inst:0104}}
\and G.        ~Comoretto                     \inst{\ref{inst:0212}}
\and W.J.      ~Cooper                        \orcit{0000-0003-3501-8967}\inst{\ref{inst:0213},\ref{inst:0030}}
\and T.        ~Cornez                        \inst{\ref{inst:0143}}
\and S.        ~Cowell                        \inst{\ref{inst:0009}}
\and F.        ~Crifo                         \inst{\ref{inst:0006}}
\and M.        ~Crosta                        \orcit{0000-0003-4369-3786}\inst{\ref{inst:0030}}
\and C.        ~Crowley                       \inst{\ref{inst:0182}}
\and C.        ~Dafonte                       \orcit{0000-0003-4693-7555}\inst{\ref{inst:0151}}
\and A.        ~Dapergolas                    \inst{\ref{inst:0068}}
\and M.        ~David                         \orcit{0000-0002-4172-3112}\inst{\ref{inst:0222}}
\and P.        ~David                         \inst{\ref{inst:0070}}
\and P.        ~de Laverny                    \inst{\ref{inst:0008}}
\and F.        ~De Luise                      \orcit{0000-0002-6570-8208}\inst{\ref{inst:0225}}
\and R.        ~De March                      \orcit{0000-0003-0567-842X}\inst{\ref{inst:0048}}
\and J.        ~De Ridder                     \orcit{0000-0001-6726-2863}\inst{\ref{inst:0144}}
\and R.        ~de Souza                      \inst{\ref{inst:0228}}
\and P.        ~de Teodoro                    \inst{\ref{inst:0015}}
\and A.        ~de Torres                     \inst{\ref{inst:0182}}
\and E.F.      ~del Peloso                    \inst{\ref{inst:0007}}
\and E.        ~del Pozo                      \inst{\ref{inst:0011}}
\and M.        ~Delbo                         \orcit{0000-0002-8963-2404}\inst{\ref{inst:0008}}
\and A.        ~Delgado                       \inst{\ref{inst:0009}}
\and H.E.      ~Delgado                       \orcit{0000-0003-1409-4282}\inst{\ref{inst:0133}}
\and J.-B.     ~Delisle                       \orcit{0000-0001-5844-9888}\inst{\ref{inst:0010}}
\and P.        ~Di Matteo                     \inst{\ref{inst:0006}}
\and S.        ~Diakite                       \inst{\ref{inst:0238}}
\and C.        ~Diener                        \inst{\ref{inst:0009}}
\and E.        ~Distefano                     \orcit{0000-0002-2448-2513}\inst{\ref{inst:0072}}
\and C.        ~Dolding                       \inst{\ref{inst:0029}}
\and D.        ~Eappachen                     \inst{\ref{inst:0242},\ref{inst:0145}}
\and B.        ~Edvardsson                    \inst{\ref{inst:0244}}
\and H.        ~Enke                          \orcit{0000-0002-2366-8316}\inst{\ref{inst:0104}}
\and P.        ~Esquej                        \orcit{0000-0001-8195-628X}\inst{\ref{inst:0246}}
\and C.        ~Fabre                         \inst{\ref{inst:0247}}
\and M.        ~Fabrizio                      \orcit{0000-0001-5829-111X}\inst{\ref{inst:0115},\ref{inst:0116}}
\and S.        ~Faigler                       \inst{\ref{inst:0250}}
\and G.        ~Fedorets                      \inst{\ref{inst:0119},\ref{inst:0252}}
\and P.        ~Fernique                      \orcit{0000-0002-3304-2923}\inst{\ref{inst:0093},\ref{inst:0254}}
\and A.        ~Fienga                        \orcit{0000-0002-4755-7637}\inst{\ref{inst:0255},\ref{inst:0070}}
\and F.        ~Figueras                      \orcit{0000-0002-3393-0007}\inst{\ref{inst:0013}}
\and C.        ~Fouron                        \inst{\ref{inst:0191}}
\and F.        ~Fragkoudi                     \inst{\ref{inst:0259}}
\and E.        ~Fraile                        \inst{\ref{inst:0246}}
\and F.        ~Franke                        \inst{\ref{inst:0261}}
\and M.        ~Gai                           \orcit{0000-0001-9008-134X}\inst{\ref{inst:0030}}
\and D.        ~Garabato                      \orcit{0000-0002-7133-6623}\inst{\ref{inst:0151}}
\and A.        ~Garcia-Gutierrez              \inst{\ref{inst:0013}}
\and M.        ~Garc\'{i}a-Torres             \orcit{0000-0002-6867-7080}\inst{\ref{inst:0265}}
\and A.        ~Garofalo                      \orcit{0000-0002-5907-0375}\inst{\ref{inst:0036}}
\and P.        ~Gavras                        \orcit{0000-0002-4383-4836}\inst{\ref{inst:0246}}
\and E.        ~Gerlach                       \orcit{0000-0002-9533-2168}\inst{\ref{inst:0014}}
\and R.        ~Geyer                         \orcit{0000-0001-6967-8707}\inst{\ref{inst:0014}}
\and P.        ~Giacobbe                      \inst{\ref{inst:0030}}
\and G.        ~Gilmore                       \orcit{0000-0003-4632-0213}\inst{\ref{inst:0009}}
\and S.        ~Girona                        \orcit{0000-0002-1975-1918}\inst{\ref{inst:0152}}
\and G.        ~Giuffrida                     \inst{\ref{inst:0115}}
\and R.        ~Gomel                         \inst{\ref{inst:0250}}
\and A.        ~Gomez                         \orcit{0000-0002-3796-3690}\inst{\ref{inst:0151}}
\and I.        ~Gonzalez-Santamaria           \orcit{0000-0002-8537-9384}\inst{\ref{inst:0151}}
\and J.J.      ~Gonz\'{a}lez-Vidal            \inst{\ref{inst:0013}}
\and M.        ~Granvik                       \orcit{0000-0002-5624-1888}\inst{\ref{inst:0119},\ref{inst:0279}}
\and R.        ~Guti\'{e}rrez-S\'{a}nchez     \inst{\ref{inst:0064}}
\and L.P.      ~Guy                           \orcit{0000-0003-0800-8755}\inst{\ref{inst:0051},\ref{inst:0212}}
\and M.        ~Hauser                        \inst{\ref{inst:0027},\ref{inst:0284}}
\and M.        ~Haywood                       \orcit{0000-0003-0434-0400}\inst{\ref{inst:0006}}
\and A.        ~Helmi                         \orcit{0000-0003-3937-7641}\inst{\ref{inst:0161}}
\and S.L.      ~Hidalgo                       \orcit{0000-0002-0002-9298}\inst{\ref{inst:0287},\ref{inst:0288}}
\and T.        ~Hilger                        \orcit{0000-0003-1646-0063}\inst{\ref{inst:0014}}
\and N.        ~H\l{}adczuk                   \inst{\ref{inst:0015}}
\and D.        ~Hobbs                         \orcit{0000-0002-2696-1366}\inst{\ref{inst:0016}}
\and G.        ~Holland                       \inst{\ref{inst:0009}}
\and H.E.      ~Huckle                        \inst{\ref{inst:0029}}
\and G.        ~Jasniewicz                    \inst{\ref{inst:0294}}
\and P.G.      ~Jonker                        \orcit{0000-0001-5679-0695}\inst{\ref{inst:0145},\ref{inst:0242}}
\and J.        ~Juaristi Campillo             \inst{\ref{inst:0007}}
\and F.        ~Julbe                         \inst{\ref{inst:0013}}
\and L.        ~Karbevska                     \inst{\ref{inst:0010}}
\and P.        ~Kervella                      \orcit{0000-0003-0626-1749}\inst{\ref{inst:0300}}
\and S.        ~Khanna                        \orcit{0000-0002-2604-4277}\inst{\ref{inst:0161}}
\and A.        ~Kochoska                      \orcit{0000-0002-9739-8371}\inst{\ref{inst:0302}}
\and M.        ~Kontizas                      \orcit{0000-0001-7177-0158}\inst{\ref{inst:0097}}
\and G.        ~Kordopatis                    \orcit{0000-0002-9035-3920}\inst{\ref{inst:0008}}
\and A.J.      ~Korn                          \orcit{0000-0002-3881-6756}\inst{\ref{inst:0098}}
\and Z.        ~Kostrzewa-Rutkowska           \inst{\ref{inst:0001},\ref{inst:0242}}
\and K.        ~Kruszy\'{n}ska                \orcit{0000-0002-2729-5369}\inst{\ref{inst:0308}}
\and S.        ~Lambert                       \orcit{0000-0001-6759-5502}\inst{\ref{inst:0066}}
\and A.F.      ~Lanza                         \orcit{0000-0001-5928-7251}\inst{\ref{inst:0072}}
\and Y.        ~Lasne                         \inst{\ref{inst:0143}}
\and J.-F.     ~Le Campion                    \inst{\ref{inst:0312}}
\and Y.        ~Le Fustec                     \inst{\ref{inst:0191}}
\and Y.        ~Lebreton                      \orcit{0000-0002-4834-2144}\inst{\ref{inst:0300},\ref{inst:0315}}
\and T.        ~Lebzelter                     \orcit{0000-0002-0702-7551}\inst{\ref{inst:0153}}
\and S.        ~Leccia                        \orcit{0000-0001-5685-6930}\inst{\ref{inst:0317}}
\and N.        ~Leclerc                       \inst{\ref{inst:0006}}
\and I.        ~Lecoeur-Taibi                 \orcit{0000-0003-0029-8575}\inst{\ref{inst:0051}}
\and S.        ~Liao                          \inst{\ref{inst:0030}}
\and E.        ~Licata                        \orcit{0000-0002-5203-0135}\inst{\ref{inst:0030}}
\and H.E.P.    ~Lindstr{\o}m                  \inst{\ref{inst:0030},\ref{inst:0323}}
\and T.A.      ~Lister                        \orcit{0000-0002-3818-7769}\inst{\ref{inst:0324}}
\and E.        ~Livanou                       \inst{\ref{inst:0097}}
\and A.        ~Lobel                         \inst{\ref{inst:0042}}
\and P.        ~Madrero Pardo                 \inst{\ref{inst:0013}}
\and S.        ~Managau                       \inst{\ref{inst:0143}}
\and R.G.      ~Mann                          \orcit{0000-0002-0194-325X}\inst{\ref{inst:0081}}
\and J.M.      ~Marchant                      \inst{\ref{inst:0330}}
\and M.        ~Marconi                       \orcit{0000-0002-1330-2927}\inst{\ref{inst:0317}}
\and M.M.S.    ~Marcos Santos                 \inst{\ref{inst:0077}}
\and S.        ~Marinoni                      \orcit{0000-0001-7990-6849}\inst{\ref{inst:0115},\ref{inst:0116}}
\and F.        ~Marocco                       \orcit{0000-0001-7519-1700}\inst{\ref{inst:0335},\ref{inst:0336}}
\and D.J.      ~Marshall                      \inst{\ref{inst:0337}}
\and L.        ~Martin Polo                   \inst{\ref{inst:0077}}
\and J.M.      ~Mart\'{i}n-Fleitas            \orcit{0000-0002-8594-569X}\inst{\ref{inst:0011}}
\and A.        ~Masip                         \inst{\ref{inst:0013}}
\and D.        ~Massari                       \orcit{0000-0001-8892-4301}\inst{\ref{inst:0036}}
\and A.        ~Mastrobuono-Battisti          \orcit{0000-0002-2386-9142}\inst{\ref{inst:0016}}
\and T.        ~Mazeh                         \orcit{0000-0002-3569-3391}\inst{\ref{inst:0250}}
\and P.J.      ~McMillan                      \orcit{0000-0002-8861-2620}\inst{\ref{inst:0016}}
\and S.        ~Messina                       \orcit{0000-0002-2851-2468}\inst{\ref{inst:0072}}
\and D.        ~Michalik                      \orcit{0000-0002-7618-6556}\inst{\ref{inst:0003}}
\and N.R.      ~Millar                        \inst{\ref{inst:0009}}
\and A.        ~Mints                         \orcit{0000-0002-8440-1455}\inst{\ref{inst:0104}}
\and D.        ~Molina                        \orcit{0000-0003-4814-0275}\inst{\ref{inst:0013}}
\and R.        ~Molinaro                      \orcit{0000-0003-3055-6002}\inst{\ref{inst:0317}}
\and L.        ~Moln\'{a}r                    \orcit{0000-0002-8159-1599}\inst{\ref{inst:0351},\ref{inst:0352},\ref{inst:0353}}
\and P.        ~Montegriffo                   \inst{\ref{inst:0036}}
\and R.        ~Mor                           \orcit{0000-0002-8179-6527}\inst{\ref{inst:0013}}
\and R.        ~Morbidelli                    \orcit{0000-0001-7627-4946}\inst{\ref{inst:0030}}
\and T.        ~Morel                         \inst{\ref{inst:0079}}
\and D.        ~Morris                        \inst{\ref{inst:0081}}
\and A.F.      ~Mulone                        \inst{\ref{inst:0048}}
\and D.        ~Munoz                         \inst{\ref{inst:0143}}
\and T.        ~Muraveva                      \orcit{0000-0002-0969-1915}\inst{\ref{inst:0036}}
\and C.P.      ~Murphy                        \inst{\ref{inst:0015}}
\and I.        ~Musella                       \orcit{0000-0001-5909-6615}\inst{\ref{inst:0317}}
\and L.        ~Noval                         \inst{\ref{inst:0143}}
\and C.        ~Ord\'{e}novic                 \inst{\ref{inst:0008}}
\and G.        ~Orr\`{u}                      \inst{\ref{inst:0048}}
\and J.        ~Osinde                        \inst{\ref{inst:0246}}
\and C.        ~Pagani                        \inst{\ref{inst:0168}}
\and I.        ~Pagano                        \orcit{0000-0001-9573-4928}\inst{\ref{inst:0072}}
\and L.        ~Palaversa                     \inst{\ref{inst:0370},\ref{inst:0009}}
\and P.A.      ~Palicio                       \orcit{0000-0002-7432-8709}\inst{\ref{inst:0008}}
\and A.        ~Panahi                        \orcit{0000-0001-5850-4373}\inst{\ref{inst:0250}}
\and M.        ~Pawlak                        \orcit{0000-0002-5632-9433}\inst{\ref{inst:0374},\ref{inst:0308}}
\and X.        ~Pe\~{n}alosa Esteller         \inst{\ref{inst:0013}}
\and A.        ~Penttil\"{ a}                 \orcit{0000-0001-7403-1721}\inst{\ref{inst:0119}}
\and A.M.      ~Piersimoni                    \orcit{0000-0002-8019-3708}\inst{\ref{inst:0225}}
\and F.-X.     ~Pineau                        \orcit{0000-0002-2335-4499}\inst{\ref{inst:0093}}
\and E.        ~Plachy                        \orcit{0000-0002-5481-3352}\inst{\ref{inst:0351},\ref{inst:0352},\ref{inst:0353}}
\and G.        ~Plum                          \inst{\ref{inst:0006}}
\and E.        ~Poggio                        \orcit{0000-0003-3793-8505}\inst{\ref{inst:0030}}
\and E.        ~Poretti                       \orcit{0000-0003-1200-0473}\inst{\ref{inst:0385}}
\and E.        ~Poujoulet                     \inst{\ref{inst:0386}}
\and A.        ~Pr\v{s}a                      \orcit{0000-0002-1913-0281}\inst{\ref{inst:0302}}
\and L.        ~Pulone                        \orcit{0000-0002-5285-998X}\inst{\ref{inst:0115}}
\and E.        ~Racero                        \inst{\ref{inst:0077},\ref{inst:0390}}
\and S.        ~Ragaini                       \inst{\ref{inst:0036}}
\and M.        ~Rainer                        \orcit{0000-0002-8786-2572}\inst{\ref{inst:0022}}
\and C.M.      ~Raiteri                       \orcit{0000-0003-1784-2784}\inst{\ref{inst:0030}}
\and N.        ~Rambaux                       \inst{\ref{inst:0070}}
\and P.        ~Ramos                         \orcit{0000-0002-5080-7027}\inst{\ref{inst:0013}}
\and M.        ~Ramos-Lerate                  \inst{\ref{inst:0396}}
\and P.        ~Re Fiorentin                  \orcit{0000-0002-4995-0475}\inst{\ref{inst:0030}}
\and S.        ~Regibo                        \inst{\ref{inst:0144}}
\and C.        ~Reyl\'{e}                     \inst{\ref{inst:0125}}
\and V.        ~Ripepi                        \orcit{0000-0003-1801-426X}\inst{\ref{inst:0317}}
\and A.        ~Riva                          \orcit{0000-0002-6928-8589}\inst{\ref{inst:0030}}
\and G.        ~Rixon                         \inst{\ref{inst:0009}}
\and N.        ~Robichon                      \orcit{0000-0003-4545-7517}\inst{\ref{inst:0006}}
\and C.        ~Robin                         \inst{\ref{inst:0143}}
\and M.        ~Roelens                       \orcit{0000-0003-0876-4673}\inst{\ref{inst:0010}}
\and L.        ~Rohrbasser                    \inst{\ref{inst:0051}}
\and M.        ~Romero-G\'{o}mez              \orcit{0000-0003-3936-1025}\inst{\ref{inst:0013}}
\and N.        ~Rowell                        \inst{\ref{inst:0081}}
\and F.        ~Royer                         \orcit{0000-0002-9374-8645}\inst{\ref{inst:0006}}
\and K.A.      ~Rybicki                       \orcit{0000-0002-9326-9329}\inst{\ref{inst:0308}}
\and G.        ~Sadowski                      \inst{\ref{inst:0020}}
\and A.        ~Sagrist\`{a} Sell\'{e}s       \orcit{0000-0001-6191-2028}\inst{\ref{inst:0007}}
\and J.        ~Sahlmann                      \orcit{0000-0001-9525-3673}\inst{\ref{inst:0246}}
\and J.        ~Salgado                       \orcit{0000-0002-3680-4364}\inst{\ref{inst:0064}}
\and E.        ~Salguero                      \inst{\ref{inst:0084}}
\and N.        ~Samaras                       \orcit{0000-0001-8375-6652}\inst{\ref{inst:0042}}
\and V.        ~Sanchez Gimenez               \inst{\ref{inst:0013}}
\and N.        ~Sanna                         \inst{\ref{inst:0022}}
\and R.        ~Santove\~{n}a                 \orcit{0000-0002-9257-2131}\inst{\ref{inst:0151}}
\and M.        ~Sarasso                       \orcit{0000-0001-5121-0727}\inst{\ref{inst:0030}}
\and M.        ~Schultheis                    \orcit{0000-0002-6590-1657}\inst{\ref{inst:0008}}
\and E.        ~Sciacca                       \orcit{0000-0002-5574-2787}\inst{\ref{inst:0072}}
\and M.        ~Segol                         \inst{\ref{inst:0261}}
\and J.C.      ~Segovia                       \inst{\ref{inst:0077}}
\and D.        ~S\'{e}gransan                 \orcit{0000-0003-2355-8034}\inst{\ref{inst:0010}}
\and D.        ~Semeux                        \inst{\ref{inst:0247}}
\and S.        ~Shahaf                        \orcit{0000-0001-9298-8068}\inst{\ref{inst:0250}}
\and H.I.      ~Siddiqui                      \orcit{0000-0003-1853-6033}\inst{\ref{inst:0428}}
\and A.        ~Siebert                       \orcit{0000-0001-8059-2840}\inst{\ref{inst:0093},\ref{inst:0254}}
\and L.        ~Siltala                       \orcit{0000-0002-6938-794X}\inst{\ref{inst:0119}}
\and E.        ~Slezak                        \inst{\ref{inst:0008}}
\and R.L.      ~Smart                         \orcit{0000-0002-4424-4766}\inst{\ref{inst:0030}}
\and E.        ~Solano                        \inst{\ref{inst:0434}}
\and F.        ~Solitro                       \inst{\ref{inst:0048}}
\and D.        ~Souami                        \orcit{0000-0003-4058-0815}\inst{\ref{inst:0300},\ref{inst:0437}}
\and J.        ~Souchay                       \inst{\ref{inst:0066}}
\and A.        ~Spagna                        \orcit{0000-0003-1732-2412}\inst{\ref{inst:0030}}
\and F.        ~Spoto                         \orcit{0000-0001-7319-5847}\inst{\ref{inst:0180}}
\and I.A.      ~Steele                        \orcit{0000-0001-8397-5759}\inst{\ref{inst:0330}}
\and H.        ~Steidelm\"{ u}ller            \inst{\ref{inst:0014}}
\and C.A.      ~Stephenson                    \inst{\ref{inst:0064}}
\and M.        ~S\"{ u}veges                  \inst{\ref{inst:0051},\ref{inst:0445},\ref{inst:0027}}
\and L.        ~Szabados                      \orcit{0000-0002-2046-4131}\inst{\ref{inst:0351}}
\and E.        ~Szegedi-Elek                  \orcit{0000-0001-7807-6644}\inst{\ref{inst:0351}}
\and F.        ~Taris                         \inst{\ref{inst:0066}}
\and G.        ~Tauran                        \inst{\ref{inst:0143}}
\and M.B.      ~Taylor                        \orcit{0000-0002-4209-1479}\inst{\ref{inst:0451}}
\and R.        ~Teixeira                      \orcit{0000-0002-6806-6626}\inst{\ref{inst:0228}}
\and W.        ~Thuillot                      \inst{\ref{inst:0070}}
\and N.        ~Tonello                       \orcit{0000-0003-0550-1667}\inst{\ref{inst:0152}}
\and F.        ~Torra                         \orcit{0000-0002-8429-299X}\inst{\ref{inst:0037}}
\and J.        ~Torra$^\dagger$               \inst{\ref{inst:0013}}
\and C.        ~Turon                         \orcit{0000-0003-1236-5157}\inst{\ref{inst:0006}}
\and N.        ~Unger                         \orcit{0000-0003-3993-7127}\inst{\ref{inst:0010}}
\and M.        ~Vaillant                      \inst{\ref{inst:0143}}
\and E.        ~van Dillen                    \inst{\ref{inst:0261}}
\and O.        ~Vanel                         \inst{\ref{inst:0006}}
\and A.        ~Vecchiato                     \orcit{0000-0003-1399-5556}\inst{\ref{inst:0030}}
\and Y.        ~Viala                         \inst{\ref{inst:0006}}
\and D.        ~Vicente                       \inst{\ref{inst:0152}}
\and S.        ~Voutsinas                     \inst{\ref{inst:0081}}
\and M.        ~Weiler                        \inst{\ref{inst:0013}}
\and T.        ~Wevers                        \orcit{0000-0002-4043-9400}\inst{\ref{inst:0009}}
\and \L{}.     ~Wyrzykowski                   \orcit{0000-0002-9658-6151}\inst{\ref{inst:0308}}
\and A.        ~Yoldas                        \inst{\ref{inst:0009}}
\and P.        ~Yvard                         \inst{\ref{inst:0261}}
\and H.        ~Zhao                          \orcit{0000-0003-2645-6869}\inst{\ref{inst:0008}}
\and J.        ~Zorec                         \inst{\ref{inst:0472}}
\and S.        ~Zucker                        \orcit{0000-0003-3173-3138}\inst{\ref{inst:0473}}
\and C.        ~Zurbach                       \inst{\ref{inst:0474}}
\and T.        ~Zwitter                       \orcit{0000-0002-2325-8763}\inst{\ref{inst:0475}}
}
\institute{
     Leiden Observatory, Leiden University, Niels Bohrweg 2, 2333 CA Leiden, The Netherlands\relax                                                                                                                                                                                                               \label{inst:0001}
\and INAF - Osservatorio astronomico di Padova, Vicolo Osservatorio 5, 35122 Padova, Italy\relax                                                                                                                                                                                                                 \label{inst:0002}
\and European Space Agency (ESA), European Space Research and Technology Centre (ESTEC), Keplerlaan 1, 2201AZ, Noordwijk, The Netherlands\relax                                                                                                                                                                  \label{inst:0003}
\and Univ. Grenoble Alpes, CNRS, IPAG, 38000 Grenoble, France\relax                                                                                                                                                                                                                                              \label{inst:0005}
\and GEPI, Observatoire de Paris, Universit\'{e} PSL, CNRS, 5 Place Jules Janssen, 92190 Meudon, France\relax                                                                                                                                                                                                    \label{inst:0006}
\and Astronomisches Rechen-Institut, Zentrum f\"{ u}r Astronomie der Universit\"{ a}t Heidelberg, M\"{ o}nchhofstr. 12-14, 69120 Heidelberg, Germany\relax                                                                                                                                                       \label{inst:0007}
\and Universit\'{e} C\^{o}te d'Azur, Observatoire de la C\^{o}te d'Azur, CNRS, Laboratoire Lagrange, Bd de l'Observatoire, CS 34229, 06304 Nice Cedex 4, France\relax                                                                                                                                            \label{inst:0008}
\and Institute of Astronomy, University of Cambridge, Madingley Road, Cambridge CB3 0HA, United Kingdom\relax                                                                                                                                                                                                    \label{inst:0009}
\and Department of Astronomy, University of Geneva, Chemin des Maillettes 51, 1290 Versoix, Switzerland\relax                                                                                                                                                                                                    \label{inst:0010}
\and Aurora Technology for European Space Agency (ESA), Camino bajo del Castillo, s/n, Urbanizacion Villafranca del Castillo, Villanueva de la Ca\~{n}ada, 28692 Madrid, Spain\relax                                                                                                                             \label{inst:0011}
\and Institut de Ci\`{e}ncies del Cosmos (ICCUB), Universitat  de  Barcelona  (IEEC-UB), Mart\'{i} i  Franqu\`{e}s  1, 08028 Barcelona, Spain\relax                                                                                                                                                              \label{inst:0013}
\and Lohrmann Observatory, Technische Universit\"{ a}t Dresden, Mommsenstra{\ss}e 13, 01062 Dresden, Germany\relax                                                                                                                                                                                               \label{inst:0014}
\and European Space Agency (ESA), European Space Astronomy Centre (ESAC), Camino bajo del Castillo, s/n, Urbanizacion Villafranca del Castillo, Villanueva de la Ca\~{n}ada, 28692 Madrid, Spain\relax                                                                                                           \label{inst:0015}
\and Lund Observatory, Department of Astronomy and Theoretical Physics, Lund University, Box 43, 22100 Lund, Sweden\relax                                                                                                                                                                                        \label{inst:0016}
\and CNES Centre Spatial de Toulouse, 18 avenue Edouard Belin, 31401 Toulouse Cedex 9, France\relax                                                                                                                                                                                                              \label{inst:0019}
\and Institut d'Astronomie et d'Astrophysique, Universit\'{e} Libre de Bruxelles CP 226, Boulevard du Triomphe, 1050 Brussels, Belgium\relax                                                                                                                                                                     \label{inst:0020}
\and F.R.S.-FNRS, Rue d'Egmont 5, 1000 Brussels, Belgium\relax                                                                                                                                                                                                                                                   \label{inst:0021}
\and INAF - Osservatorio Astrofisico di Arcetri, Largo Enrico Fermi 5, 50125 Firenze, Italy\relax                                                                                                                                                                                                                \label{inst:0022}
\and Laboratoire d'astrophysique de Bordeaux, Univ. Bordeaux, CNRS, B18N, all{\'e}e Geoffroy Saint-Hilaire, 33615 Pessac, France\relax                                                                                                                                                                           \label{inst:0024}
\and Max Planck Institute for Astronomy, K\"{ o}nigstuhl 17, 69117 Heidelberg, Germany\relax                                                                                                                                                                                                                     \label{inst:0027}
\and Mullard Space Science Laboratory, University College London, Holmbury St Mary, Dorking, Surrey RH5 6NT, United Kingdom\relax                                                                                                                                                                                \label{inst:0029}
\and INAF - Osservatorio Astrofisico di Torino, via Osservatorio 20, 10025 Pino Torinese (TO), Italy\relax                                                                                                                                                                                                       \label{inst:0030}
\and University of Turin, Department of Physics, Via Pietro Giuria 1, 10125 Torino, Italy\relax                                                                                                                                                                                                                  \label{inst:0033}
\and INAF - Osservatorio di Astrofisica e Scienza dello Spazio di Bologna, via Piero Gobetti 93/3, 40129 Bologna, Italy\relax                                                                                                                                                                                    \label{inst:0036}
\and DAPCOM for Institut de Ci\`{e}ncies del Cosmos (ICCUB), Universitat  de  Barcelona  (IEEC-UB), Mart\'{i} i  Franqu\`{e}s  1, 08028 Barcelona, Spain\relax                                                                                                                                                   \label{inst:0037}
\and Royal Observatory of Belgium, Ringlaan 3, 1180 Brussels, Belgium\relax                                                                                                                                                                                                                                      \label{inst:0042}
\and ALTEC S.p.a, Corso Marche, 79,10146 Torino, Italy\relax                                                                                                                                                                                                                                                     \label{inst:0048}
\and Department of Astronomy, University of Geneva, Chemin d'Ecogia 16, 1290 Versoix, Switzerland\relax                                                                                                                                                                                                          \label{inst:0051}
\and Sednai S\`{a}rl, Geneva, Switzerland\relax                                                                                                                                                                                                                                                                  \label{inst:0052}
\and Gaia DPAC Project Office, ESAC, Camino bajo del Castillo, s/n, Urbanizacion Villafranca del Castillo, Villanueva de la Ca\~{n}ada, 28692 Madrid, Spain\relax                                                                                                                                                \label{inst:0061}
\and Telespazio Vega UK Ltd for European Space Agency (ESA), Camino bajo del Castillo, s/n, Urbanizacion Villafranca del Castillo, Villanueva de la Ca\~{n}ada, 28692 Madrid, Spain\relax                                                                                                                        \label{inst:0064}
\and SYRTE, Observatoire de Paris, Universit\'{e} PSL, CNRS,  Sorbonne Universit\'{e}, LNE, 61 avenue de l’Observatoire 75014 Paris, France\relax                                                                                                                                                              \label{inst:0066}
\and National Observatory of Athens, I. Metaxa and Vas. Pavlou, Palaia Penteli, 15236 Athens, Greece\relax                                                                                                                                                                                                       \label{inst:0068}
\and IMCCE, Observatoire de Paris, Universit\'{e} PSL, CNRS, Sorbonne Universit{\'e}, Univ. Lille, 77 av. Denfert-Rochereau, 75014 Paris, France\relax                                                                                                                                                           \label{inst:0070}
\and INAF - Osservatorio Astrofisico di Catania, via S. Sofia 78, 95123 Catania, Italy\relax                                                                                                                                                                                                                     \label{inst:0072}
\and Serco Gesti\'{o}n de Negocios for European Space Agency (ESA), Camino bajo del Castillo, s/n, Urbanizacion Villafranca del Castillo, Villanueva de la Ca\~{n}ada, 28692 Madrid, Spain\relax                                                                                                                 \label{inst:0077}
\and Institut d'Astrophysique et de G\'{e}ophysique, Universit\'{e} de Li\`{e}ge, 19c, All\'{e}e du 6 Ao\^{u}t, B-4000 Li\`{e}ge, Belgium\relax                                                                                                                                                                  \label{inst:0079}
\and CRAAG - Centre de Recherche en Astronomie, Astrophysique et G\'{e}ophysique, Route de l'Observatoire Bp 63 Bouzareah 16340 Algiers, Algeria\relax                                                                                                                                                           \label{inst:0080}
\and Institute for Astronomy, University of Edinburgh, Royal Observatory, Blackford Hill, Edinburgh EH9 3HJ, United Kingdom\relax                                                                                                                                                                                \label{inst:0081}
\and ATG Europe for European Space Agency (ESA), Camino bajo del Castillo, s/n, Urbanizacion Villafranca del Castillo, Villanueva de la Ca\~{n}ada, 28692 Madrid, Spain\relax                                                                                                                                    \label{inst:0084}
\and ETSE Telecomunicaci\'{o}n, Universidade de Vigo, Campus Lagoas-Marcosende, 36310 Vigo, Galicia, Spain\relax                                                                                                                                                                                                 \label{inst:0089}
\and Universit\'{e} de Strasbourg, CNRS, Observatoire astronomique de Strasbourg, UMR 7550,  11 rue de l'Universit\'{e}, 67000 Strasbourg, France\relax                                                                                                                                                          \label{inst:0093}
\and Kavli Institute for Cosmology Cambridge, Institute of Astronomy, Madingley Road, Cambridge, CB3 0HA\relax                                                                                                                                                                                                   \label{inst:0096}
\and Department of Astrophysics, Astronomy and Mechanics, National and Kapodistrian University of Athens, Panepistimiopolis, Zografos, 15783 Athens, Greece\relax                                                                                                                                                \label{inst:0097}
\and Observational Astrophysics, Division of Astronomy and Space Physics, Department of Physics and Astronomy, Uppsala University, Box 516, 751 20 Uppsala, Sweden\relax                                                                                                                                         \label{inst:0098}
\and Leibniz Institute for Astrophysics Potsdam (AIP), An der Sternwarte 16, 14482 Potsdam, Germany\relax                                                                                                                                                                                                        \label{inst:0104}
\and CENTRA, Faculdade de Ci\^{e}ncias, Universidade de Lisboa, Edif. C8, Campo Grande, 1749-016 Lisboa, Portugal\relax                                                                                                                                                                                          \label{inst:0107}
\and Department of Informatics, Donald Bren School of Information and Computer Sciences, University of California, 5019 Donald Bren Hall, 92697-3440 CA Irvine, United States\relax                                                                                                                              \label{inst:0108}
\and Dipartimento di Fisica e Astronomia ""Ettore Majorana"", Universit\`{a} di Catania, Via S. Sofia 64, 95123 Catania, Italy\relax                                                                                                                                                                             \label{inst:0110}
\and CITIC, Department of Nautical Sciences and Marine Engineering, University of A Coru\~{n}a, Campus de Elvi\~{n}a S/N, 15071, A Coru\~{n}a, Spain\relax                                                                                                                                                       \label{inst:0113}
\and INAF - Osservatorio Astronomico di Roma, Via Frascati 33, 00078 Monte Porzio Catone (Roma), Italy\relax                                                                                                                                                                                                     \label{inst:0115}
\and Space Science Data Center - ASI, Via del Politecnico SNC, 00133 Roma, Italy\relax                                                                                                                                                                                                                           \label{inst:0116}
\and Department of Physics, University of Helsinki, P.O. Box 64, 00014 Helsinki, Finland\relax                                                                                                                                                                                                                   \label{inst:0119}
\and Finnish Geospatial Research Institute FGI, Geodeetinrinne 2, 02430 Masala, Finland\relax                                                                                                                                                                                                                    \label{inst:0120}
\and Institut UTINAM CNRS UMR6213, Universit\'{e} Bourgogne Franche-Comt\'{e}, OSU THETA Franche-Comt\'{e} Bourgogne, Observatoire de Besan\c{c}on, BP1615, 25010 Besan\c{c}on Cedex, France\relax                                                                                                               \label{inst:0125}
\and STFC, Rutherford Appleton Laboratory, Harwell, Didcot, OX11 0QX, United Kingdom\relax                                                                                                                                                                                                                       \label{inst:0127}
\and HE Space Operations BV for European Space Agency (ESA), Keplerlaan 1, 2201AZ, Noordwijk, The Netherlands\relax                                                                                                                                                                                              \label{inst:0131}
\and Dpto. de Inteligencia Artificial, UNED, c/ Juan del Rosal 16, 28040 Madrid, Spain\relax                                                                                                                                                                                                                     \label{inst:0133}
\and Applied Physics Department, Universidade de Vigo, 36310 Vigo, Spain\relax                                                                                                                                                                                                                                   \label{inst:0137}
\and Thales Services for CNES Centre Spatial de Toulouse, 18 avenue Edouard Belin, 31401 Toulouse Cedex 9, France\relax                                                                                                                                                                                          \label{inst:0143}
\and Instituut voor Sterrenkunde, KU Leuven, Celestijnenlaan 200D, 3001 Leuven, Belgium\relax                                                                                                                                                                                                                    \label{inst:0144}
\and Department of Astrophysics/IMAPP, Radboud University, P.O.Box 9010, 6500 GL Nijmegen, The Netherlands\relax                                                                                                                                                                                                 \label{inst:0145}
\vfill
\and CITIC - Department of Computer Science and Information Technologies, University of A Coru\~{n}a, Campus de Elvi\~{n}a S/N, 15071, A Coru\~{n}a, Spain\relax                                                                                                                                                 \label{inst:0151}
\and Barcelona Supercomputing Center (BSC) - Centro Nacional de Supercomputaci\'{o}n, c/ Jordi Girona 29, Ed. Nexus II, 08034 Barcelona, Spain\relax                                                                                                                                                             \label{inst:0152}
\and University of Vienna, Department of Astrophysics, T\"{ u}rkenschanzstra{\ss}e 17, A1180 Vienna, Austria\relax                                                                                                                                                                                               \label{inst:0153}
\and European Southern Observatory, Karl-Schwarzschild-Str. 2, 85748 Garching, Germany\relax                                                                                                                                                                                                                     \label{inst:0154}
\and Kapteyn Astronomical Institute, University of Groningen, Landleven 12, 9747 AD Groningen, The Netherlands\relax                                                                                                                                                                                             \label{inst:0161}
\and School of Physics and Astronomy, University of Leicester, University Road, Leicester LE1 7RH, United Kingdom\relax                                                                                                                                                                                          \label{inst:0168}
\and Center for Research and Exploration in Space Science and Technology, University of Maryland Baltimore County, 1000 Hilltop Circle, Baltimore MD, USA\relax                                                                                                                                                  \label{inst:0176}
\and GSFC - Goddard Space Flight Center, Code 698, 8800 Greenbelt Rd, 20771 MD Greenbelt, United States\relax                                                                                                                                                                                                    \label{inst:0177}
\and EURIX S.r.l., Corso Vittorio Emanuele II 61, 10128, Torino, Italy\relax                                                                                                                                                                                                                                     \label{inst:0179}
\and Harvard-Smithsonian Center for Astrophysics, 60 Garden St., MS 15, Cambridge, MA 02138, USA\relax                                                                                                                                                                                                           \label{inst:0180}
\and HE Space Operations BV for European Space Agency (ESA), Camino bajo del Castillo, s/n, Urbanizacion Villafranca del Castillo, Villanueva de la Ca\~{n}ada, 28692 Madrid, Spain\relax                                                                                                                        \label{inst:0182}
\and CAUP - Centro de Astrofisica da Universidade do Porto, Rua das Estrelas, Porto, Portugal\relax                                                                                                                                                                                                              \label{inst:0183}
\and SISSA - Scuola Internazionale Superiore di Studi Avanzati, via Bonomea 265, 34136 Trieste, Italy\relax                                                                                                                                                                                                      \label{inst:0188}
\and Telespazio for CNES Centre Spatial de Toulouse, 18 avenue Edouard Belin, 31401 Toulouse Cedex 9, France\relax                                                                                                                                                                                               \label{inst:0191}
\and University of Turin, Department of Computer Sciences, Corso Svizzera 185, 10149 Torino, Italy\relax                                                                                                                                                                                                         \label{inst:0196}
\and Dpto. de Matem\'{a}tica Aplicada y Ciencias de la Computaci\'{o}n, Univ. de Cantabria, ETS Ingenieros de Caminos, Canales y Puertos, Avda. de los Castros s/n, 39005 Santander, Spain\relax                                                                                                                 \label{inst:0199}
\and Centro de Astronom\'{i}a - CITEVA, Universidad de Antofagasta, Avenida Angamos 601, Antofagasta 1270300, Chile\relax                                                                                                                                                                                        \label{inst:0209}
\and Vera C Rubin Observatory,  950 N. Cherry Avenue, Tucson, AZ 85719, USA\relax                                                                                                                                                                                                                                \label{inst:0212}
\and Centre for Astrophysics Research, University of Hertfordshire, College Lane, AL10 9AB, Hatfield, United Kingdom\relax                                                                                                                                                                                       \label{inst:0213}
\and University of Antwerp, Onderzoeksgroep Toegepaste Wiskunde, Middelheimlaan 1, 2020 Antwerp, Belgium\relax                                                                                                                                                                                                   \label{inst:0222}
\and INAF - Osservatorio Astronomico d'Abruzzo, Via Mentore Maggini, 64100 Teramo, Italy\relax                                                                                                                                                                                                                   \label{inst:0225}
\and Instituto de Astronomia, Geof\`{i}sica e Ci\^{e}ncias Atmosf\'{e}ricas, Universidade de S\~{a}o Paulo, Rua do Mat\~{a}o, 1226, Cidade Universitaria, 05508-900 S\~{a}o Paulo, SP, Brazil\relax                                                                                                              \label{inst:0228}
\and M\'{e}socentre de calcul de Franche-Comt\'{e}, Universit\'{e} de Franche-Comt\'{e}, 16 route de Gray, 25030 Besan\c{c}on Cedex, France\relax                                                                                                                                                                \label{inst:0238}
\and SRON, Netherlands Institute for Space Research, Sorbonnelaan 2, 3584CA, Utrecht, The Netherlands\relax                                                                                                                                                                                                      \label{inst:0242}
\and Theoretical Astrophysics, Division of Astronomy and Space Physics, Department of Physics and Astronomy, Uppsala University, Box 516, 751 20 Uppsala, Sweden\relax                                                                                                                                           \label{inst:0244}
\and RHEA for European Space Agency (ESA), Camino bajo del Castillo, s/n, Urbanizacion Villafranca del Castillo, Villanueva de la Ca\~{n}ada, 28692 Madrid, Spain\relax                                                                                                                                          \label{inst:0246}
\and ATOS for CNES Centre Spatial de Toulouse, 18 avenue Edouard Belin, 31401 Toulouse Cedex 9, France\relax                                                                                                                                                                                                     \label{inst:0247}
\and School of Physics and Astronomy, Tel Aviv University, Tel Aviv 6997801, Israel\relax                                                                                                                                                                                                                        \label{inst:0250}
\and Astrophysics Research Centre, School of Mathematics and Physics, Queen's University Belfast, Belfast BT7 1NN, UK\relax                                                                                                                                                                                      \label{inst:0252}
\and Centre de Donn\'{e}es Astronomique de Strasbourg, Strasbourg, France\relax                                                                                                                                                                                                                                  \label{inst:0254}
\and Universit\'{e} C\^{o}te d'Azur, Observatoire de la C\^{o}te d'Azur, CNRS, Laboratoire G\'{e}oazur, Bd de l'Observatoire, CS 34229, 06304 Nice Cedex 4, France\relax                                                                                                                                         \label{inst:0255}
\and Max-Planck-Institut f\"{ u}r Astrophysik, Karl-Schwarzschild-Stra{\ss}e 1, 85748 Garching, Germany\relax                                                                                                                                                                                                    \label{inst:0259}
\and APAVE SUDEUROPE SAS for CNES Centre Spatial de Toulouse, 18 avenue Edouard Belin, 31401 Toulouse Cedex 9, France\relax                                                                                                                                                                                      \label{inst:0261}
\and \'{A}rea de Lenguajes y Sistemas Inform\'{a}ticos, Universidad Pablo de Olavide, Ctra. de Utrera, km 1. 41013, Sevilla, Spain\relax                                                                                                                                                                         \label{inst:0265}
\and Onboard Space Systems, Lule\aa{} University of Technology, Box 848, S-981 28 Kiruna, Sweden\relax                                                                                                                                                                                                           \label{inst:0279}
\and TRUMPF Photonic Components GmbH, Lise-Meitner-Stra{\ss}e 13,  89081 Ulm, Germany\relax                                                                                                                                                                                                                      \label{inst:0284}
\and IAC - Instituto de Astrofisica de Canarias, Via L\'{a}ctea s/n, 38200 La Laguna S.C., Tenerife, Spain\relax                                                                                                                                                                                                 \label{inst:0287}
\and Department of Astrophysics, University of La Laguna, Via L\'{a}ctea s/n, 38200 La Laguna S.C., Tenerife, Spain\relax                                                                                                                                                                                        \label{inst:0288}
\and Laboratoire Univers et Particules de Montpellier, CNRS Universit\'{e} Montpellier, Place Eug\`{e}ne Bataillon, CC72, 34095 Montpellier Cedex 05, France\relax                                                                                                                                               \label{inst:0294}
\and LESIA, Observatoire de Paris, Universit\'{e} PSL, CNRS, Sorbonne Universit\'{e}, Universit\'{e} de Paris, 5 Place Jules Janssen, 92190 Meudon, France\relax                                                                                                                                                 \label{inst:0300}
\and Villanova University, Department of Astrophysics and Planetary Science, 800 E Lancaster Avenue, Villanova PA 19085, USA\relax                                                                                                                                                                               \label{inst:0302}
\and Astronomical Observatory, University of Warsaw,  Al. Ujazdowskie 4, 00-478 Warszawa, Poland\relax                                                                                                                                                                                                           \label{inst:0308}
\and Laboratoire d'astrophysique de Bordeaux, Univ. Bordeaux, CNRS, B18N, all\'{e}e Geoffroy Saint-Hilaire, 33615 Pessac, France\relax                                                                                                                                                                           \label{inst:0312}
\and Universit\'{e} Rennes, CNRS, IPR (Institut de Physique de Rennes) - UMR 6251, 35000 Rennes, France\relax                                                                                                                                                                                                    \label{inst:0315}
\and INAF - Osservatorio Astronomico di Capodimonte, Via Moiariello 16, 80131, Napoli, Italy\relax                                                                                                                                                                                                               \label{inst:0317}
\and Niels Bohr Institute, University of Copenhagen, Juliane Maries Vej 30, 2100 Copenhagen {\O}, Denmark\relax                                                                                                                                                                                                  \label{inst:0323}
\and Las Cumbres Observatory, 6740 Cortona Drive Suite 102, Goleta, CA 93117, USA\relax                                                                                                                                                                                                                          \label{inst:0324}
\and Astrophysics Research Institute, Liverpool John Moores University, 146 Brownlow Hill, Liverpool L3 5RF, United Kingdom\relax                                                                                                                                                                                \label{inst:0330}
\and IPAC, Mail Code 100-22, California Institute of Technology, 1200 E. California Blvd., Pasadena, CA 91125, USA\relax                                                                                                                                                                                         \label{inst:0335}
\and Jet Propulsion Laboratory, California Institute of Technology, 4800 Oak Grove Drive, M/S 169-327, Pasadena, CA 91109, USA\relax                                                                                                                                                                             \label{inst:0336}
\vfill
\and IRAP, Universit\'{e} de Toulouse, CNRS, UPS, CNES, 9 Av. colonel Roche, BP 44346, 31028 Toulouse Cedex 4, France\relax                                                                                                                                                                                      \label{inst:0337}
\and Konkoly Observatory, Research Centre for Astronomy and Earth Sciences, MTA Centre of Excellence, Konkoly Thege Mikl\'{o}s \'{u}t 15-17, 1121 Budapest, Hungary\relax                                                                                                                                        \label{inst:0351}
\and MTA CSFK Lend\"{ u}let Near-Field Cosmology Research Group, Konkoly Thege Mikl\'os \'ut 15–17, 1121 Budapest, Hungary\relax                         \label{inst:0352}
\and ELTE E\"{ o}tv\"{ o}s Lor\'{a}nd University, Institute of Physics, 1117, P\'{a}zm\'{a}ny P\'{e}ter s\'{e}t\'{a}ny 1A, Budapest, Hungary\relax                                                                                                                                                               \label{inst:0353}
\and Ru{\dj}er Bo\v{s}kovi\'{c} Institute, Bijeni\v{c}ka cesta 54, 10000 Zagreb, Croatia\relax                                                                                                                                                                                                                   \label{inst:0370}
\and Institute of Theoretical Physics, Faculty of Mathematics and Physics, Charles University in Prague, Czech Republic\relax                                                                                                                                                                                    \label{inst:0374}
\and INAF - Osservatorio Astronomico di Brera, via E. Bianchi 46, 23807 Merate (LC), Italy\relax                                                                                                                                                                                                                 \label{inst:0385}
\and AKKA for CNES Centre Spatial de Toulouse, 18 avenue Edouard Belin, 31401 Toulouse Cedex 9, France\relax                                                                                                                                                                                                     \label{inst:0386}
\and Departmento de F\'{i}sica de la Tierra y Astrof\'{i}sica, Universidad Complutense de Madrid, 28040 Madrid, Spain\relax                                                                                                                                                                                      \label{inst:0390}
\and Vitrociset Belgium for European Space Agency (ESA), Camino bajo del Castillo, s/n, Urbanizacion Villafranca del Castillo, Villanueva de la Ca\~{n}ada, 28692 Madrid, Spain\relax                                                                                                                            \label{inst:0396}
\and Department of Astrophysical Sciences, 4 Ivy Lane, Princeton University, Princeton NJ 08544, USA\relax                                                                                                                                                                                                       \label{inst:0428}
\and Departamento de Astrof\'{i}sica, Centro de Astrobiolog\'{i}a (CSIC-INTA), ESA-ESAC. Camino Bajo del Castillo s/n. 28692 Villanueva de la Ca\~{n}ada, Madrid, Spain\relax                                                                                                                                    \label{inst:0434}
\and naXys, University of Namur, Rempart de la Vierge, 5000 Namur, Belgium\relax                                                                                                                                                                                                                                 \label{inst:0437}
\and EPFL - Ecole Polytechnique f\'{e}d\'{e}rale de Lausanne, Institute of Mathematics, Station 8 EPFL SB MATH SDS, Lausanne, Switzerland\relax                                                                                                                                                                  \label{inst:0445}
\and H H Wills Physics Laboratory, University of Bristol, Tyndall Avenue, Bristol BS8 1TL, United Kingdom\relax                                                                                                                                                                                                  \label{inst:0451}
\and Sorbonne Universit\'{e}, CNRS, UMR7095, Institut d'Astrophysique de Paris, 98bis bd. Arago, 75014 Paris, France\relax                                                                                                                                                                                       \label{inst:0472}
\and Porter School of the Environment and Earth Sciences, Tel Aviv University, Tel Aviv 6997801, Israel\relax                                                                                                                                                                                                    \label{inst:0473}
\and Laboratoire Univers et Particules de Montpellier, Universit\'{e} Montpellier, Place Eug\`{e}ne Bataillon, CC72, 34095 Montpellier Cedex 05, France\relax                                                                                                                                                    \label{inst:0474}
\and Faculty of Mathematics and Physics, University of Ljubljana, Jadranska ulica 19, 1000 Ljubljana, Slovenia\relax                                                                                                                                                                                             \label{inst:0475}
}

\date{Received ; accepted }

\abstract{We present the early installment of the third {\gaia} data release, \edr{3}, consisting of astrometry and
photometry for $1.8$ billion sources brighter than magnitude 21, complemented with the list of radial velocities from
\gdr{2}.}
{A summary of the contents of \edr{3} is presented, accompanied by a discussion on the differences with respect to
\gdr{2} and an overview of the main limitations which are present in the survey. Recommendations are made on the
responsible use of \edr{3} results.}
{The raw data collected with the {\gaia} instruments during the first 34 months of the mission have been processed by
the {\gaia} Data Processing and Analysis Consortium (DPAC) and turned into this early third data release, which
represents a major advance with respect to \gdr{2} in terms of astrometric and photometric precision, accuracy, and
homogeneity.}
{\edr{3} contains celestial positions and the apparent brightness in $G$ for approximately $1.8$ billion sources. For
    $1.5$ billion of those sources, parallaxes, proper motions, and the {\bpminrp} colour are also available. The
    passbands for $G$, \gbp, and \grp\ are provided as part of the release. For ease of use, the $7$ million radial
    velocities from \gdr{2} are included in this release, after the removal of a small number of spurious values. New
    radial velocities will appear as part of \gdr{3}. Finally, \edr{3} represents an updated materialisation of the
    celestial reference frame (CRF) in the optical, the \gcrf{3}, which is based solely on extragalactic sources. The
    creation of the source list for \edr{3} includes enhancements that make it more robust with respect to high proper
    motion stars, and the disturbing effects of spurious and partially resolved sources. The source list is largely the
same as that for \gdr{2}, but it does feature new sources and there are some notable changes. The source list will not
change for \gdr{3}.}
{\edr{3} represents a significant advance over \gdr{2}, with parallax precisions increased by 30 per cent, proper motion
    precisions increased by a factor of 2, and the systematic errors in the astrometry suppressed by $30$--$40$\% for
    the parallaxes and by a factor $\sim2.5$ for the proper motions. The photometry also features increased precision,
    but above all much better homogeneity across colour, magnitude, and celestial position. A single passband for
$G$, \gbp, and {\grp} is valid over the entire magnitude and colour range, with no systematics above
the 1\% level.}


\maketitle

\titlerunning{{\gaia} Early Data Release 3: Summary} 
\authorrunning{{\gaia} Collaboration}

\section{Introduction}
\label{sec:intro}

We present the first installment of the third intermediate {\gaia} data release, {\gaia} Early Data Release 3 (\edr{3}),
which is based on the data collected during the first 34 months of the mission. This early part of \gdr{3} consists of
an updated source list, astrometry, and broad band photometry in the $G$, {\gbp}, and {\grp} bands. In addition, an
updated list of radial velocities from \gdr{2}, cleaned from spurious values, is included. \edr{3} represents a
significant improvement in both the precision and accuracy of the astrometry and broad-band photometry. The factor two
improvement in proper motion precision provides new views of the fine structure of Galactic phase space. The suppression
of systematic errors enables, for the first time, a measurement in the optical of the acceleration of the solar system
barycentre with respect to the rest frame of distant extragalactic sources \citep{EDR3-DPACP-134}, which is a beautiful
confirmation of the superb astrometric quality of \edr{3}. Likewise, the photometry is significantly improved over
\gdr{2}; it is much more homogeneous over the sky, as well as over source brightness and colour, where a single passband
for $G$, \gbp, and {\grp} can now be used over the entire magnitude and colour range, with no systematics above the 1\%
level.

The full \gdr{3} release is expected in 2022 and will enrich the current release with the following: updated and new
radial velocities; astrophysical parameters for sources based on the blue and red prism photometer (BP and RP) spectra,
as well as spectra from the radial velocity spectrograph (RVS); the mean BP and RP prism- and RVS-spectra for a subset
of sources; an extended catalogue of variable stars; the first catalogue of binary stars, including eclipsing,
spectroscopic, and astrometric binaries; astrometry for a larger sample of solar system objects, and reflectance spectra
for a small subset of asteroids; quasi-stellar object (QSO) host and galaxy morphological characterisation; and the
light curves for all sources in a field centred on M31. We stress here that the source list, astrometry, and broad-band
photometry will not be updated from \edr{3} to \gdr{3}, both releases being based on the same number of input
observations.

This paper is structured as follows. In \secref{sec:dataprocessing} we provide a short overview of the improvements and
additions to the data processing that led to the production of \edr{3}. We summarise the contents of the early
installment of the third data release in \secref{sec:overview} and comment on the quality of this release in
\secref{sec:performance}. In \secref{sec:changes} we discuss the major differences between \edr{3} and \gdr{2}. In
\secref{sec:use} we comment on the completeness of \edr{3} and some of the known limitations which the user of
the data should keep in mind. Additional guidance on the use of \edr{3} is provided in \secref{sec:guidance}. In
\secref{sec:access} we provide updates to the {\gaia} data access facilities and documentation available to the
astronomical community. {\gaia} started collecting scientific data in July 2014 \citep{2016A&A...595A...1G} and is
currently in its extended mission phase, the nominal 60 month mission having been concluded on July 16, 2019.  We
conclude with a look ahead at the extended {\gaia} mission and the next data releases in \secref{sec:conclusions}.
Throughout the paper we make reference to other {\gaia} Collaboration and {\gaia} Data Processing and Analysis
Consortium (DPAC) papers that provide more details on the data processing and validation for \edr{3}. All these papers
(together with the present article) can be found in the Astronomy \& Astrophysics Special issue on \edr{3}.

\section{Data processing for \edr{3}}
\label{sec:dataprocessing}

As described in detail in \cite{2016A&A...595A...1G}, {\gaia} measurements are collected with three instruments. The
astrometric instrument collects images in {\gaia}'s white-light $G$-band (330--1050~nm); the blue (BP) and red (RP)
prism photometers collect low resolution spectrophotometric measurements of source spectral energy distributions over
the wavelength ranges 330--680 nm and 640--1050 nm, respectively; and the radial velocity spectrometer (RVS) collects
medium resolution ($R\sim11\,700$) spectra over the wavelength range 845--872 nm centred on the Calcium triplet region
\citep{2018A&A...616A...5C}. 

We repeat here for convenience the way events on board {\gaia}, including the data collection, are timed. The times are
given in terms of the on board mission timeline (OBMT) which is generated by the {\gaia} on board clock. By convention
OBMT is expressed in units of six hour ($21\,600$~s) spacecraft revolutions \citep{2016A&A...595A...1G}. The approximate
relation between OBMT (in revolutions) and barycentric coordinate time (TCB, in Julian years) at {\gaia} is
\begin{equation}
  \text{TCB} \simeq \text{J}2015.0 + (\text{OBMT} - 1717.6256~\text{rev})/(1461~\text{rev yr}^{-1})\,.
\end{equation}
The 34 month time interval covered by the observations used for \edr{3} starts at OBMT 1078.3795~rev = J2014.5624599~TCB
(approximately July~25, 2014, 10:30:00~UTC), and ends at OBMT 5230.0880~rev = J2017.4041495~TCB (approximately
May~28, 2017, 08:45:00~UTC). This time interval contains gaps caused by both spacecraft events and by on-ground data
processing problems. This leads to gaps in the data collection or stretches of time over which the input data cannot be
used. Which data are considered unusable varies across the {\gaia} data processing systems (here astrometry and
photometry), and as a consequence the effective amount of input data used differs from one system to the other. We refer
to the specific \edr{3} data processing papers (listed below) for the details.

The pre-processing for all {\gaia} instruments is described in \cite{2018A&A...616A..15H} and includes the removal of
the effects of non-uniformity of the charge-coupled device (CCD) bias levels. A summary of the entire data processing
system for {\gaia} is given in \cite{2016A&A...595A...1G}. The sub-sections below summarise the major improvements in
the data processing for \edr{3} with respect to \gdr{2}.

\subsection{Source list}
\label{sec:sourcelist}

A given data processing cycle for {\gaia} starts with the creation of the list of sources that will be treated. The
series of CCD measurements recorded as a source travels across the focal plane \citep[referred to collectively as a
``transit'',][]{EDR3-DPACP-124}, are grouped and assigned to known sources on the sky or to newly ``created'' sources,
corresponding to groups of transits at a celestial position where previously no source was catalogued. The starting
point for creating the source list is the previous {\gaia} data release, or the Initial {\gaia} Source List
\citep{2014A&A...570A..87S} in the case of \gdr{1}. As pointed out in \cite{2018A&A...616A...1G} the source list may
evolve from one release to the next due to the merging of groups of transits previously assigned to two or more sources,
the splitting of a group of transits into two or more sources, or changing the list of transits assigned to a source.
The changes in the source list from \gdr{1} to \gdr{2} were significant but from \gdr{2} to \edr{3} the source list has
largely stabilised, the changes being at the 2--3 per cent level overall. 

A full account of how the source list is created for \edr{3} (and \gdr{3}) can be found in \cite{EDR3-DPACP-124}, who
note the following significant improvements with respect to \gdr{2}. The identification of spurious on-board detections
(caused among others by bright star diffraction spikes, bright cosmic rays, or major planets in the solar system
transiting across or near a telescope field of view) has been improved which leads to a much cleaner list of sources and
associated transits.

The algorithm that groups together transits and assigns them to sources has been improved with respect to the treatment
of high-proper motion stars and variable stars. High proper motion stars are seen as groups of observations stretched
out over the sky which were mistaken for multiple sources in previous releases. They are now reliably recognised as
belonging to the same source. The grouping of transits and their association to sources contains a stage where the
magnitude of the source observed during a transit is taken into account. For highly variable sources this can lead to a
splitting of the transits over multiple sources. This type of error is now prevented through a post-processing step
which can recognise clusters of detections very close together on the sky, but disjoint in time, as belonging to the
same (variable) source.

A more comprehensive analysis and cleaning of the observation-to-source matching results led to less sources with highly
significant negative parallaxes or too large parallaxes \citep[see appendix C in][]{2018A&A...616A...2L}, which also
removes spuriously high proper motions. The treatment of close source pairs was improved to deal with the pairs with
separations below $400$~mas which were erroneously considered duplicate sources in \gdr{2}. These now appear as two
sources in \edr{3}. The separation limit below which two sources are considered duplicates was lowered to $180$~mas.

For quantitative information on the above please refer to \cite{EDR3-DPACP-124}. We stress here that the source list
for \edr{3} and \gdr{3} will be the same and thus the process described in \cite{EDR3-DPACP-124} applies to both
releases.

\subsection{Astrometric data processing improvements}

The astrometric processing for \edr{3} is described in \cite{EDR3-DPACP-128}, who note the following major improvements
with respect to the processing for \gdr{2}. The basic inputs for the Astrometric Global Iterative Solution
\citep[AGIS;][]{Lindegren2012} are the source image locations in the {\gaia} CCD pixel stream, translated to observation
times and across-scan locations. The image locations are determined in a process called ``image parameter determination''
\citep[IPD;][]{EDR3-DPACP-73}, by fitting a point spread function (PSF) or line spread function (LSF) to the 2D or 1D
observation windows containing the image samples. The modelling of the PSF and LSF has been very much improved with the
introduction of time and source colour dependencies, among many other enhancements. The details are described in
\cite{EDR3-DPACP-73}. It is shown in \cite{EDR3-DPACP-128} that for sources fainter than $G=13$ the chromatic effects on
image locations are almost completely removed during the image parameter determination stage, thus mostly eliminating an
important source of systematic errors. Eventually for future {\gaia} data releases this should remove the need for
colour terms in the astrometric calibrations. The colour of a source is included in the form of the effective wave
number {\nueff}, which is derived from the flux as a function of wavelength in the BP and RP prism spectra.

For the first time the image parameter determination and astrometric solution were iterated. Specifically, after the
first IPD run an AGIS solution was produced, which was then used to improve the PSF and LSF calibrations for a second
round of IPD (which benefits from improved source positions from the first AGIS run), followed by the \edr{3} production
run of AGIS. This leads to a more self-consistent set of image locations and source astrometric parameters. In addition,
the image fluxes estimated as part of the IPD are improved, which in turn improves the $G$-band photometric processing.
A diagram illustrating this data processing flow can be found in \cite{EDR3-DPACP-128}.

The astrometric calibration model was improved and extended. Several new dependencies were introduced to better handle
the locations of saturated images, the effects of charge transfer inefficiency, and imperfections in the PSF and LSF
models that leave residual effects at the sub-pixel level, and as a function of the rate at which sources move across
the telescope field of view, perpendicular to the scan direction (caused by the spin axis precession of {\gaia}).

The \edr{3} astrometric processing includes a model (Velocity error and effective Basic Angle Calibration, VBAC) that
can calibrate out the effects of spin-related instrument distortions, in particular distortions over time scales
comparable to the six hour spin period of {\gaia}. An important component of these distortions are the basic angle
variations, of which the term proportional to the cosine of the spacecraft spin phase $\Omega$ can lead to a global
parallax bias if left uncorrected \citep{Butkevich2017}. The VBAC model introduces additive corrections to the basic
angle variation corrections calculated on the basis of the basic angle monitor measurements
\cite[cf.][]{2018A&A...616A...2L}. \cite{EDR3-DPACP-128} describe the successful attempt to fit the coefficient of the
$\cos\Omega$ term, which leads to a reduction in the size of the overall parallax zero point, which for \edr{3} is
$-17$~\muas\ (compared to $-29$~\muas\ for \gdr{2}), as estimated from quasar parallaxes.

An additional calibration model, which handles telescope focal length and optical distortion variations over smaller
scales than handled by VBAC, was introduced to further reduce instrument distortion related systematics. Lastly, an
ad-hoc correction was introduced to ensure that the bright star reference frame has no net spin with respect to the
reference frame defined by quasars, an issue that was described in detail in \cite{2018A&A...616A...2L} and
\cite{2020A&A...633A...1L, 2020A&A...637C...5L}.

Overall the \edr{3} astrometry shows a reduction in the median uncertainties at $G=15$ by a factor $0.71$ for the positions
and parallaxes and $0.44$ for the proper motions. At the bright end ($G<12$) the gain is larger, a factor $0.43$ for
positions and parallaxes and $0.27$ for the proper motions, thanks to much improved calibrations. The overall parallax
zero point has improved as mentioned above, and the variance over the sky in the systematic errors (as estimated
from quasars) has been reduced by 30--40\% for the parallaxes and by a factor of $\sim2.5$ for proper motions.
\cite{EDR3-DPACP-128, EDR3-DPACP-132} provide many more details (see also \secref{sec:astrometryusage}). 

\subsection{Photometric data processing improvements}

The photometric data processing for \edr{3} is described in \cite{EDR3-DPACP-117}, where only the processing for the
broad band photometry is considered. The processing and calibration of the spectra will be described in forthcoming
papers \citep{EDR3-DPACP-118, EDR3-DPACP-119, EDR3-DPACP-120}. The BP and RP spectra are still being validated
internally to DPAC at the time of the \edr{3} release, by employing them in the astrophysical characterisation of
sources. The spectra will be published as part of \gdr{3} (for a subset of sources only). \cite{EDR3-DPACP-117} describe
the following improvements to the broad-band photometric processing.

The $G$-band photometry benefits from the improvements implemented for the astrometric instrument image parameter
determination stage. As described in \cite{EDR3-DPACP-73}, this includes a better PSF and LSF modelling, better
treatment of saturated images, the masking out of suspected disturbing sources and a more precise determination of the
background flux for each observation window. This leads to more accurate and robust $G$-band flux estimates.

The broad band photometry benefits from a detailed evaluation of the observations potentially affected by neighbouring
sources in crowded fields. Although the crowding effects were not corrected, the crowding evaluation led to a cleaner
list of internal calibration sources. The background flux in BP and RP due to stray light and astronomical sources is
better determined, with higher spatial resolution to follow smaller scale variations.

\begin{table}[t!]
  \caption{Number of sources of a certain type, or the number of sources for which a given data product is
  available in \edr{3}.}
  \label{tab:edr3stats}
  \centering
  \begin{tabular}{lr}
    \hline\hline
    \noalign{\smallskip}
    Data product or source type & Number of sources \\
    \noalign{\smallskip}
    \hline
    \noalign{\smallskip}
    Total & \edrtotal \\
    \noalign{\smallskip}
    5-parameter astrometry & \edrfiveptot \\
    6-parameter astrometry & \edrsixptot \\
    2-parameter astrometry & \edrtwoptot \\
    \noalign{\smallskip}
    \gcrf{3} sources & \edrgcrftot \\
    ICRF3 sources for frame orientation & \edricrfconsidered \\
    \gcrf{3} sources for frame spin & \edrspinconsidered \\
    \noalign{\smallskip}
    $G$-band & \edrwithgtot \\
    \gbp-band & \edrwithbptot \\
    \grp-band & \edrwithrptot \\
    \noalign{\smallskip}
    Radial velocity & \edrvradtot \\
    \noalign{\smallskip}
    \hline
  \end{tabular}
\end{table}

The range of time over which observations free from telescope throughput losses \citep[due to
contamination;][]{2016A&A...595A...1G} are available was much extended. This allowed for better sky coverage of internal
calibrator sources and more flexibility to select an optimal set of calibrators, well distributed in colour and
magnitude. The external calibration used to determine the passbands is based on a much larger set of calibrators,
covering a wider spectral range, where in \gdr{2} only the spectrophotometric standard stars \citep{2012MNRAS.426.1767P}
were used which limited the aspects of the passbands that could be determined reliably.

These improvements, and the larger set of input observations, have led to broad-band photometry which is
significantly better than \gdr{2} photometry in both precision and accuracy. Especially at the bright end ($G<13$) large
gains were made, and many of the systematic effects reported for \gdr{2}, such as imprints from the zodiacal light and
the scan law features \citep{2018A&A...616A...4E, 2018A&A...616A..17A}, have been removed or greatly suppressed
\citep[cf.][]{EDR3-DPACP-126}. In addition the problem that two passbands are needed to describe the \gdr{2} $G$-band
photometry \citep{2018A&A...619A.180M} has been resolved, with only one passband needed for \edr{3} for each of $G$,
\gbp, and \grp.

Despite the improvements in the astrometry and photometry, several limitations remain in \edr{3} which require taking
care when using the data. In \secref{sec:use} we summarise the known limitations of the present {\gaia} data release and
point out, where relevant, the causes. In \secref{sec:guidance}, and in \cite{EDR3-DPACP-128} (astrometry),
\cite{EDR3-DPACP-117} (photometry), and \cite{EDR3-DPACP-126} (catalogue validation) we provide additional guidance on
the use of \edr{3} results. The reader is strongly encouraged to read these papers and the online
documentation\footnote{\url{https://gea.esac.esa.int/archive/documentation/GEDR3/index.html}} to understand the
    limitations in detail.

\section{Overview of the contents of \edr{3}}
\label{sec:overview}

\begin{figure*}[t]
  \sidecaption
  \includegraphics[width=12cm]{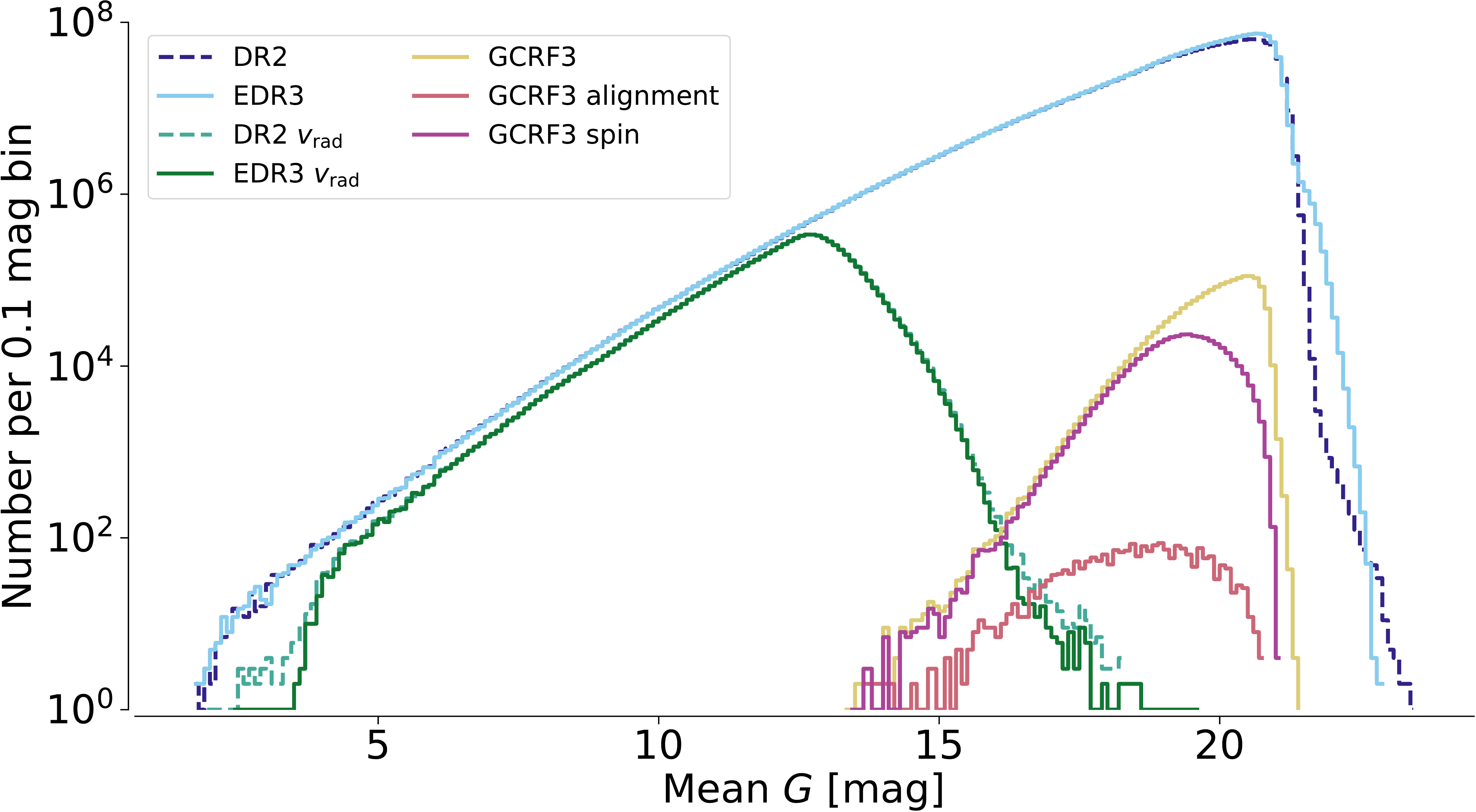}
  \caption{Distribution of the mean values of $G$ for all \edr{3} sources shown as histograms with $0.1$~mag wide bins.
  The distribution of the \gdr{2} sources (dashed lines, for the full catalogue and for the radial velocity sample) is
  included for comparison. The other histograms are for the main \edr{3} components as indicated in the
  legend.\label{fig:gmaghistos}}
\end{figure*}

\edr{3} contains new astrometry and broad-band photometry, as well as radial velocities from \gdr{2}.  Basic statistics
on the source numbers and the overall distribution in $G$ can be found in \tabsref{tab:edr3stats} and
\ref{tab:edr3magperc}. The overall quality of \edr{3} results in terms of the typically achieved uncertainties is
summarised in \tabref{tab:qualitystats}. The contents of the main components of the release, of which the magnitude
distributions are shown in \figsref{fig:gmaghistos} and \ref{fig:gmaghistoastrometry}, are summarised in the following.

\begin{table}[t!]
  \caption{Distribution of the \edr{3} sources in $G$-band magnitude.}
  \label{tab:edr3magperc}
  \centering
  \begin{tabular}{lrrrrr}
    \hline\hline
    \noalign{\smallskip}
    \multicolumn{5}{c}{Magnitude distribution percentiles ($G$)} \\
    Percentile & All & 5-p & 6-p & 2-p & \vrad \\
    \noalign{\smallskip}
    \hline
    \noalign{\smallskip}
    $0.135$\%  & 11.7 & 10.6 & 15.1 & 15.7 &  6.7 \\
    $2.275$\%  & 15.1 & 13.7 & 17.4 & 18.8 &  9.2 \\
    $15.866$\% & 17.9 & 16.3 & 18.9 & 20.1 & 11.2 \\
    $50$\%     & 19.7 & 18.1 & 19.9 & 20.7 & 12.4 \\
    $84.134$\% & 20.6 & 19.4 & 20.5 & 21.0 & 13.2 \\
    $97.725$\% & 21.0 & 20.4 & 20.8 & 21.2 & 14.1 \\
    $99.865$\% & 21.5 & 20.8 & 20.9 & 21.7 & 15.1 \\[3pt]
    \noalign{\smallskip}
    \hline
  \end{tabular}
  \tablefoot{The distribution percentiles are shown for all sources and for those with a 5-p, 6-p, and 2-p astrometric
  solution, respectively, as well as the sources for which a radial velocity is available in \edr{3}.}
\end{table}

\subsection{Astrometric data set}
\label{sec:astrometry}

The astrometric data in \edr{3} comprises three subsets:

\textsf{\emph{5-parameter solutions (5-p).}} For these sources the colour information ({\nueff} derived during the processing
for \gdr{2}) was of high enough quality to assume that any chromatic effects were removed during the IPD stage, thus
allowing for a standard 5-parameter ($\alpha$, $\delta$, $\varpi$, \pmra, \pmdec) astrometric solution.

\textsf{\emph{6-parameter solutions (6-p).}} For these sources no colour information of sufficient quality was available, thus
forcing an estimate of the image locations with a PSF or LSF model for a default source colour.  This means that
chromatic effects have to be accounted for in the astrometric solution by estimating {\nueff} for the source along with
the astrometric parameters. Thus for these sources the 5 astrometric parameters and the so-called pseudo-colour are
published along with a $6\times6$ covariance matrix listing, in addition to the astrometric covariance matrix entries,
the uncertainty on the pseudo-colour and the correlations between the estimated colour and the astrometric parameters.

\textsf{\emph{2-parameter solutions (2-p).}} As for previous releases there are sources for which insufficient data of the
required quality is available to justify publishing a full 5-p or 6-p solution. For these sources a 5- or 6-parameter
solution is in fact made, but with so-called galactic priors on the parallaxes and proper motions \citep[the fall-back
solution]{2015A&A...583A..68M, 2018A&A...616A...2L}. Only the positions and their covariance matrix are published for
these sources. In principle all sources at $G>21$ have a 2-p solution.  However this boundary is not strict because the
limit in $G$ was decided on the basis of the \gdr{2} value for the magnitude or that based on the on-board estimate.
Hence there are 5-p and 6-p solutions at $G>21$, and sources with 2-p solutions at $G\leq21$ for which in principle a
5-p or 6-p solution was possible. See \cite{EDR3-DPACP-128} for details on how the decision was taken to use the
fall-back solution.

\begin{table*}[t]
  \caption{Basic performance statistics for \edr{3}.\label{tab:qualitystats}}
  \centering
  \begin{tabular}{lr}   
    \hline\hline        
    \noalign{\smallskip}
    Data product or source type                                               & Typical uncertainty\\
    \noalign{\smallskip}
    \hline
    \noalign{\smallskip}
    Five-parameter astrometry (position)                                      & $0.01$--$0.02$~mas at $G<15$\\
                                                                              & $0.05$~mas at $G=17$\\
                                                                              & $0.4$~mas at $G=20$\\
                                                                              & $1.0$~mas at $G=21$\\
    \noalign{\smallskip}
    Five-parameter astrometry (parallax)                                      & $0.02$--$0.03$~mas at $G<15$\\
                                                                              & $0.07$~mas at $G=17$\\
                                                                              & $0.5$~mas at $G=20$\\
                                                                              & $1.3$~mas at $G=21$\\
    \noalign{\smallskip}
    Five-parameter astrometry (proper motion)                                 & $0.02$--$0.03$~mas~yr$^{-1}$ at $G<15$\\
                                                                              & $0.07$~~mas~yr$^{-1}$ at $G=17$\\
                                                                              & $0.5$~mas~yr$^{-1}$ at $G=20$\\
                                                                              & $1.4$~mas~yr$^{-1}$ at $G=21$\\
    \noalign{\smallskip}
    Six-parameter astrometry (position)                                       & $0.02$--$0.03$~mas at $G<15$\\
                                                                              & $0.08$~mas at $G=17$\\
                                                                              & $0.4$~mas at $G=20$\\
                                                                              & $1.0$~mas at $G=21$\\
    \noalign{\smallskip}
    Six-parameter astrometry (parallax)                                       & $0.02$--$0.04$~mas at $G<15$\\
                                                                              & $0.1$~mas at $G=17$\\
                                                                              & $0.5$~mas at $G=20$\\
                                                                              & $1.4$~mas at $G=21$\\
    \noalign{\smallskip}
    Six-parameter astrometry (proper motion)                                  & $0.02$--$0.04$~mas~yr$^{-1}$ at $G<15$\\
                                                                              & $0.1$~~mas~yr$^{-1}$ at $G=17$\\
                                                                              & $0.6$~mas~yr$^{-1}$ at $G=20$\\
                                                                              & $1.5$~mas~yr$^{-1}$ at $G=21$\\
    \noalign{\smallskip}
    Two-parameter astrometry (position only)                                  & $1$--$3$~mas\\
    \noalign{\smallskip}
    Systematic astrometric errors (averaged over the sky)                     & $<0.05$~mas\\
    \noalign{\smallskip}
    \gcrf{3} alignment with ICRF                                              & $0.01$~mas at $G=19$\\
    \gcrf{3} rotation with respect to ICRF                                    & $<0.01$~mas~yr$^{-1}$ at $G=19$\\
    \noalign{\smallskip}
    Mean $G$-band photometry                                                  & $0.3$~mmag at $G<13$\\
                                                                              & $1$~mmag at $G=17$\\
                                                                              & $6$~mmag at $G=20$\\
    \noalign{\smallskip}
    Mean \gbp-band photometry                                                 & $0.9$~mmag at $G<13$\\
                                                                              & $12$~mmag at $G=17$\\
                                                                              & $108$~mmag at $G=20$\\
    \noalign{\smallskip}
    Mean \grp-band photometry                                                 & $0.6$~mmag at $G<13$\\
                                                                              & $6$~mmag at $G=17$\\
                                                                              & $52$~mmag at $G=20$\\
    \hline
  \end{tabular}
  \tablefoot{The astrometric uncertainties and the \gcrf{3} alignment and rotation limits refer to epoch J2016.0 TCB.}
\end{table*}

\begin{figure}[t]
  \includegraphics[width=\linewidth]{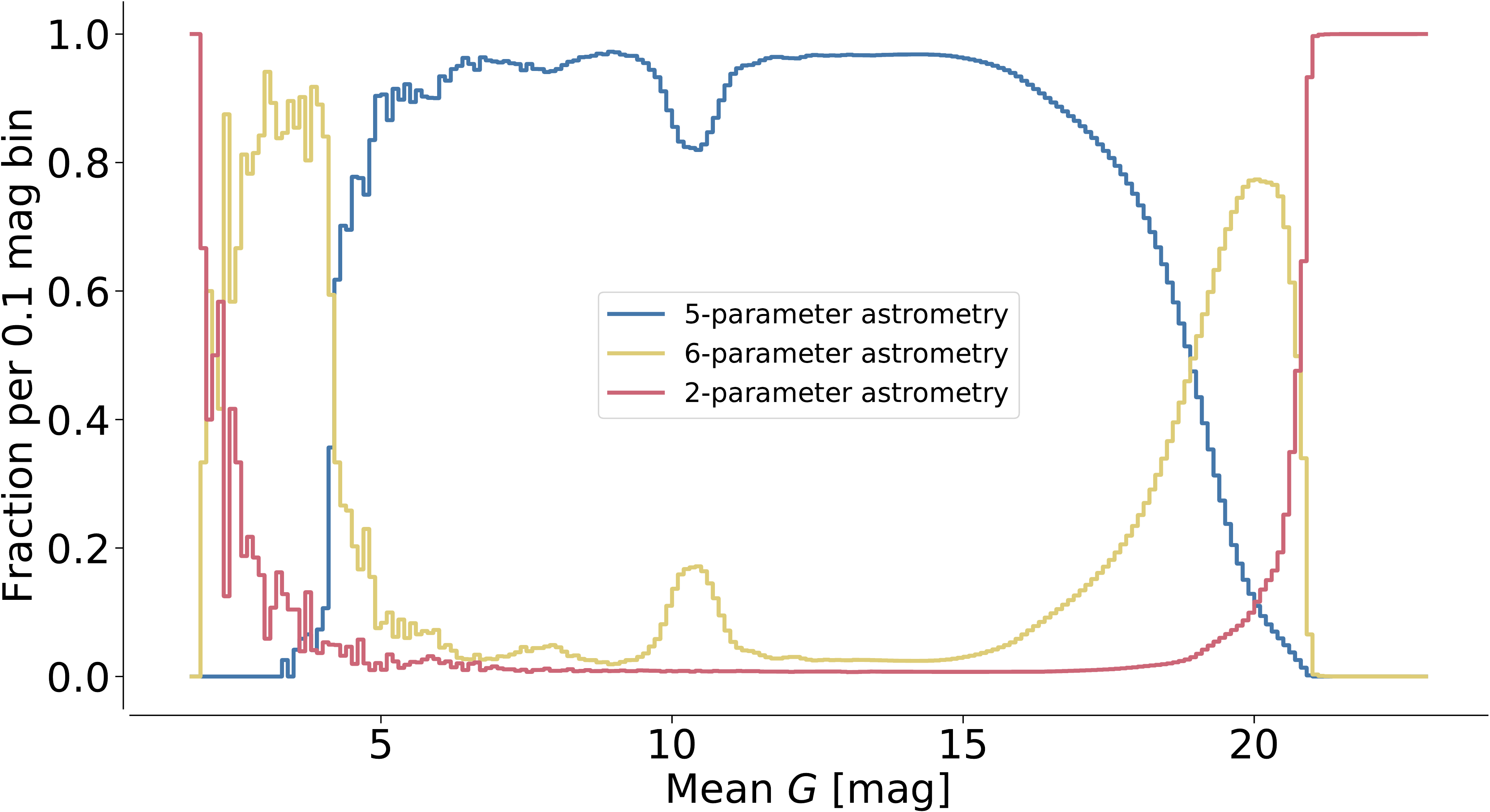}
  \caption{Distribution of the fraction of astrometric solution types as a function of $G$-band
  magnitude.\label{fig:gmaghistoastrometry}}
\end{figure}

The three solution types can be identified through the \texttt{astrometric\_params\_solved} field in the \edr{3} archive
which equals 3, 31, and 95, respectively for 2-p, 5-p, and 6-p astrometric solutions. \figref{fig:gmaghistoastrometry}
shows the distribution of the three solution types over magnitude. We note the prevalence of 6-p solutions at $G<4$ and the
relative increase in 6-p solutions around $G=11$. In both cases this reflects the source list evolution at these
magnitudes, where for a large fraction of sources the change in source identifier (ID) prevented looking up the colour
calculated for the corresponding source in \gdr{2}. The 100\% fraction of sources with 2-p solutions at $G>21$ is by
construction \citep{EDR3-DPACP-128}.

Along with the astrometry, new data quality indicators are published as part of \edr{3}. The renormalised unit weight
error, introduced after \gdr{2} was published
\citep[RUWE;][]{ruwe}\footnote{\url{https://www.cosmos.esa.int/web/gaia/dr2-known-issues\#AstrometryConsiderations}}, is
now part of the release. New quality indicators, related to the image parameter determination process, provide indications
whether a source is one of a close pair (possibly a binary) or whether the data suffers from nearby disturbing sources.
The indicators are as follows:

\textsf{\emph{ipd\_gof\_harmonic\_amplitude.}} This statistic measures the amplitude of the variation of the image parameter
determination goodness of fit (reduced $\chi^2$) as function of the position angle of the scan direction. A large
amplitude indicates that the source is double, in which case the phase (next item) indicates the position angle of the
pair, modulo 180 degrees.

\textsf{\emph{ipd\_gof\_harmonic\_phase.}} This statistic measures the phase of the variation of the IPD goodness of fit as
function of the position angle of the scan direction

\textsf{\emph{ipd\_frac\_multi\_peak.}} This field provides information on the observation windows used for the astrometric
processing of this source. It provides the fraction of windows for which the IPD algorithm has identified a double peak,
meaning that the detection may be a resolved double star (either an optical pair or a physical binary).

\textsf{\emph{ipd\_frac\_odd\_win.}} Percentage of transits with truncated windows or multiple gates applied to a window. A
high percentage indicates that a source is disturbed due to nearby sources in a crowded field or due to a nearby bright
($G<13$) source.

The reference epoch for all sources is J2016.0 (TCB). This epoch near the middle of the observation interval included in
\edr{3} was chosen to minimise correlations between the position and proper motion parameters. This epoch is $0.5$~year
later than the reference epoch for \gdr{2}, which should be accounted for when comparing source positions between the
two releases.

As in previous releases all sources were treated as single stars when solving for the astrometric parameters. For a
binary the parameters may thus refer to either component, or to the photocentre of the system, and the proper motion
represents the mean motion of the component, or photocentre, over the 2.8~years of data included in the solution.
Depending on the orbital motion, this could be significantly different from the proper motion of the same object in
\gdr{2}, and significantly different from the proper motion of the centre of mass of the binary.

The positions and proper motions are provided with respect to the third realisation of the {\gaia} celestial reference
frame (\gcrf{3}) which is aligned with the International Celestial Reference Frame (ICRF) to about $0.01$~mas
root-mean-square (RMS) at epoch J2016.0 (TCB), and non-rotating with respect to the ICRF to within $0.01$~\masyr\ RMS.
The \gcrf{3} is materialised by {\edrgcrftot} QSOs and aligned to the ICRF through a subset of {\edricrfconsidered}
QSOs. The construction and properties of the \gcrf{3} and the comparison to the ICRF are described in
\cite{EDR3-DPACP-133}.

\subsection{Photometric data set}
\label{sec:photometry}

The photometric data set contains the broad band photometry in the $G$, {\gbp}, and {\grp} bands. The mean value of the
$G$-band fluxes is reported for practically all sources while for about 85 per cent of the sources the mean values of
the {\gbp} and {\grp} fluxes are provided. For a small fraction of the sources any of the three bands may be missing
(see \secref{sec:photometrylimitations}). As for \gdr{2}, the photometric data processing consisted of three categories,
``Gold'', ``Silver'', and ``Bronze'', which represent decreasing quality levels of the photometric calibration achieved, where
in the case of the Bronze sources no colour information is available. The photometric processing category of each source
is indicated in the released catalogue by a numeric field (\texttt{phot\_proc\_mode}) assuming values 0, 1, and 2 for
gold, silver, and bronze sources respectively. At the bright end the photometric uncertainties are dominated by
calibration effects which are estimated to contribute $2.0$, $3.1$, and $1.8$ mmag RMS per CCD observation, respectively
for $G$, {\gbp}, and {\grp} \citep{EDR3-DPACP-117}.

Two new data quality indicators are included with the photometry which allow filtering of sources according to the
probability that their photometry is affected by crowding effects. The fields \texttt{phot\_bp\_n\_blended\_transits}
and \texttt{phot\_bp\_n\_contaminated\_transits} (and similar for RP) indicate the number of transits for a given source
believed to be affected by ``blending'' or ``contamination''. The former refers to the presence of another source in the
same observation window or very nearby, and the latter refers to the presence of sources in the wider neighbourhood of
the target source, which are bright enough to contribute flux to the observation window. The ratio between the numbers
in these fields and the total number of transits (\texttt{phot\_bp\_n\_obs} or \texttt{phot\_rp\_n\_obs}) can be used to
identify sources for which the photometry is possibly less reliable \citep[see][for a more detailed
discussion]{EDR3-DPACP-117}.

\subsection{Radial velocity data set}
\label{sec:radialvelocities}

The radial velocity spectrograph data processing relies on the preliminary astrometric solution \citep[AGIS-3.1
in][]{EDR3-DPACP-128} in order to have sufficiently accurate source positions to fix the correct wavelength scale of the
RVS spectra. This means that the RVS processing can only start later during a processing cycle. In the planning of the
\edr{3} and \gdr{3} releases it was not possible to accommodate the full sequence of RVS processing, radial velocity
estimation, and validation, to allow new radial velocities to be published as part of \edr{3}. These will instead appear
with \gdr{3} in 2022. At the time of writing, the \gdr{3} RVS processing is finished and the results are undergoing
internal validation through the use of epoch radial velocities in the non-single star pipeline, and of RVS spectra in
the astrophysical characterisation of {\gaia} sources.

In order to keep \edr{3} maximally useful it was decided to copy \gdr{2} radial velocities to this release. The
evolution of the source list \citep{EDR3-DPACP-124} necessitated a careful tracing of \gdr{2} sources to their \edr{3}
counterparts, before assigning the DR2 radial velocities to the latter. The opportunity was used to also clean up the
list of radial velocities from the potentially spurious values identified by \cite{2019MNRAS.486.2618B}. The process is
described in detail in \cite{EDR3-DPACP-121}. The result is that {\edrvradtot} out of {\gdrtwovradnum} \gdr{2} radial
velocities were transferred to \edr{3}, where 97\% of the \edr{3} sources with a radial velocity have the same source ID
as in \gdr{2}. The radial velocities that were discarded correspond to sources that could not be traced to \edr{3} or
were shown, from a comparison with \gdr{3} preliminary radial velocities and investigations of the raw observations, to
have unreliable or erroneous radial velocities. The histograms presented in \cite{EDR3-DPACP-121} show that in particular in
the tails of the distribution (at $|\vrad|>600$~\kms) radial velocities have been removed. It is also noteworthy that of
the $\sim70\,000$ potentially spurious radial velocities identified by \cite{2019MNRAS.486.2618B}, 96\% were retained as
reliable in \edr{3}. We note that all radial velocity related fields in the \edr{3} archive are prefixed with
``\texttt{dr2\_}'', leading to \texttt{dr2\_radial\_velocity}, \texttt{dr2\_radial\_velocity\_error}, etc. For the
detailed characteristics of the radial velocity data set (precision, accuracy, limitations) please refer to the relevant
\gdr{2} papers \citep{2018A&A...616A...6S, 2019A&A...622A.205K, 2018A&A...616A...1G}.

The distribution of the astrometric, photometric, and radial velocity data sets in $G$ is shown in
\figref{fig:gmaghistos}, where for comparison the distribution for \gdr{2} is also shown. We note the improved
completeness at the faint end, at magnitudes close to $G=21$. With respect to \gdr{2} there are noticeably fewer radial
velocities at $G>15$ and at $G<4$, the latter due to the source list evolution at the bright end. The distribution of
the \gcrf{3} sources shows a sharp drop at $G=21$ which reflects that only QSOs at $G<21$ were used for the construction
of the reference frame. The magnitude distribution is also shown for the \gcrf{3} sources used for ensuring that the
spin of the reference frame is zero, and for the \gcrf{3} sources used for the alignment of the reference frame
(\edricrfconsidered\ ICRF3 sources with counterparts in \edr{3}).

\section{Scientific performance and potential of \edr{3}}
\label{sec:performance}

\edr{3} is accompanied by four papers that provide basic demonstrations of the scientific quality of the results
included in this release. The following topics are treated. The \edr{3} proper motions of quasars reveal a systematic
pattern that can be ascribed to the acceleration of the solar system barycentre with respect to the rest frame of
distant extragalactic sources. The value and direction of the acceleration can be determined from these data. That this
measurement is now possible is testament to the much improved quality of the astrometry, in particular thanks to the
suppression of systematic errors \citep{EDR3-DPACP-134}. The {\gaia} catalogue of nearby stars presents and
characterises a carefully selected sample of sources located within 100~pc from the sun \citep{EDR3-DPACP-81}. A study
of the Galactic anti-centre region illustrates the increased richness in phase space unveiled by the more precise
\edr{3} proper motions \citep{EDR3-DPACP-113}. The structure and properties of the Magellanic Clouds are studied in
\cite{EDR3-DPACP-109}.

We note in the following a couple of specific areas of improvement in the \edr{3} astrometry and photometry compared to
\gdr{2}. For further insights into the increased data quality see \cite{EDR3-DPACP-126} and the papers cited above.

The overall completeness of the catalogue at the faint end has increased somewhat as can be appreciated from
\figref{fig:gmaghistos}. At the bright end \edr{3} has a similar level of incompleteness as \gdr{2}. The spatial
resolution of \edr{3} has slightly improved with respect to \gdr{2}. This can be seen from the number counts of source
pairs as a function of their separation presented in \cite{EDR3-DPACP-126}. The counts drop below the expected line for a
random source distribution at $\sim1.5$~arcsec ($2.2$~arcsec for \gdr{2}). The improved resolution can also be seen in
the increased completeness of visual pairs from the Washington Double Star catalogue \citep{EDR3-DPACP-126}. The
completeness in close source pairs decreases rapidly below about $0.7$~arcsec which is to be expected as the treatment
of sources in crowded fields has not fundamentally changed between \gdr{2} and \edr{3}, even if the crowding effects were
better characterised during the data processing. 

The better treatment of high proper motion stars at the source list creation stage and the generally cleaner
source list has led to a much more reliable sample of high proper motion stars in \edr{3}. This is demonstrated in
\cite{EDR3-DPACP-126} where the diagram of proper motion vs.\ parallax shows a striking improvement from \gdr{2} to
\edr{3}, with most of the high proper motion stars now located on the positive parallax side of the $500$~\kms\
tangential velocity locus. The suppression of spurious parallaxes has also removed a lot of unrealistically high proper
motions for stars with negative parallaxes.

\section{Changes from \gdr{2} to \edr{3}}
\label{sec:changes}

In \cite{2018A&A...616A...1G} it was emphasised that \gdr{2} should be treated as independent from \gdr{1} due to the
evolution of the source list and the photometric system. This still holds when using \edr{3} instead of \gdr{2}, the two
releases should be treated as independent and in particular the user of the data is warned against making source by
source comparisons between the two releases. Comparison should only be done at the statistical level for well defined
samples of sources.

We repeat here the change in reference epoch from J2015.5 for \gdr{2} to J2016.0 for \edr{3}, which should be taken into
account when propagating \edr{3} source positions into the future or past. The photometric system has changed in terms
of the better internal calibration leading to much smaller magnitude and colour terms, and a new set of passbands is
presented in \cite{EDR3-DPACP-117}. Synthetic photometry for the prediction of {\gaia} observations should be updated
with the new passbands.

As stressed in \cite{EDR3-DPACP-124} the source lists of the {\gaia} data releases should be treated as completely
independent. Although extensive efforts are made to ensure that a physical source retains its identifier across
releases, changes in the identifier associated to a source will occur in a small fraction of cases.  Ideally, a given
\gdr{2} source with its associated transits appears in \edr{3} (and \gdr{3}) with the same source ID and the same
transits, plus the new transits added for \edr{3}. This would allow a simple matching of the sources across the two
releases through the source ID.  However, for the reasons mentioned in \secref{sec:sourcelist} \citep[and elaborated
in][]{EDR3-DPACP-124}, this should {\em not} be done, instead the DR2 to DR3 match table in the {\gaia} archive
(\texttt{dr2\_neighbourhood}) should be used to trace sources from \gdr{2} to \edr{3}. This prevents mistakes in
cross-identifications of sources due to the source list evolution.

Another form of source list evolution is in the transits assigned to a given source ID. For a small fraction of sources
a significant fraction of the transits that were assigned to the source for \gdr{2} may have been reassigned to other
sources or discarded altogether. This means that the source character may change from \gdr{2} to \edr{3}. The \edr{3}
archive contains the following fields to help in understanding why a source with the same source ID in \gdr{2} and
\edr{3} may have significantly different source parameters (astrometry, photometry, etc):
\begin{description}
    \item[\texttt{\bfseries matched\_transits}] The total number of field of view transits $m$ matched to a given source ID.
    \item[\texttt{\bfseries new\_matched\_transits}] The number of transits $n$ newly included in the transit list of an
        existing source ID.
    \item[\texttt{\bfseries matched\_transits\_removed}]  The number of transits $r$ removed from the transit list of an existing
        source ID.
\end{description}
The fraction of transits retained for a source ID from \gdr{2} to \edr{3} is given by $(m-n)/(m-n+r)$. Fig.~15 in
\cite{EDR3-DPACP-124} shows this fraction as a function of $G$-band magnitude, which is lower than 100\% for a
significant number of source IDs at $G\lesssim12$ and $G\gtrsim19$. For source IDs with large changes in the transit
list one should be careful when making comparisons between \gdr{2} and \edr{3}.

As an example we look a bit more closely at the $2179$ sources for which the $G$-band magnitude in \gdr{2} is less than
5. These can be traced to \edr{3} with the following Astronomical Data Query Language (ADQL) query in the {\gaia} archive:
\begin{lstlisting}[language=SQL]
select dr2.source_id, dr2.phot_g_mean_mag, edr3.*
from gaiadr2.gaia_source as dr2
join gaiaedr3.dr2_neighbourhood as edr3
on dr2.source_id=edr3.dr2_source_id
where dr2.phot_g_mean_mag<5
order by dr2.source_id asc
\end{lstlisting}
This results in a list of $2208$ matches from a \gdr{2} to a \edr{3} source. In 34 cases there are two possible matches
in \edr{3} for the \gdr{2} source, where in all such cases one of the matches is rather far ($>0.9$~arcsec) from the
target position. This leaves 2174 ``best'' matches for very bright \gdr{2} sources in \edr{3}, hence 5 sources in \gdr{2}
at $G_\mathrm{DR2}<5$ have no counterpart in \edr{3}. Of the matched sources about 40\% changed source ID, among which a
prominent example is $\beta$~Pictoris (\gdr{2} 4792774797545105664 $\rightarrow$ \edr{3} 4792774797545800832).

These drastic changes occur mostly at the bright end as shown in Fig.~1 in \cite{EDR3-DPACP-126}. For all other
magnitudes the changes in source ID occur for a few per cent of the sources, except over the range $9<G<12$ where up to
almost 10 per cent of the sources changed ID.

\section{Using \edr{3} data: Completeness and limitations}
\label{sec:use}

\edr{3} represents a significant improvement in {\gaia} astrometry and broad-band photometry, but as pointed out in
\cite{EDR3-DPACP-73}, \cite{EDR3-DPACP-124}, \cite{EDR3-DPACP-128}, and \cite{EDR3-DPACP-117}, there are still many
improvements to be made to the data processing. This implies that there are limitations which should be kept in mind
when using \edr{3} data. Next, we describe how the data were filtered between the data processing and
release publication stages and what the main limitations are that the user should be aware of.

\subsection{\edr{3}: Validation and source filtering}
\label{sec:validation}

For details on how the quality of the \edr{3} data were assessed we refer to the astrometric and photometric processing
papers \citep{EDR3-DPACP-128, EDR3-DPACP-117} for validation at the pipeline level, while a more global view of the
data quality is provided in \cite{EDR3-DPACP-126}. Here we describe the main filtering that was applied before accepting
a source for publication.

\subsubsection{Astrometry}
\label{sec:astrometryfilters}

The filtering of the astrometric data set was very similar to the procedure used for \gdr{2}. The results were filtered
by requiring that a source was observed by {\gaia} at least five times (five focal plane transits), and that the
semi-major axis of the position uncertainty ellipse is less than 100~mas. In contrast to \gdr{2}, no filtering
on astrometric excess noise was done. The parallax and proper motions are determined only for sources satisfying the
requirement that they are brighter than $G=21$, the number of ``visibility periods'' used is at least 9 (a visibility
period represents a group of observations separated from other such groups by at least four days), and the
semi-major axis of the 5-dimensional uncertainty ellipse is below a magnitude dependent threshold.  We refer to
\cite{EDR3-DPACP-128} for the details. For sources that do not meet these requirements only the positions are reported
in \edr{3}. We remind the reader that the sources with parallaxes and proper motions fall into the two categories of 5-p
and 6-p astrometry solutions (see \secref{sec:astrometry}). For source pairs closer together than $0.18$~arcsec only one
source was retained \citep[detailed criteria in][]{EDR3-DPACP-128}, which is then marked as a \texttt{duplicate\_source}
in the \edr{3} archive.

\subsubsection{Photometry}
\label{sec:photometryfilters}

Unlike the previous releases, sources were not discarded from \edr{3} if no $G$-band photometry was available. There
are in fact some $5.5$ million sources in \edr{3} without a value for $G$ in the published catalogue. For these sources the
values of $G$ could only be estimated after the processing and validation stages were finished and they will be provided
through a separate channel (see \secref{sec:photometryusage}). The criteria to publish photometry for a source are: the
$G$-band is only provided for sources with $\text{\tt phot\_g\_n\_obs}\geq10$; the \gbp-band is only provided for sources
with $\text{\tt phot\_bp\_n\_obs}\geq2$; the \grp-band is only provided for sources with $\text{\tt phot\_rp\_n\_obs}\geq2$.
We note that any of the photometric bands can be absent for a given source. No filtering on the flux excess factor was
done (in contrast to \gdr{2}).

\subsection{Survey completeness}
\label{sec:completeness}

\figref{fig:gmaghistos} shows that the completeness of the {\gaia} survey has improved slightly with respect to \gdr{2}
at the faint end, between $G\approx19$ and $G\approx21$. The fraction of bright stars missing at $G<7$ is unchanged with
respect to \gdr{2}. The brightest star in \edr{3} is at $G=1.73$. 

The large-scale completeness limit, as estimated by the 99th percentile of the $G$ magnitude distribution
\citep{EDR3-DPACP-126}, varies between $G\sim20$ at low galactic latitudes ($b\lesssim30$) and around the Magellanic
Clouds, to $G\sim22$ at higher latitudes. The survey limit variations over the sky show clear imprints of the {\gaia}
scanning pattern. 

In crowded regions the capability to observe all stars is reduced \citep{2016A&A...595A...1G}. In combination with the
still limited data treatment in crowded areas this means that the survey limit in regions with densities above a few
hundred thousand stars per square degree can be substantially brighter than $G=20$.  \cite{EDR3-DPACP-126} show that the
completeness as measured on OGLE fields is $100$\% up to source densities of $2\times10^5$~deg$^{-2}$, while at higher
densities the completeness has improved with respect to \gdr{2}, staying close to $100$\% up to $6\times10^5$
stars~deg$^{-2}$ and dropping to $50$\% at densities over $8\times10^5$~deg$^{-2}$. \cite{EDR3-DPACP-126} also studied
the completeness in a sample of 26 globular clusters which were observed previously with the Hubble Space Telescope. On
average they find a completeness at $G=17$ of $\sim80$\% for densities below $5\times10^5$ stars~deg$^{-2}$, $\sim50$\%
at $5\times10^5$--$2\times10^6$ stars~deg$^{-2}$, and $\sim15$\% at $2\times10^6$--$2\times10^7$ stars~deg$^{-2}$, with
strong variations across the clusters and between the cores and the outer regions. In the very densest regions the
incompleteness can be so severe as to give the appearance of holes in the source distribution.

As described in \secref{sec:performance}, the effective angular resolution of the \edr{3} source list has slightly
improved with respect to \gdr{2}, with incompleteness in close pairs of stars starting below about $1.5$ arcsec. Refer
to \cite{EDR3-DPACP-126} for details. \cite{EDR3-DPACP-126} note that among the source pairs with separations between
$0.18$ and $0.4$ arcseconds there may be many spurious solutions.

\subsection{Limitations}
\label{sec:limitations}

\subsubsection{Astrometry}
\label{sec:astrometrylimitations}

The major gain in the precision of parallaxes and proper motions from \gdr{2} to \edr{3} is complemented by a
significant reduction in the systematic errors, which is evident from the reduced variance in the parallax and proper
motion bias variations over the sky, and confirmed by the beautiful measurement of the acceleration of the solar system
barycentre with respect to the distant universe. Nevertheless the following characteristics of the astrometry should be
kept in mind.

The two solution types, 5-parameter and 6-parameter, behave differently in terms of uncertainties and systematics, with
the 6-p astrometry in general being less precise. For the 5-p solutions the astrometric uncertainties are underestimated
by $\sim5$\% at the faint end ($G>16$) and by up to $\sim30$\% at the bright end ($G<14$). For the 6-p solutions the
numbers are $\sim20$\% and up to $\sim40$\%, respectively. The underestimation of the uncertainties increases in crowded
areas such as the Large Magellanic Cloud, and for sources with indications that they may have companions or be
part of a partially resolved double. The details can be found in \cite{EDR3-DPACP-126}.

By examining the distribution of negative parallaxes, \cite{EDR3-DPACP-126} estimate that among the sources with
formally high quality parallaxes ($\varpi/\sigma_\varpi>5$) some $1.6$\% of the $1.5$ billion 5-p and 6-p astrometric
solutions are spurious, meaning that the listed parallax may be significantly in error despite the formally high
precision. \cite{EDR3-DPACP-126} show that the fraction of spurious solutions is strongly dependent on magnitude and
source density on the sky. For faint sources ($G\gtrsim17$ for 6-p astrometric solutions and $G\gtrsim19$ for 5-p
solutions) and in crowded regions the fractions of spurious solutions can reach 10 per cent or more. It should be
stressed that the spurious astrometric solutions in \edr{3} produce smaller errors on the astrometric parameters than
was the case for \gdr{2}.

The global parallax zero point for \edr{3}, as measured from quasars, is $-17$~\muas. The RMS angular (i.e.\ source to
source) covariances of the parallaxes and proper motions on small scales are $\sim26$~\muas\ and $\sim33$~\muasyr,
respectively. Details on the angular covariances can be found in \cite{EDR3-DPACP-128}. The parallax zero point (and
the proper motion systematics) varies as a function of magnitude, colour, and celestial position. This is described in
detail in \cite{EDR3-DPACP-132}.

\subsubsection{Photometry}
\label{sec:photometrylimitations}

The increased precision and homogeneity of the \edr{3} broad band photometry make it harder to assess the external
accuracy of the photometry. Nevertheless \cite{EDR3-DPACP-117} show that the discontinuities that appeared at $G=13$ and
$G=16$, when comparing the \gdr{2} photometry to APASS \citep{2015AAS...22533616H,apass9} and SDSS DR15
\citep{2019ApJS..240...23A}, have disappeared in \edr{3}. As shown in \cite{EDR3-DPACP-126}, the same is true of the
strong saturation effects in $G$ at the bright end, and the significant variation of the $G$-band zero point with
magnitude present in \gdr{2}. The effects are now below the $0.01$ magnitude level for most sources. We stress again
that a single passband now suffices for all three bands, $G$, \gbp, and \grp.  The following issues with the photometry
were revealed following internal investigations, all described in detail in \cite{EDR3-DPACP-117} and
\cite{EDR3-DPACP-126}.

For faint red sources the flux in the BP band is overestimated which leads to these sources appearing much bluer in
\bpminrp\ than they should be. This can be recognised for example in open cluster colour magnitude diagrams as a
blue-ward turn of the lower main sequence in $G$ vs.\ \bpminrp. This issue is caused by the rejection of observations
with $G$-band fluxes below $1$~e$^{-}$~s$^{-1}$, where the rejection was also applied to the BP and RP observations.
This does not cause problems for $G$ and \grp, but at the faint end leads to overestimated BP fluxes.

During the internal validation of the \edr{3} photometry a small tail of sources going as faint as $G\approx25.5$ was
noticed. Such faint sources will never be observed by {\gaia}, even when taking into account the fuzziness of the
nominal $G=20.7$ survey limit. The problem was traced to sources with unreliable colours for which the application of
the internal photometric calibration failed \citep{EDR3-DPACP-117, EDR3-DPACP-126}. As a result it was decided to remove
from the \edr{3} catalogue the unreliable fluxes and magnitudes, which means there are sources for which any of the
three bands could be missing. All in all there are $5\,455\,339$ sources for which no $G$-band flux is available in the
main \edr{3} catalogue. For these sources the $G$-band flux was estimated ad-hoc by calibrating the sources assuming
default colours. The values are available as a separate table through the \edr{3} ``known issues''
pages\footnote{\url{https://www.cosmos.esa.int/web/gaia/edr3-known-issues}}. For $54\,125$ of the sources without
$G$-band fluxes this ad-hoc calibration was not possible. An indication of their brightness will be provided based on
the on-board magnitude estimate.

At the image parameter determination stage (which precedes the astrometric and photometric data processing) the $G$-band
fluxes (and locations) of sources for which no reliable colour information was available (from \gdr{2}) were estimated
with a PSF or LSF model for a default source colour. This concerns the sources for which in the astrometry 6-p solutions
were derived, which mitigated the remaining chromatic effect in the source image locations, as well as sources with 2-p
solutions. Such a colour effect is also present in the fluxes but this was unfortunately not accounted for in the
photometric processing. However, it is possible to correct the published $G$-band photometry for sources with 6-p
solutions ($\text{\tt astrometric\_params\_solved}=95$) to bring them onto the photometric system of the 5-p sources.
The correction formula is presented in \cite{EDR3-DPACP-117} and should also be applied to sources with 2-p astrometric
solutions ($\text{\tt astrometric\_params\_solved}=3$). We stress that this issue is {\em not} related to the
photometric pass-bands. 

For bright and extremely blue sources ($G<13$, $\gbp-\grp<-0.1$) there is a residual trend of about 5~mmag/mag for
sources in the range $8<G<13$, when comparing the \edr{3} magnitudes to synthetic magnitudes derived from BP and RP
spectra. This is only seen in $G$ and is probably related to deficiencies in the PSF modelling for bright stars. At
$G<8$ the residuals are dominated by saturation effects.

\section{Using \edr{3} data: Additional guidance}
\label{sec:guidance}

Here we provide some further advice on the use of the \edr{3} data. This concerns issues specific to this release. The
papers listed in \secref{sec:performance} provide extensive examples of how to use \edr{3} data responsibly, and we
remind the reader of the need to be careful when estimating distances from parallaxes with relatively large
uncertainties \citep{2018A&A...616A...9L}. We note again, as stressed in \secref{sec:changes}, that tracing
sources from \gdr{2} to \edr{3} should not be done by blindly matching source IDs. The \gdr{2} to \edr{3} match
table (\texttt{dr2\_neighbourhood}) should be used for this purpose.

\subsection{Astrometry}
\label{sec:astrometryusage}

\cite{EDR3-DPACP-126} make the following recommendation for dealing with spurious astrometric solutions. Even when
selecting only a sample of high-quality parallaxes, for example $\varpi/\sigma_\varpi>5$, one should select also the
corresponding sample with negative parallaxes ($\varpi/\sigma_\varpi<-5$) in order to ascertain what fraction of the
positive parallaxes may in fact be spurious. The parameter \texttt{ipd\_gof\_harmonic\_amplitude} is useful in
identifying spurious solutions as shown in \cite{EDR3-DPACP-126}, where values above $0.1$ in combination with a
\texttt{ruwe} larger than $1.4$ are indicative of resolved doubles, which are still not correctly handled in the
astrometric processing, and may cause spurious solutions.

For the sources with 6-p solutions there will in many cases be independent colour information available from photometric
or spectroscopic observations, which may provide a superior estimate for \nueff\ than the value estimated during the
\edr{3} astrometric data processing. In these cases it is possible to incorporate the better colour information to
update the astrometry for the 6-p solution to more precise values. The formulae for calculating the updated astrometry
are given in the appendix of \cite{EDR3-DPACP-128}, where it is demonstrated that for significant correlations
(coefficients larger than $0.3$) between the pseudo-colour and the astrometric parameters, real gains in precision and
accuracy can be expected.

\cite{EDR3-DPACP-132} present an extensive investigation of the parallax zero point variations as a function of source
brightness, colour, and celestial position. The samples used in this investigation are quasars, the Large Magellanic
Cloud, red clump stars -- sources for which the parallax is precisely known from independent estimates -- and wide pairs
of co-moving stars for which the parallaxes should be the same. As a result of the detailed characterisation of the zero
point variations, a correction recipe is presented, separately for sources with 5-p and 6-p astrometry, which allows
removing the parallax bias as a function of source magnitude, colour, and ecliptic latitude. It should be stressed that
this is a tentative recipe, primarily intended as an illustration of how corrections could be derived for other samples
of sources for which precise independent distance information is available. The recipe is not intended to be applied
blindly and has not been applied to the published \edr{3} parallaxes. Python code to apply the recipe will be made
available as part of the \edr{3} access facilities\footnote{\url{https://gitlab.com/icc-ub/public/gaiadr3_zeropoint}}.

\subsection{Photometry}
\label{sec:photometryusage}

\cite{EDR3-DPACP-117} provide guidance on the use of the \edr{3} photometry which we summarise here. Further insights
into the photometric data are presented by \cite{EDR3-DPACP-126}. 

The \gbp\ flux of faint sources is likely to be biased (Sect.~\ref{sec:photometrylimitations}) and one can
elect to filter on the value of \gbp, retaining only the brighter bias-free sources. This introduces undesirable
selection effects and a better alternative may be to use the $(G-\grp)$ colour, for example when studying the lower main
sequence. 

The $G$-band photometry for sources with 2-p or 6-p astrometric solutions should be corrected to account for the use of
a default colour at the flux estimation stage, the correction formula is presented in \cite{EDR3-DPACP-117}. In Appendix
\ref{app:gcorrection} we show how to calculate the corrections on the fly as part of a \edr{3} archive query, and also
present Python code that can be used for the same purpose. We note that out of the 344 million 2-p sources present in
\edr{3}, about 20 million have an astrometric solution in which the actual source colour was used instead of a default
colour. This means that for these 20 million 2-p sources the $G$-band correction should actually not be applied. These
sources are mostly faint, with 96\% at magnitudes $G>20$, and for 75\% of these 20 million sources the correction that
is (wrongly) applied amounts to less than 4~milli-magnitudes. It was thus decided not to make a special effort to
exclude these sources from the correction recipe presented in Appendix~\ref{app:gcorrection}. Should a user of the
\edr{3} data wish to undo the wrong correction for one or more of these 20 million sources, the list of source IDs and
applied corrections can be provided on request.

For sources where the $G$ magnitude is missing, the value can be looked up in separate tables to be provided through the
\edr{3} ``known issues'' web pages. While this may seem unnecessarily cumbersome, this choice was deliberately made to
ensure that it is very clear to the user that the $G$-band magnitudes for these sources are from a very different origin
(from an ad-hoc calibration, and for a small number of sources from the on-board magnitude estimate) and not directly
comparable to the main catalogue photometry.

As was the case for \gdr{2}, at the bright end ($G<8$ for $G$ and $G\lesssim4$ for \gbp\ and \grp) the magnitudes should
be corrected for saturation effects. The correction formulae can be found in the appendix of \cite{EDR3-DPACP-117}.

An important point made in \cite{EDR3-DPACP-117} is that the flux excess factor in \edr{3} is much more representative
of astrophysical inconsistencies between the fluxes in BP and RP with respect to the flux in $G$, for example due to the
extended nature of a source or its non-standard (non-stellar) spectral energy distribution (although
\cite{EDR3-DPACP-126} show that some artefacts in the photometry can still be traced in the flux excess factor). It is
thus not possible to easily identify problematic photometry through the flux excess factor, and using this quantity in
the construction of samples should be done with care. Refer to \cite{EDR3-DPACP-117} for detailed guidance. They present
a corrected version of the flux excess factor which is recommended for use instead of the raw
\texttt{phot\_bp\_rp\_excess\_factor} value listed in \edr{3}. The corrected version can be calculated from the formula
presented in section 6 of \cite{EDR3-DPACP-117}. Appendix \ref{app:excessfact} shows how to include the correction of
the flux excess factor within an ADQL query, and presents Python code to achieve the same.

As described in \secref{sec:photometry}, \edr{3} contains fields from which a metric can be constructed that indicates how
likely it is that the photometry of a given source is affected by crowding. This metric should be used with some
care, which is explained further in \cite{EDR3-DPACP-117}.

Finally, due to the evolution of the source list and the improvements in the photometry we strongly discourage
comparisons between \edr{3} and \gdr{2} photometry, in particular on a source by source basis. Comparisons at the sample
level are likely to reveal mostly differences due to changes in the photometric system and errors in \gdr{2}.

\section{\edr{3} access facilities}
\label{sec:access}

The \edr{3} data will be available through the archive hosted by ESA\footnote{\url{https://archives.esac.esa.int/gaia}},
with the facilities as described in \cite{2016A&A...595A...2G} and \cite{2018A&A...616A...1G}.  The data is also
accessible at the partner and affiliated data centres in Europe, the United States, Japan, Australia, and South Africa.
These data centres provide their own access facilities, but do not necessarily host all data contained in the ESA
{\gaia} archive. We note the following enhancements and changes.

The pointing of the {\gaia} telescopes as a function of time is available as the table \texttt{commanded\_scan\_law}.
The pointing for the entire 34 month period covering \edr{3} is available at 10 second intervals, and allows one to
reconstruct how often and at what scan angles a given position on the sky was observed by {\gaia}. We note that the {\em
commanded} pointing is provided  which may deviate from the actual attitude of {\gaia} by up to 30 arcsec. In addition
gaps in the data collection due to spacecraft or on-ground problems are not accounted for.

The {\gaia} Universe Model Snapshot \citep[GUMS,][]{2012A&A...543A.100R} and the corresponding simulated {\gaia}
catalogue \citep[GOG,][]{gog} are now available as part of the \edr{3} archive in the tables
\texttt{gaia\_universe\_model} and \texttt{gaia\_source\_simulation}, respectively. They correspond to version 20 of
GUMS and GOG and are described in detail in the on-line documentation
\footnote{\url{https://gea.esac.esa.int/archive/documentation/GEDR3/Data_processing/chap_simulated/}}. We note that a few
issues in the simulations could not be corrected on time. Notably, young star kinematics were wrongly set, such that
their astrometry should be corrected by the user before using the simulation. Essentially, the stars with ages less than
$0.15$~Gyr should follow the Milky Way rotation curve, while they do not (their mean rotation velocity was erroneously
set to 0). In addition, RR~Lyrae stars are missing from GUMS, and the number of outliers in GOG, both in astrometry and
photometry, is larger than expected from the simulated errors.

The astrometric performance predictions for \gdr{4} and beyond have been updated, based on an extrapolation of the
\edr{3} performance. The new predictions will appear on the {\gaia} science performance
pages\footnote{\url{https://www.cosmos.esa.int/web/gaia/science-performance}}. The tables \texttt{agn\_cross\_id} and
\texttt{frame\_rotator\_source} provide the source IDs of the \gcrf{3} sources.

Pre-computed cross matches to other large surveys are provided.  We recommend using these cross-matches because they have
been carefully validated and their use facilitates reproducing analyses of \edr{3} data combined with other survey data.
The details of the cross-match procedure are provided in \cite{EDR3-DPACP-129} \citep[see also][]{2017A&A...607A.105M,
2019A&A...621A.144M}. Pre-computed cross-matches are provided for the following surveys: {\hip} \citep[new
reduction,][]{book:newhip}; {\tyc}-2 \citep{2000A&A...355L..27H}, merged with the Tycho Double Star Catalogue
\citep{2002A&A...384..180F}; 2MASS \citep[point source and extended catalogue merged,][]{2006AJ....131.1163S}; SDSS DR13
\citep{2017ApJS..233...25A}; Pan-STARRS1 \citep{2016arXiv161205560C}; SkyMapper DR2 \citep{2019PASA...36...33O}; AllWise
\citep{2010AJ....140.1868W}; URAT1 \citep{2015AJ....150..101Z}; GSC2.3 \citep{2008AJ....136..735L}; APASS DR9
\citep{apass9, 2015AAS...22533616H}; and RAVE DR5 \citep{2017AJ....153...75K}.

\section{Conclusions}
\label{sec:conclusions}

\edr{3} represents another significant advance in the series of data releases resulting from the {\gaia} mission. Based
on 34 months of input data, this release features major improvements in the astrometry and broad-band photometry, where
for the first time the astrometry and photometry benefit from iterative processing between the determination of image
locations and fluxes, and the astrometric solution. Other significant improvements are the increased robustness and
stability of the source list and a much more sophisticated modelling of the PSF and LSF for the astrometric instrument.
Next to the significant increase in the precision of the astrometry and photometry, the suppression of systematic errors
is a major component of the improvements.

\edr{3} represents the first installment of the \gdr{3} release planned for publication in 2022. \gdr{3} will feature new
data products of which the BP, RP, and RVS spectra (to be released for a subset of sources) and the non-single star
catalogue represent qualitative changes in character with respect to \gdr{2}. The planned contents are: astrometry and
broad-band photometry from \edr{3} will remain unchanged for \gdr{3}, and the same is true for the source list; an
expanded radial velocity survey (some 30 million stars brighter than $\grvs\sim14$); astrophysical parameter estimates
based on the parallaxes, broad-band photometry, and BP, RP, and RVS spectra, where the latter are a new element enabling
a richer astrophysical characterisation of sources; an order of magnitude larger sample of variable stars, with their
light curves, classifications, and astrophysical properties; a non-single star (mostly binary stars) catalogue based on
the analysis of epoch astrometry, epoch radial velocities and the light curves of eclipsing binaries; analyses of
extended objects (galaxies, and QSO hosts); epoch astrometry and photometry, as well as orbits, for an expanded list of
over $100\,000$ solar system objects; mean BP, RP, and RVS spectra, for a subset of astrophysically well-characterised
sources; reflectance spectra for $\sim5000$ asteroids derived from BP and RP spectra; the {\gaia} Andromeda Photometric
Survey (GAPS), which consists of the broad-band photometric time-series for {\em all} sources in a $5.5^\circ$ radius
field centred on M31.

There is thus much to look forward to in \gdr{3}, with the GAPS data set providing a taste of what is to come in \gdr{4}.
We stress that epoch astrometry and epoch radial velocities will {\em not} appear in \gdr{3} (except for solar system
object epoch astrometry).

Looking ahead, the {\gaia} spacecraft is currently in good overall health. The spacecraft operations since the
appearance of \gdr{2} have largely been smooth, with little to no degradation of the detectors in {\gaia}'s
focal plane, except for the steadily increasing radiation damage. The latter is however still well below the anticipated
levels before the launch of {\gaia}. The last decontamination of the telescopes and focal plane took place in August
2016 and no further decontamination was needed since. The current evolution of the throughput of the telescopes suggests
that also in the future no further decontamination is needed. The only limiting factor to the lifetime of {\gaia} as a
high precision astrometry mission is the amount of propellant for the micro-propulsion system. This is predicted to be
exhausted in early 2025, after which time the attitude and spin rate of {\gaia} can no longer be maintained at the
levels of precision needed for the astrometry. With this end-of-life date for {\gaia} in mind, and the end of the
nominal mission planned for mid-2019, the process of applying for an extended mission was started already in 2016. The
nominal {\gaia} mission ended on July 16, 2019 and {\gaia} has been in extended mission operations since that date. The
mission extension is formally approved to the end of 2022 at the time of writing, with good hopes of the mission
continuing to its estimated end-of-life. This would bring the total mission lifetime to 10 years, implying a 40 per cent
improvement on the precision of all data products with respect to a five year mission, and a factor of almost three
improvement for the proper motions.

In this context the community can look forward to two major data releases, \gdr{4} and \gdr{5}, both incorporating data
from the extended {\gaia} mission. \gdr{4} will be based on 66 months of input data (which is already in hand), while
\gdr{5} will include all data collected over the entire (nominal + extended) {\gaia} mission. The extra half year of
data from the extended mission included in the \gdr{4} data processing is motivated by the wish to include part of the
one year period between July 16, 2019 and July 29, 2020 when {\gaia} was operated with a reversed direction of the
precession of the spin axis around the direction to the sun. This was introduced to break the degeneracy between the
across-scan rate at which sources move across the focal plane and their parallax factor \citep[see appendix B in][for
more details]{EDR3-DPACP-128}.  Including the first 6 months of the reverse precession period in the inputs for \gdr{4}
is expected to already significantly mitigate the effects of this degeneracy.

The main new features of \gdr{4} are the publication of a list of exoplanets discovered with {\gaia} and the publication
of all the time series data, meaning epoch astrometry, broad-band photometry, radial velocities, as well as BP, RP, and
RVS spectra for {\em all} sources. This will be a significant expansion in the volume of data released. We leave to the
imagination of the reader the expanded scientific possibilities.


%
%

\begin{acknowledgements}

This work presents results from the European Space Agency (ESA) space mission \gaia. \gaia\ data are being processed by
the \gaia\ Data Processing and Analysis Consortium (DPAC). Funding for the DPAC is provided by national institutions, in
particular the institutions participating in the \gaia\ MultiLateral Agreement (MLA). The \gaia\ mission website is
\url{https://www.cosmos.esa.int/gaia}. The \gaia\ archive website is \url{https://archives.esac.esa.int/gaia}.

The \gaia\ mission and data processing have financially been supported by, in alphabetical order by country:
     the Algerian Centre de Recherche en Astronomie, Astrophysique et G\'{e}ophysique of Bouzareah Observatory;
     the Austrian Fonds zur F\"{o}rderung der wissenschaftlichen Forschung (FWF) Hertha Firnberg Programme through
        grants T359, P20046, and P23737;
     the BELgian federal Science Policy Office (BELSPO) through various PROgramme de D\'eveloppement d'Exp\'eriences
        scientifiques (PRODEX) grants and the Polish Academy of Sciences - Fonds Wetenschappelijk Onderzoek through grant
        VS.091.16N, and the Fonds de la Recherche Scientifique (FNRS);
     the Brazil-France exchange programmes Funda\c{c}\~{a}o de Amparo \`{a} Pesquisa do Estado de S\~{a}o Paulo
        (FAPESP) and Coordena\c{c}\~{a}o de Aperfeicoamento de Pessoal de N\'{\i}vel Superior (CAPES) - Comit\'{e}
        Fran\c{c}ais d'Evaluation de la Coop\'{e}ration Universitaire et Scientifique avec le Br\'{e}sil (COFECUB);
     the National Science Foundation of China (NSFC) through grants 11573054 and 11703065 and the China Scholarship
        Council through grant 201806040200;  
     the Tenure Track Pilot Programme of the Croatian Science Foundation and the \'{E}cole Polytechnique
        F\'{e}d\'{e}rale de Lausanne and the project TTP-2018-07-1171 'Mining the Variable Sky', with the funds of the
        Croatian-Swiss Research Programme;
     the Czech-Republic Ministry of Education, Youth, and Sports through grant LG 15010 and INTER-EXCELLENCE grant
        LTAUSA18093, and the Czech Space Office through ESA PECS contract 98058;
     the Danish Ministry of Science;
     the Estonian Ministry of Education and Research through grant IUT40-1;
     the European Commission’s Sixth Framework Programme through the European Leadership in Space Astrometry
        (\href{https://www.cosmos.esa.int/web/gaia/elsa-rtn-programme}{ELSA}) Marie Curie Research Training Network
        (MRTN-CT-2006-033481), through Marie Curie project PIOF-GA-2009-255267 (Space AsteroSeismology \& RR Lyrae stars,
        SAS-RRL), and through a Marie Curie Transfer-of-Knowledge (ToK) fellowship (MTKD-CT-2004-014188); the European
        Commission's Seventh Framework Programme through grant FP7-606740 (FP7-SPACE-2013-1) for the \gaia\ European Network
        for Improved data User Services (\href{https://gaia.ub.edu/twiki/do/view/GENIUS/}{GENIUS}) and through grant 264895
        for the \gaia\ Research for European Astronomy Training
        (\href{https://www.cosmos.esa.int/web/gaia/great-programme}{GREAT-ITN}) network;
     the European Research Council (ERC) through grants 320360 and 647208 and through the European Union’s Horizon 2020
        research and innovation and excellent science programmes through Marie Sk{\l}odowska-Curie grant 745617 as well as
        grants 670519 (Mixing and Angular Momentum tranSport of massIvE stars -- MAMSIE), 687378 (Small Bodies: Near and
        Far), 682115 (Using the Magellanic Clouds to Understand the Interaction of Galaxies), and 695099 (A sub-percent
        distance scale from binaries and Cepheids -- CepBin);
     the European Science Foundation (ESF), in the framework of the \gaia\ Research for European Astronomy Training
        Research Network Programme (\href{https://www.cosmos.esa.int/web/gaia/great-programme}{GREAT-ESF});
     the European Space Agency (ESA) in the framework of the \gaia\ project, through the Plan for European Cooperating
        States (PECS) programme through grants for Slovenia, through contracts C98090 and 4000106398/12/NL/KML for Hungary,
        and through contract 4000115263/15/NL/IB for Germany;
     the Academy of Finland and the Magnus Ehrnrooth Foundation;
     the French Centre National d’Etudes Spatiales (CNES), the Agence Nationale de la Recherche (ANR) through grant
        ANR-10-IDEX-0001-02 for the 'Investissements d'avenir' programme, through grant ANR-15-CE31-0007 for project
        'Modelling the Milky Way in the Gaia era' (MOD4Gaia), through grant ANR-14-CE33-0014-01 for project 'The Milky Way
        disc formation in the Gaia era' (ARCHEOGAL), and through grant ANR-15-CE31-0012-01 for project 'Unlocking the
        potential of Cepheids as primary distance calibrators' (UnlockCepheids), the Centre National de la Recherche
        Scientifique (CNRS) and its SNO Gaia of the Institut des Sciences de l’Univers (INSU), the 'Action
        F\'{e}d\'{e}ratrice Gaia' of the Observatoire de Paris, the R\'{e}gion de Franche-Comt\'{e}, and the Programme
        National de Gravitation, R\'{e}f\'{e}rences, Astronomie, et M\'{e}trologie (GRAM) of CNRS/INSU with the Institut
        National Polytechnique (INP) and the Institut National de Physique nucléaire et de Physique des Particules (IN2P3)
        co-funded by CNES;
     the German Aerospace Agency (Deutsches Zentrum f\"{u}r Luft- und Raumfahrt e.V., DLR) through grants 50QG0501,
        50QG0601, 50QG0602, 50QG0701, 50QG0901, 50QG1001, 50QG1101, 50QG1401, 50QG1402, 50QG1403, 50QG1404, and 50QG1904 and
        the Centre for Information Services and High Performance Computing (ZIH) at the Technische Universit\"{a}t (TU)
        Dresden for generous allocations of computer time;
     the Hungarian Academy of Sciences through the Lend\"{u}let Programme grants LP2014-17 and LP2018-7 and through the
        Premium Postdoctoral Research Programme (L.~Moln\'{a}r), and the Hungarian National Research, Development, and
        Innovation Office (NKFIH) through grant KH\_18-130405;
     the Science Foundation Ireland (SFI) through a Royal Society - SFI University Research Fellowship (M.~Fraser);
     the Israel Science Foundation (ISF) through grant 848/16;
     the Agenzia Spaziale Italiana (ASI) through contracts I/037/08/0, I/058/10/0, 2014-025-R.0, 2014-025-R.1.2015, and
        2018-24-HH.0 to the Italian Istituto Nazionale di Astrofisica (INAF), contract 2014-049-R.0/1/2 to INAF for the
        Space Science Data Centre (SSDC, formerly known as the ASI Science Data Center, ASDC), contracts I/008/10/0,
        2013/030/I.0, 2013-030-I.0.1-2015, and 2016-17-I.0 to the Aerospace Logistics Technology Engineering Company (ALTEC
        S.p.A.), INAF, and the Italian Ministry of Education, University, and Research (Ministero dell'Istruzione,
        dell'Universit\`{a} e della Ricerca) through the Premiale project 'MIning The Cosmos Big Data and Innovative Italian
        Technology for Frontier Astrophysics and Cosmology' (MITiC);
     the Netherlands Organisation for Scientific Research (NWO) through grant NWO-M-614.061.414, through a VICI grant
        (A.~Helmi), and through a Spinoza prize (A.~Helmi), and the Netherlands Research School for Astronomy (NOVA);
     the Polish National Science Centre through HARMONIA grant 2018/06/M/ST9/00311, DAINA grant 2017/27/L/ST9/03221,
        and PRELUDIUM grant 2017/25/N/ST9/01253, and the Ministry of Science and Higher Education (MNiSW) through grant
        DIR/WK/2018/12;
     the Portugese Funda\c{c}\~ao para a Ci\^{e}ncia e a Tecnologia (FCT) through grants SFRH/BPD/74697/2010 and
        SFRH/BD/128840/2017 and the Strategic Programme UID/FIS/00099/2019 for CENTRA;
     the Slovenian Research Agency through grant P1-0188;
     the Spanish Ministry of Economy (MINECO/FEDER, UE) through grants ESP2016-80079-C2-1-R, ESP2016-80079-C2-2-R,
        RTI2018-095076-B-C21, RTI2018-095076-B-C22, BES-2016-078499, and BES-2017-083126 and the Juan de la Cierva
        formaci\'{o}n 2015 grant FJCI-2015-2671, the Spanish Ministry of Education, Culture, and Sports through grant
        FPU16/03827, the Spanish Ministry of Science and Innovation (MICINN) through grant AYA2017-89841P for project
        'Estudio de las propiedades de los f\'{o}siles estelares en el entorno del Grupo Local' and through grant
        TIN2015-65316-P for project 'Computaci\'{o}n de Altas Prestaciones VII', the Severo Ochoa Centre of Excellence
        Programme of the Spanish Government through grant SEV2015-0493, the Institute of Cosmos Sciences University of
        Barcelona (ICCUB, Unidad de Excelencia ’Mar\'{\i}a de Maeztu’) through grants MDM-2014-0369 and CEX2019-000918-M,
        the University of Barcelona's official doctoral programme for the development of an R+D+i project through an Ajuts
        de Personal Investigador en Formaci\'{o} (APIF) grant, the Spanish Virtual Observatory through project
        AyA2017-84089, the Galician Regional Government, Xunta de Galicia, through grants ED431B-2018/42 and
        ED481A-2019/155, support received from the Centro de Investigaci\'{o}n en Tecnolog\'{\i}as de la Informaci\'{o}n y
        las Comunicaciones (CITIC) funded by the Xunta de Galicia, the Xunta de Galicia and the Centros Singulares de
        Investigaci\'{o}n de Galicia for the period 2016-2019 through CITIC, the European Union through the European
        Regional Development Fund (ERDF) / Fondo Europeo de Desenvolvemento Rexional (FEDER) for the Galicia 2014-2020
        Programme through grant ED431G-2019/01, the Red Espa\~{n}ola de Supercomputaci\'{o}n (RES) computer resources at
        MareNostrum, the Barcelona Supercomputing Centre - Centro Nacional de Supercomputaci\'{o}n (BSC-CNS) through
        activities AECT-2016-1-0006, AECT-2016-2-0013, AECT-2016-3-0011, and AECT-2017-1-0020, the Departament
        d'Innovaci\'{o}, Universitats i Empresa de la Generalitat de Catalunya through grant 2014-SGR-1051 for project
        'Models de Programaci\'{o} i Entorns d'Execuci\'{o} Parallels' (MPEXPAR), and Ramon y Cajal Fellowship
        RYC2018-025968-I;
     the Swedish National Space Agency (SNSA/Rymdstyrelsen);
     the Swiss State Secretariat for Education, Research, and Innovation through the Mesures d’Accompagnement, the
        Swiss Activit\'es Nationales Compl\'ementaires, and the Swiss National Science Foundation;
     the United Kingdom Particle Physics and Astronomy Research Council (PPARC), the United Kingdom Science and
        Technology Facilities Council (STFC), and the United Kingdom Space Agency (UKSA) through the following grants to the
        University of Bristol, the University of Cambridge, the University of Edinburgh, the University of Leicester, the
        Mullard Space Sciences Laboratory of University College London, and the United Kingdom Rutherford Appleton
        Laboratory (RAL): PP/D006511/1, PP/D006546/1, PP/D006570/1, ST/I000852/1, ST/J005045/1, ST/K00056X/1, ST/K000209/1,
        ST/K000756/1, ST/L006561/1, ST/N000595/1, ST/N000641/1, ST/N000978/1, ST/N001117/1, ST/S000089/1, ST/S000976/1,
        ST/S001123/1, ST/S001948/1, ST/S002103/1, and ST/V000969/1.

This work made use of the following software: 
Astropy, a community-developed core Python package for Astronomy \citep[\url{http://www.astropy.org}]{astropy:2013, astropy:2018}, 
IPython \citep[\url{https://ipython.org/}]{ipython:2007},
Jupyter (\url{https://jupyter.org/}), 
Matplotlib \citep[\url{https://matplotlib.org}]{matplotlib:2007},
SciPy \citep[\url{https://www.scipy.org}]{scipy:2020}, 
NumPy \citep[\url{https://numpy.org}]{numpy:2020},
and TOPCAT \citep[\url{http://www.starlink.ac.uk/topcat/}]{topcat:2005}.

This work has made use of NASA’s Astrophysics Data System.

We thank the referee, Andy Casey, for a careful reading of the manuscript.

\end{acknowledgements}

%
%

\bibliographystyle{aa} 
\bibliography{refsedrthree} 

\begin{thebibliography}{62}
\expandafter\ifx\csname natexlab\endcsname\relax\def\natexlab#1{#1}\fi

\bibitem[{{Aguado} {et~al.}(2019){Aguado}, {Ahumada}, {Almeida}, {Anderson},
  {Andrews}, {Anguiano}, {Aquino Ort{\'\i}z}, {Arag{\'o}n-Salamanca},
  {Argudo-Fern{\'a}ndez}, {Aubert}, {Avila-Reese}, {Badenes}, {Barboza
  Rembold}, {Barger}, {Barrera-Ballesteros}, {Bates}, {Bautista}, {Beaton},
  {Beers}, {Belfiore}, {Bernardi}, {Bershady}, {Beutler}, {Bird}, {Bizyaev},
  {Blanc}, {Blanton}, {Blomqvist}, {Bolton}, {Boquien}, {Borissova}, {Bovy},
  {Brand t}, {Brinkmann}, {Brownstein}, {Bundy}, {Burgasser}, {Byler}, {Cano
  Diaz}, {Cappellari}, {Carrera}, {Cervantes Sodi}, {Chen}, {Cherinka}, {Choi},
  {Chung}, {Coffey}, {Comerford}, {Comparat}, {Covey}, {da Silva Ilha}, {da
  Costa}, {Dai}, {Damke}, {Darling}, {Davies}, {Dawson}, {de Sainte Agathe},
  {Deconto Machado}, {Del Moro}, {De Lee}, {Diamond-Stanic}, {Dom{\'\i}nguez
  S{\'a}nchez}, {Donor}, {Drory}, {du Mas des Bourboux}, {Duckworth}, {Dwelly},
  {Ebelke}, {Emsellem}, {Escoffier}, {Fern{\'a}ndez-Trincado}, {Feuillet},
  {Fischer}, {Fleming}, {Fraser-McKelvie}, {Freischlad}, {Frinchaboy}, {Fu},
  {Galbany}, {Garcia-Dias}, {Garc{\'\i}a-Hern{\'a}ndez}, {Garma Oehmichen},
  {Geimba Maia}, {Gil-Mar{\'\i}n}, {Grabowski}, {Gu}, {Guo}, {Ha},
  {Harrington}, {Hasselquist}, {Hayes}, {Hearty}, {Hernandez Toledo}, {Hicks},
  {Hogg}, {Holley-Bockelmann}, {Holtzman}, {Hsieh}, {Hunt}, {Hwang},
  {Ibarra-Medel}, {Jimenez Angel}, {Johnson}, {Jones}, {J{\"o}nsson},
  {Kinemuchi}, {Kollmeier}, {Krawczyk}, {Kreckel}, {Kruk}, {Lacerna}, {Lan},
  {Lane}, {Law}, {Lee}, {Li}, {Lian}, {Lin}, {Lin}, {Lintott}, {Long},
  {Longa-Pe{\~n}a}, {Mackereth}, {de la Macorra}, {Majewski}, {Malanushenko},
  {Manchado}, {Maraston}, {Mariappan}, {Marinelli}, {Marques-Chaves},
  {Masseron}, {Masters}, {McDermid}, {Medina Pe{\~n}a}, {Meneses-Goytia},
  {Merloni}, {Merrifield}, {Meszaros}, {Minniti}, {Minsley}, {Muna}, {Myers},
  {Nair}, {Correa do Nascimento}, {Newman}, {Nitschelm}, {Olmstead}, {Oravetz},
  {Oravetz}, {Ortega Minakata}, {Pace}, {Padilla}, {Palicio}, {Pan}, {Pan},
  {Parikh}, {Parker}, {Peirani}, {Penny}, {Percival}, {Perez-Fournon},
  {Peterken}, {Pinsonneault}, {Prakash}, {Raddick}, {Raichoor}, {Riffel},
  {Riffel}, {Rix}, {Robin}, {Roman-Lopes}, {Rose}, {Ross}, {Rossi}, {Rowlands},
  {Rubin}, {S{\'a}nchez}, {S{\'a}nchez-Gallego}, {Sayres}, {Schaefer},
  {Schiavon}, {Schimoia}, {Schlafly}, {Schlegel}, {Schneider}, {Schultheis},
  {Seo}, {Shamsi}, {Shao}, {Shen}, {Shetty}, {Simonian}, {Smethurst}, {Sobeck},
  {Souter}, {Spindler}, {Stark}, {Stassun}, {Steinmetz}, {Storchi-Bergmann},
  {Stringfellow}, {Su{\'a}rez}, {Sun}, {Taghizadeh-Popp}, {Talbot}, {Tayar},
  {Thakar}, {Thomas}, {Tissera}, {Tojeiro}, {Troup}, {Unda-Sanzana},
  {Valenzuela}, {Vargas-Maga{\~n}a}, {V{\'a}zquez-Mata}, {Wake}, {Weaver},
  {Weijmans}, {Westfall}, {Wild}, {Wilson}, {Woods}, {Yan}, {Yang}, {Zamora},
  {Zasowski}, {Zhang}, {Zheng}, {Zheng}, {Zhu}, {Zinn}, \&
  {Zou}}]{2019ApJS..240...23A}
{Aguado}, D.~S., {Ahumada}, R., {Almeida}, A., {et~al.} 2019, \apjs, 240, 23

\bibitem[{{Albareti} {et~al.}(2017){Albareti}, {Allende Prieto}, {Almeida},
  {Anders}, {Anderson}, {Andrews}, {Arag{\'o}n-Salamanca},
  {Argudo-Fern{\'a}ndez}, {Armengaud}, {Aubourg}, {Avila-Reese}, {Badenes},
  {Bailey}, {Barbuy}, {Barger}, {Barrera-Ballesteros}, {Bartosz}, {Basu},
  {Bates}, {Battaglia}, {Baumgarten}, {Baur}, {Bautista}, {Beers}, {Belfiore},
  {Bershady}, {Bertran de Lis}, {Bird}, {Bizyaev}, {Blanc}, {Blanton},
  {Blomqvist}, {Bolton}, {Borissova}, {Bovy}, {Brand t}, {Brinkmann},
  {Brownstein}, {Bundy}, {Burtin}, {Busca}, {Orlando Camacho Chavez}, {Cano
  D{\'\i}az}, {Cappellari}, {Carrera}, {Chen}, {Cherinka}, {Cheung},
  {Chiappini}, {Chojnowski}, {Chuang}, {Chung}, {Cirolini}, {Clerc}, {Cohen},
  {Comerford}, {Comparat}, {Correa do Nascimento}, {Cousinou}, {Covey},
  {Crane}, {Croft}, {Cunha}, {Darling}, {Davidson}, {Dawson}, {Da Costa}, {Da
  Silva Ilha}, {Deconto Machado}, {Delubac}, {De Lee}, {De la Macorra}, {De la
  Torre}, {Diamond-Stanic}, {Donor}, {Downes}, {Drory}, {Du}, {Du Mas des
  Bourboux}, {Dwelly}, {Ebelke}, {Eigenbrot}, {Eisenstein}, {Elsworth},
  {Emsellem}, {Eracleous}, {Escoffier}, {Evans}, {Falc{\'o}n-Barroso}, {Fan},
  {Favole}, {Fernandez-Alvar}, {Fernand ez-Trincado}, {Feuillet}, {Fleming},
  {Font-Ribera}, {Freischlad}, {Frinchaboy}, {Fu}, {Gao}, {Garcia},
  {Garcia-Dias}, {Garcia-Hern{\'a}ndez}, {Garcia P{\'e}rez}, {Gaulme}, {Ge},
  {Geisler}, {Gillespie}, {Gil Marin}, {Girardi}, {Goddard}, {Gomez Maqueo
  Chew}, {Gonzalez-Perez}, {Grabowski}, {Green}, {Grier}, {Grier}, {Guo},
  {Guy}, {Hagen}, {Hall}, {Harding}, {Harley}, {Hasselquist}, {Hawley},
  {Hayes}, {Hearty}, {Hekker}, {Hernandez Toledo}, {Ho}, {Hogg},
  {Holley-Bockelmann}, {Holtzman}, {Holzer}, {Hu}, {Huber}, {Hutchinson},
  {Hwang}, {Ibarra-Medel}, {Ivans}, {Ivory}, {Jaehnig}, {Jensen}, {Johnson},
  {Jones}, {Jullo}, {Kallinger}, {Kinemuchi}, {Kirkby}, {Klaene}, {Kneib},
  {Kollmeier}, {Lacerna}, {Lane}, {Lang}, {Laurent}, {Law}, {Leauthaud}, {Le
  Goff}, {Li}, {Li}, {Li}, {Li}, {Liang}, {Liang}, {Lima}, {Lin}, {Lin}, {Lin},
  {Liu}, {Long}, {Lucatello}, {MacDonald}, {MacLeod}, {Mackereth}, {Mahadevan},
  {Maia}, {Maiolino}, {Majewski}, {Malanushenko}, {Malanushenko}, {Mallmann},
  {Manchado}, {Maraston}, {Marques-Chaves}, {Martinez Valpuesta}, {Masters},
  {Mathur}, {McGreer}, {Merloni}, {Merrifield}, {M{\'e}sz{\'a}ros}, {Meza},
  {Miglio}, {Minchev}, {Molaverdikhani}, {Montero-Dorta}, {Mosser}, {Muna},
  {Myers}, {Nair}, {Nandra}, {Ness}, {Newman}, {Nichol}, {Nidever},
  {Nitschelm}, {O'Connell}, {Oravetz}, {Oravetz}, {Pace}, {Padilla},
  {Palanque-Delabrouille}, {Pan}, {Parejko}, {Paris}, {Park}, {Peacock},
  {Peirani}, {Pellejero-Ibanez}, {Penny}, {Percival}, {Percival},
  {Perez-Fournon}, {Petitjean}, {Pieri}, {Pinsonneault}, {Pisani}, {Prada},
  {Prakash}, {Price-Jones}, {Raddick}, {Rahman}, {Raichoor}, {Barboza Rembold},
  {Reyna}, {Rich}, {Richstein}, {Ridl}, {Riffel}, {Riffel}, {Rix}, {Robin},
  {Rockosi}, {Rodr{\'\i}guez-Torres}, {Rodrigues}, {Roe}, {Roman Lopes},
  {Rom{\'a}n-Z{\'u}{\~n}iga}, {Ross}, {Rossi}, {Ruan}, {Ruggeri}, {Runnoe},
  {Salazar-Albornoz}, {Salvato}, {Sanchez}, {Sanchez}, {Sanchez-Gallego},
  {Santiago}, {Schiavon}, {Schimoia}, {Schlafly}, {Schlegel}, {Schneider},
  {Sch{\"o}nrich}, {Schultheis}, {Schwope}, {Seo}, {Serenelli}, {Sesar},
  {Shao}, {Shetrone}, {Shull}, {Silva Aguirre}, {Skrutskie}, {Slosar}, {Smith},
  {Smith}, {Sobeck}, {Somers}, {Souto}, {Stark}, {Stassun}, {Steinmetz},
  {Stello}, {Storchi Bergmann}, {Strauss}, {Streblyanska}, {Stringfellow},
  {Suarez}, {Sun}, {Taghizadeh-Popp}, {Tang}, {Tao}, {Tayar}, {Tembe},
  {Thomas}, {Tinker}, {Tojeiro}, {Tremonti}, {Troup}, {Trump}, {Unda-Sanzana},
  {Valenzuela}, {Van den Bosch}, {Vargas-Maga{\~n}a}, {Vazquez}, {Villanova},
  {Vivek}, {Vogt}, {Wake}, {Walterbos}, {Wang}, {Wang}, {Weaver}, {Weijmans},
  {Weinberg}, {Westfall}, {Whelan}, {Wilcots}, {Wild}, {Williams}, {Wilson},
  {Wood-Vasey}, {Wylezalek}, {Xiao}, {Yan}, {Yang}, {Ybarra}, {Yeche}, {Yuan},
  {Zakamska}, {Zamora}, {Zasowski}, {Zhang}, {Zhao}, {Zhao}, {Zheng}, {Zheng},
  {Zhou}, {Zhu}, {Zinn}, \& {Zou}}]{2017ApJS..233...25A}
{Albareti}, F.~D., {Allende Prieto}, C., {Almeida}, A., {et~al.} 2017, \apjs,
  233, 25

\bibitem[{{Arenou} {et~al.}(2018){Arenou}, {Luri}, {Babusiaux}, {Fabricius},
  {Helmi}, {Muraveva}, {Robin}, {Spoto}, {Vallenari}, {Antoja},
  {Cantat-Gaudin}, {Jordi}, {Leclerc}, {Reyl{\'e}}, {Romero-G{\'o}mez}, {Shih},
  {Soria}, {Barache}, {Bossini}, {Bragaglia}, {Breddels}, {Fabrizio},
  {Lambert}, {Marrese}, {Massari}, {Moitinho}, {Robichon}, {Ruiz-Dern},
  {Sordo}, {Veljanoski}, {Eyer}, {Jasniewicz}, {Pancino}, {Soubiran}, {Spagna},
  {Tanga}, {Turon}, \& {Zurbach}}]{2018A&A...616A..17A}
{Arenou}, F., {Luri}, X., {Babusiaux}, C., {et~al.} 2018, \aap, 616, A17

\bibitem[{{Astropy Collaboration} {et~al.}(2018){Astropy Collaboration},
  {Price-Whelan}, {Sip{\H{o}}cz}, {G{\"u}nther}, {Lim}, {Crawford}, {Conseil},
  {Shupe}, {Craig}, {Dencheva}, {Ginsburg}, {Vand erPlas}, {Bradley},
  {P{\'e}rez-Su{\'a}rez}, {de Val-Borro}, {Aldcroft}, {Cruz}, {Robitaille},
  {Tollerud}, {Ardelean}, {Babej}, {Bach}, {Bachetti}, {Bakanov}, {Bamford},
  {Barentsen}, {Barmby}, {Baumbach}, {Berry}, {Biscani}, {Boquien}, {Bostroem},
  {Bouma}, {Brammer}, {Bray}, {Breytenbach}, {Buddelmeijer}, {Burke},
  {Calderone}, {Cano Rodr{\'\i}guez}, {Cara}, {Cardoso}, {Cheedella}, {Copin},
  {Corrales}, {Crichton}, {D'Avella}, {Deil}, {Depagne}, {Dietrich}, {Donath},
  {Droettboom}, {Earl}, {Erben}, {Fabbro}, {Ferreira}, {Finethy}, {Fox},
  {Garrison}, {Gibbons}, {Goldstein}, {Gommers}, {Greco}, {Greenfield},
  {Groener}, {Grollier}, {Hagen}, {Hirst}, {Homeier}, {Horton}, {Hosseinzadeh},
  {Hu}, {Hunkeler}, {Ivezi{\'c}}, {Jain}, {Jenness}, {Kanarek}, {Kendrew},
  {Kern}, {Kerzendorf}, {Khvalko}, {King}, {Kirkby}, {Kulkarni}, {Kumar},
  {Lee}, {Lenz}, {Littlefair}, {Ma}, {Macleod}, {Mastropietro}, {McCully},
  {Montagnac}, {Morris}, {Mueller}, {Mumford}, {Muna}, {Murphy}, {Nelson},
  {Nguyen}, {Ninan}, {N{\"o}the}, {Ogaz}, {Oh}, {Parejko}, {Parley}, {Pascual},
  {Patil}, {Patil}, {Plunkett}, {Prochaska}, {Rastogi}, {Reddy Janga},
  {Sabater}, {Sakurikar}, {Seifert}, {Sherbert}, {Sherwood-Taylor}, {Shih},
  {Sick}, {Silbiger}, {Singanamalla}, {Singer}, {Sladen}, {Sooley},
  {Sornarajah}, {Streicher}, {Teuben}, {Thomas}, {Tremblay}, {Turner},
  {Terr{\'o}n}, {van Kerkwijk}, {de la Vega}, {Watkins}, {Weaver}, {Whitmore},
  {Woillez}, {Zabalza}, \& {Astropy Contributors}}]{astropy:2018}
{Astropy Collaboration}, {Price-Whelan}, A.~M., {Sip{\H{o}}cz}, B.~M., {et~al.}
  2018, \aj, 156, 123

\bibitem[{{Astropy Collaboration} {et~al.}(2013){Astropy Collaboration},
  {Robitaille}, {Tollerud}, {Greenfield}, {Droettboom}, {Bray}, {Aldcroft},
  {Davis}, {Ginsburg}, {Price-Whelan}, {Kerzendorf}, {Conley}, {Crighton},
  {Barbary}, {Muna}, {Ferguson}, {Grollier}, {Parikh}, {Nair}, {Unther},
  {Deil}, {Woillez}, {Conseil}, {Kramer}, {Turner}, {Singer}, {Fox}, {Weaver},
  {Zabalza}, {Edwards}, {Azalee Bostroem}, {Burke}, {Casey}, {Crawford},
  {Dencheva}, {Ely}, {Jenness}, {Labrie}, {Lim}, {Pierfederici}, {Pontzen},
  {Ptak}, {Refsdal}, {Servillat}, \& {Streicher}}]{astropy:2013}
{Astropy Collaboration}, {Robitaille}, T.~P., {Tollerud}, E.~J., {et~al.} 2013,
  \aap, 558, A33

\bibitem[{{Boubert} {et~al.}(2019){Boubert}, {Strader}, {Aguado}, {Seabroke},
  {Koposov}, {Sanders}, {Swihart}, {Chomiuk}, \& {Evans}}]{2019MNRAS.486.2618B}
{Boubert}, D., {Strader}, J., {Aguado}, D., {et~al.} 2019, \mnras, 486, 2618

\bibitem[{{Butkevich} {et~al.}(2017){Butkevich}, {Klioner}, {Lindegren},
  {Hobbs}, \& {van Leeuwen}}]{Butkevich2017}
{Butkevich}, A.~G., {Klioner}, S.~A., {Lindegren}, L., {Hobbs}, D., \& {van
  Leeuwen}, F. 2017, \aap, 603, A45

\bibitem[{{Carrasco} {et~al.}(2021){Carrasco}, {Weiler}, {Jordi}, {Fabricius},
  {De Angeli}, {Evans}, {van Leeuwen}, {Riello}, \&
  {Montegriffo}}]{EDR3-DPACP-119}
{Carrasco}, J.~M., {Weiler}, M., {Jordi}, C., {et~al.} 2021, arXiv e-prints,
  arXiv:2106.01752

\bibitem[{{Chambers} {et~al.}(2016){Chambers}, {Magnier}, {Metcalfe},
  {Flewelling}, {Huber}, {Waters}, {Denneau}, {Draper}, {Farrow}, {Finkbeiner},
  {Holmberg}, {Koppenhoefer}, {Price}, {Saglia}, {Schlafly}, {Smartt},
  {Sweeney}, {Wainscoat}, {Burgett}, {Grav}, {Heasley}, {Hodapp}, {Jedicke},
  {Kaiser}, {Kudritzki}, {Luppino}, {Lupton}, {Monet}, {Morgan}, {Onaka},
  {Stubbs}, {Tonry}, {Banados}, {Bell}, {Bender}, {Bernard}, {Botticella},
  {Casertano}, {Chastel}, {Chen}, {Chen}, {Cole}, {Deacon}, {Frenk},
  {Fitzsimmons}, {Gezari}, {Goessl}, {Goggia}, {Goldman}, {Grebel}, {Hambly},
  {Hasinger}, {Heavens}, {Heckman}, {Henderson}, {Henning}, {Holman}, {Hopp},
  {Ip}, {Isani}, {Keyes}, {Koekemoer}, {Kotak}, {Long}, {Lucey}, {Liu},
  {Martin}, {McLean}, {Morganson}, {Murphy}, {Nieto-Santisteban}, {Norberg},
  {Peacock}, {Pier}, {Postman}, {Primak}, {Rae}, {Rest}, {Riess}, {Riffeser},
  {Rix}, {Roser}, {Schilbach}, {Schultz}, {Scolnic}, {Szalay}, {Seitz},
  {Shiao}, {Small}, {Smith}, {Soderblom}, {Taylor}, {Thakar}, {Thiel},
  {Thilker}, {Urata}, {Valenti}, {Walter}, {Watters}, {Werner}, {White},
  {Wood-Vasey}, \& {Wyse}}]{2016arXiv161205560C}
{Chambers}, K.~C., {Magnier}, E.~A., {Metcalfe}, N., {et~al.} 2016, ArXiv
  e-prints [\eprint[arXiv]{1612.05560}]

\bibitem[{{Cropper} {et~al.}(2018){Cropper}, {Katz}, {Sartoretti}, {Prusti},
  {de Bruijne}, {Chassat}, {Charvet}, {Boyadjian}, {Perryman}, {Sarri}, {Gare},
  {Erdmann}, {Munari}, {Zwitter}, {Wilkinson}, {Arenou}, {Vallenari},
  {G{\'o}mez}, {Panuzzo}, {Seabroke}, {Allende Prieto}, {Benson}, {Marchal},
  {Huckle}, {Smith}, {Dolding}, {Jan{\ss}en}, {Viala}, {Blomme}, {Baker},
  {Boudreault}, {Crifo}, {Soubiran}, {Fr{\'e}mat}, {Jasniewicz}, {Guerrier},
  {Guy}, {Turon}, {Jean-Antoine-Piccolo}, {Th{\'e}venin}, {David}, {Gosset}, \&
  {Damerdji}}]{2018A&A...616A...5C}
{Cropper}, M., {Katz}, D., {Sartoretti}, P., {et~al.} 2018, \aap, 616, A5

\bibitem[{{De Angeli et al.}(2021)}]{EDR3-DPACP-118}
{De Angeli et al.} 2021, \aap\ in prep.

\bibitem[{{Evans} {et~al.}(2018){Evans}, {Riello}, {De Angeli}, {Carrasco},
  {Montegriffo}, {Fabricius}, {Jordi}, {Palaversa}, {Diener}, {Busso},
  {Cacciari}, {van Leeuwen}, {Burgess}, {Davidson}, {Harrison}, {Hodgkin},
  {Pancino}, {Richards}, {Altavilla}, {Balaguer-N{\'u}{\~n}ez}, {Barstow},
  {Bellazzini}, {Brown}, {Castellani}, {Cocozza}, {De Luise}, {Delgado},
  {Ducourant}, {Galleti}, {Gilmore}, {Giuffrida}, {Holl}, {Kewley}, {Koposov},
  {Marinoni}, {Marrese}, {Osborne}, {Piersimoni}, {Portell}, {Pulone},
  {Ragaini}, {Sanna}, {Terrett}, {Walton}, {Wevers}, \&
  {Wyrzykowski}}]{2018A&A...616A...4E}
{Evans}, D.~W., {Riello}, M., {De Angeli}, F., {et~al.} 2018, \aap, 616, A4

\bibitem[{{Fabricius} {et~al.}(2002){Fabricius}, {H{\o}g}, {Makarov}, {Mason},
  {Wycoff}, \& {Urban}}]{2002A&A...384..180F}
{Fabricius}, C., {H{\o}g}, E., {Makarov}, V.~V., {et~al.} 2002, \aap, 384, 180

\bibitem[{{Fabricius} {et~al.}(2021){Fabricius}, {Luri}, {Arenou}, {Babusiaux},
  {Helmi}, {Muraveva}, {Reyl{\'e}}, {Spoto}, {Vallenari}, {Antoja}, {Balbinot},
  {Barache}, {Bauchet}, {Bragaglia}, {Busonero}, {Cantat-Gaudin}, {Carrasco},
  {Diakit{\'e}}, {Fabrizio}, {Figueras}, {Garcia-Gutierrez}, {Garofalo},
  {Jordi}, {Kervella}, {Khanna}, {Leclerc}, {Licata}, {Lambert}, {Marrese},
  {Masip}, {Ramos}, {Robichon}, {Robin}, {Romero-G{\'o}mez}, {Rubele}, \&
  {Weiler}}]{EDR3-DPACP-126}
{Fabricius}, C., {Luri}, X., {Arenou}, F., {et~al.} 2021, \aap, 649, A5

\bibitem[{{Gaia Collaboration} {et~al.}(2021{\natexlab{a}}){Gaia
  Collaboration}, {Antoja}, {McMillan}, {Kordopatis}, {Ramos}, {Helmi},
  {Balbinot}, {Cantat-Gaudin}, {Chemin}, {Figueras}, {Jordi}, {Khanna},
  {Romero-G{\'o}mez}, {Seabroke}, {Brown}, {Vallenari}, {Prusti}, {de Bruijne},
  {Babusiaux}, {Biermann}, {Creevey}, {Evans}, {Eyer}, {Hutton}, {Jansen},
  {Klioner}, {Lammers}, {Lindegren}, {Luri}, {Mignard}, {Panem}, {Pourbaix},
  {Randich}, {Sartoretti}, {Soubiran}, {Walton}, {Arenou}, {Bailer-Jones},
  {Bastian}, {Cropper}, {Drimmel}, {Katz}, {Lattanzi}, {van Leeuwen}, {Bakker},
  {Casta{\~n}eda}, {De Angeli}, {Ducourant}, {Fabricius}, {Fouesneau},
  {Fr{\'e}mat}, {Guerra}, {Guerrier}, {Guiraud}, {Jean-Antoine Piccolo},
  {Masana}, {Messineo}, {Mowlavi}, {Nicolas}, {Nienartowicz}, {Pailler},
  {Panuzzo}, {Riclet}, {Roux}, {Sordo}, {Tanga}, {Th{\'e}venin},
  {Gracia-Abril}, {Portell}, {Teyssier}, {Altmann}, {Andrae}, {Bellas-Velidis},
  {Benson}, {Berthier}, {Blomme}, {Brugaletta}, {Burgess}, {Busso}, {Carry},
  {Cellino}, {Cheek}, {Clementini}, {Damerdji}, {Davidson}, {Delchambre},
  {Dell'Oro}, {Fern{\'a}ndez-Hern{\'a}ndez}, {Galluccio}, {Garc{\'\i}a-Lario},
  {Garcia-Reinaldos}, {Gonz{\'a}lez-N{\'u}{\~n}ez}, {Gosset}, {Haigron},
  {Halbwachs}, {Hambly}, {Harrison}, {Hatzidimitriou}, {Heiter},
  {Hern{\'a}ndez}, {Hestroffer}, {Hodgkin}, {Holl}, {Jan{\ss}en}, {Jevardat de
  Fombelle}, {Jordan}, {Krone-Martins}, {Lanzafame}, {L{\"o}ffler}, {Lorca},
  {Manteiga}, {Marchal}, {Marrese}, {Moitinho}, {Mora}, {Muinonen}, {Osborne},
  {Pancino}, {Pauwels}, {Recio-Blanco}, {Richards}, {Riello}, {Rimoldini},
  {Robin}, {Roegiers}, {Rybizki}, {Sarro}, {Siopis}, {Smith}, {Sozzetti},
  {Ulla}, {Utrilla}, {van Leeuwen}, {van Reeven}, {Abbas}, {Abreu Aramburu},
  {Accart}, {Aerts}, {Aguado}, {Ajaj}, {Altavilla}, {{\'A}lvarez}, {{\'A}lvarez
  Cid-Fuentes}, {Alves}, {Anderson}, {Varela}, {Audard}, {Baines}, {Baker},
  {Balaguer-N{\'u}{\~n}ez}, {Balog}, {Barache}, {Barbato}, {Barros}, {Barstow},
  {Bartolom{\'e}}, {Bassilana}, {Bauchet}, {Baudesson-Stella}, {Becciani},
  {Bellazzini}, {Bernet}, {Bertone}, {Bianchi}, {Blanco-Cuaresma}, {Boch},
  {Bombrun}, {Bossini}, {Bouquillon}, {Bragaglia}, {Bramante}, {Breedt},
  {Bressan}, {Brouillet}, {Bucciarelli}, {Burlacu}, {Busonero}, {Butkevich},
  {Buzzi}, {Caffau}, {Cancelliere}, {C{\'a}novas}, {Carballo}, {Carlucci},
  {Carnerero}, {Carrasco}, {Casamiquela}, {Castellani}, {Castro-Ginard},
  {Castro Sampol}, {Chaoul}, {Charlot}, {Chiavassa}, {Cioni}, {Comoretto},
  {Cooper}, {Cornez}, {Cowell}, {Crifo}, {Crosta}, {Crowley}, {Dafonte},
  {Dapergolas}, {David}, {David}, {de Laverny}, {De Luise}, {De March}, {De
  Ridder}, {de Souza}, {de Teodoro}, {de Torres}, {del Peloso}, {del Pozo},
  {Delgado}, {Delgado}, {Delisle}, {Di Matteo}, {Diakite}, {Diener},
  {Distefano}, {Dolding}, {Eappachen}, {Enke}, {Esquej}, {Fabre}, {Fabrizio},
  {Faigler}, {Fedorets}, {Fernique}, {Fienga}, {Fouron}, {Fragkoudi}, {Fraile},
  {Franke}, {Gai}, {Garabato}, {Garcia-Gutierrez}, {Garc{\'\i}a-Torres},
  {Garofalo}, {Gavras}, {Gerlach}, {Geyer}, {Giacobbe}, {Gilmore}, {Girona},
  {Giuffrida}, {Gomez}, {Gonzalez-Santamaria}, {Gonz{\'a}lez-Vidal}, {Granvik},
  {Guti{\'e}rrez-S{\'a}nchez}, {Guy}, {Hauser}, {Haywood}, {Hidalgo}, {Hilger},
  {H{\l}adczuk}, {Hobbs}, {Holland}, {Huckle}, {Jasniewicz}, {Jonker},
  {Juaristi Campillo}, {Julbe}, {Karbevska}, {Kervella}, {Kochoska},
  {Kontizas}, {Korn}, {Kostrzewa-Rutkowska}, {Kruszy{\'n}ska}, {Lambert},
  {Lanza}, {Lasne}, {Le Campion}, {Le Fustec}, {Lebreton}, {Lebzelter},
  {Leccia}, {Leclerc}, {Lecoeur-Taibi}, {Liao}, {Licata}, {Lindstr{\o}m},
  {Lister}, {Livanou}, {Lobel}, {Madrero Pardo}, {Managau}, {Mann}, {Marchant},
  {Marconi}, {Marcos Santos}, {Marinoni}, {Marocco}, {Marshall}, {Martin Polo},
  {Mart{\'\i}n-Fleitas}, {Masip}, {Massari}, {Mastrobuono-Battisti}, {Mazeh},
  {Messina}, {Michalik}, {Millar}, {Mints}, {Molina}, {Molinaro}, {Moln{\'a}r},
  {Montegriffo}, {Mor}, {Morbidelli}, {Morel}, {Morris}, {Mulone}, {Munoz},
  {Muraveva}, {Murphy}, {Musella}, {Noval}, {Ord{\'e}novic}, {Orr{\`u}},
  {Osinde}, {Pagani}, {Pagano}, {Palaversa}, {Palicio}, {Panahi}, {Pawlak},
  {Pe{\~n}alosa Esteller}, {Penttil{\"a}}, {Piersimoni}, {Pineau}, {Plachy},
  {Plum}, {Poggio}, {Poretti}, {Poujoulet}, {Pr{\v{s}}a}, {Pulone}, {Racero},
  {Ragaini}, {Rainer}, {Raiteri}, {Rambaux}, {Ramos-Lerate}, {Re Fiorentin},
  {Regibo}, {Reyl{\'e}}, {Ripepi}, {Riva}, {Rixon}, {Robichon}, {Robin},
  {Roelens}, {Rohrbasser}, {Rowell}, {Royer}, {Rybicki}, {Sadowski},
  {Sagrist{\`a} Sell{\'e}s}, {Sahlmann}, {Salgado}, {Salguero}, {Samaras},
  {Sanchez Gimenez}, {Sanna}, {Santove{\~n}a}, {Sarasso}, {Schultheis},
  {Sciacca}, {Segol}, {Segovia}, {S{\'e}gransan}, {Semeux}, {Siddiqui},
  {Siebert}, {Siltala}, {Slezak}, {Smart}, {Solano}, {Solitro}, {Souami},
  {Souchay}, {Spagna}, {Spoto}, {Steele}, {Steidelm{\"u}ller}, {Stephenson},
  {S{\"u}veges}, {Szabados}, {Szegedi-Elek}, {Taris}, {Tauran}, {Taylor},
  {Teixeira}, {Thuillot}, {Tonello}, {Torra}, {Torra}, {Turon}, {Unger},
  {Vaillant}, {van Dillen}, {Vanel}, {Vecchiato}, {Viala}, {Vicente},
  {Voutsinas}, {Weiler}, {Wevers}, {Wyrzykowski}, {Yoldas}, {Yvard}, {Zhao},
  {Zorec}, {Zucker}, {Zurbach}, \& {Zwitter}}]{EDR3-DPACP-113}
{Gaia Collaboration}, {Antoja}, T., {McMillan}, P.~J., {et~al.}
  2021{\natexlab{a}}, \aap, 649, A8

\bibitem[{{Gaia Collaboration} {et~al.}(2018){Gaia Collaboration}, {Brown},
  {Vallenari}, {Prusti}, {de Bruijne}, {Babusiaux}, {Bailer-Jones}, {Biermann},
  {Evans}, {Eyer}, {Jansen}, {Jordi}, {Klioner}, {Lammers}, {Lindegren},
  {Luri}, {Mignard}, {Panem}, {Pourbaix}, {Randich}, {Sartoretti}, {Siddiqui},
  {Soubiran}, {van Leeuwen}, {Walton}, {Arenou}, {Bastian}, {Cropper},
  {Drimmel}, {Katz}, {Lattanzi}, {Bakker}, {Cacciari}, {Casta{\~n}eda},
  {Chaoul}, {Cheek}, {De Angeli}, {Fabricius}, {Guerra}, {Holl}, {Masana},
  {Messineo}, {Mowlavi}, {Nienartowicz}, {Panuzzo}, {Portell}, {Riello},
  {Seabroke}, {Tanga}, {Th{\'e}venin}, {Gracia-Abril}, {Comoretto},
  {Garcia-Reinaldos}, {Teyssier}, {Altmann}, {Andrae}, {Audard},
  {Bellas-Velidis}, {Benson}, {Berthier}, {Blomme}, {Burgess}, {Busso},
  {Carry}, {Cellino}, {Clementini}, {Clotet}, {Creevey}, {Davidson}, {De
  Ridder}, {Delchambre}, {Dell'Oro}, {Ducourant},
  {Fern{\'a}ndez-Hern{\'a}ndez}, {Fouesneau}, {Fr{\'e}mat}, {Galluccio},
  {Garc{\'\i}a-Torres}, {Gonz{\'a}lez-N{\'u}{\~n}ez}, {Gonz{\'a}lez-Vidal},
  {Gosset}, {Guy}, {Halbwachs}, {Hambly}, {Harrison}, {Hern{\'a}ndez},
  {Hestroffer}, {Hodgkin}, {Hutton}, {Jasniewicz}, {Jean-Antoine-Piccolo},
  {Jordan}, {Korn}, {Krone-Martins}, {Lanzafame}, {Lebzelter}, {L{\"o}ffler},
  {Manteiga}, {Marrese}, {Mart{\'\i}n-Fleitas}, {Moitinho}, {Mora}, {Muinonen},
  {Osinde}, {Pancino}, {Pauwels}, {Petit}, {Recio-Blanco}, {Richards},
  {Rimoldini}, {Robin}, {Sarro}, {Siopis}, {Smith}, {Sozzetti}, {S{\"u}veges},
  {Torra}, {van Reeven}, {Abbas}, {Abreu Aramburu}, {Accart}, {Aerts},
  {Altavilla}, {{\'A}lvarez}, {Alvarez}, {Alves}, {Anderson}, {Andrei},
  {Anglada Varela}, {Antiche}, {Antoja}, {Arcay}, {Astraatmadja}, {Bach},
  {Baker}, {Balaguer-N{\'u}{\~n}ez}, {Balm}, {Barache}, {Barata}, {Barbato},
  {Barblan}, {Barklem}, {Barrado}, {Barros}, {Barstow}, {Bartholom{\'e}
  Mu{\~n}oz}, {Bassilana}, {Becciani}, {Bellazzini}, {Berihuete}, {Bertone},
  {Bianchi}, {Bienaym{\'e}}, {Blanco-Cuaresma}, {Boch}, {Boeche}, {Bombrun},
  {Borrachero}, {Bossini}, {Bouquillon}, {Bourda}, {Bragaglia}, {Bramante},
  {Breddels}, {Bressan}, {Brouillet}, {Br{\"u}semeister}, {Brugaletta},
  {Bucciarelli}, {Burlacu}, {Busonero}, {Butkevich}, {Buzzi}, {Caffau},
  {Cancelliere}, {Cannizzaro}, {Cantat-Gaudin}, {Carballo}, {Carlucci},
  {Carrasco}, {Casamiquela}, {Castellani}, {Castro-Ginard}, {Charlot},
  {Chemin}, {Chiavassa}, {Cocozza}, {Costigan}, {Cowell}, {Crifo}, {Crosta},
  {Crowley}, {Cuypers}, {Dafonte}, {Damerdji}, {Dapergolas}, {David}, {David},
  {de Laverny}, {De Luise}, {De March}, {de Martino}, {de Souza}, {de Torres},
  {Debosscher}, {del Pozo}, {Delbo}, {Delgado}, {Delgado}, {Di Matteo},
  {Diakite}, {Diener}, {Distefano}, {Dolding}, {Drazinos}, {Dur{\'a}n},
  {Edvardsson}, {Enke}, {Eriksson}, {Esquej}, {Eynard Bontemps}, {Fabre},
  {Fabrizio}, {Faigler}, {Falc{\~a}o}, {Farr{\`a}s Casas}, {Federici},
  {Fedorets}, {Fernique}, {Figueras}, {Filippi}, {Findeisen}, {Fonti},
  {Fraile}, {Fraser}, {Fr{\'e}zouls}, {Gai}, {Galleti}, {Garabato},
  {Garc{\'\i}a-Sedano}, {Garofalo}, {Garralda}, {Gavel}, {Gavras}, {Gerssen},
  {Geyer}, {Giacobbe}, {Gilmore}, {Girona}, {Giuffrida}, {Glass}, {Gomes},
  {Granvik}, {Gueguen}, {Guerrier}, {Guiraud}, {Guti{\'e}rrez-S{\'a}nchez},
  {Haigron}, {Hatzidimitriou}, {Hauser}, {Haywood}, {Heiter}, {Helmi}, {Heu},
  {Hilger}, {Hobbs}, {Hofmann}, {Holland}, {Huckle}, {Hypki}, {Icardi},
  {Jan{\ss}en}, {Jevardat de Fombelle}, {Jonker}, {Juh{\'a}sz}, {Julbe},
  {Karampelas}, {Kewley}, {Klar}, {Kochoska}, {Kohley}, {Kolenberg},
  {Kontizas}, {Kontizas}, {Koposov}, {Kordopatis}, {Kostrzewa-Rutkowska},
  {Koubsky}, {Lambert}, {Lanza}, {Lasne}, {Lavigne}, {Le Fustec}, {Le
  Poncin-Lafitte}, {Lebreton}, {Leccia}, {Leclerc}, {Lecoeur-Taibi},
  {Lenhardt}, {Leroux}, {Liao}, {Licata}, {Lindstr{\o}m}, {Lister}, {Livanou},
  {Lobel}, {L{\'o}pez}, {Managau}, {Mann}, {Mantelet}, {Marchal}, {Marchant},
  {Marconi}, {Marinoni}, {Marschalk{\'o}}, {Marshall}, {Martino}, {Marton},
  {Mary}, {Massari}, {Matijevi{\v{c}}}, {Mazeh}, {McMillan}, {Messina},
  {Michalik}, {Millar}, {Molina}, {Molinaro}, {Moln{\'a}r}, {Montegriffo},
  {Mor}, {Morbidelli}, {Morel}, {Morris}, {Mulone}, {Muraveva}, {Musella},
  {Nelemans}, {Nicastro}, {Noval}, {O'Mullane}, {Ord{\'e}novic},
  {Ord{\'o}{\~n}ez-Blanco}, {Osborne}, {Pagani}, {Pagano}, {Pailler},
  {Palacin}, {Palaversa}, {Panahi}, {Pawlak}, {Piersimoni}, {Pineau}, {Plachy},
  {Plum}, {Poggio}, {Poujoulet}, {Pr{\v{s}}a}, {Pulone}, {Racero}, {Ragaini},
  {Rambaux}, {Ramos-Lerate}, {Regibo}, {Reyl{\'e}}, {Riclet}, {Ripepi}, {Riva},
  {Rivard}, {Rixon}, {Roegiers}, {Roelens}, {Romero-G{\'o}mez}, {Rowell},
  {Royer}, {Ruiz-Dern}, {Sadowski}, {Sagrist{\`a} Sell{\'e}s}, {Sahlmann},
  {Salgado}, {Salguero}, {Sanna}, {Santana-Ros}, {Sarasso}, {Savietto},
  {Schultheis}, {Sciacca}, {Segol}, {Segovia}, {S{\'e}gransan}, {Shih},
  {Siltala}, {Silva}, {Smart}, {Smith}, {Solano}, {Solitro}, {Sordo}, {Soria
  Nieto}, {Souchay}, {Spagna}, {Spoto}, {Stampa}, {Steele},
  {Steidelm{\"u}ller}, {Stephenson}, {Stoev}, {Suess}, {Surdej}, {Szabados},
  {Szegedi-Elek}, {Tapiador}, {Taris}, {Tauran}, {Taylor}, {Teixeira},
  {Terrett}, {Teyssand ier}, {Thuillot}, {Titarenko}, {Torra Clotet}, {Turon},
  {Ulla}, {Utrilla}, {Uzzi}, {Vaillant}, {Valentini}, {Valette}, {van Elteren},
  {Van Hemelryck}, {van Leeuwen}, {Vaschetto}, {Vecchiato}, {Veljanoski},
  {Viala}, {Vicente}, {Vogt}, {von Essen}, {Voss}, {Votruba}, {Voutsinas},
  {Walmsley}, {Weiler}, {Wertz}, {Wevers}, {Wyrzykowski}, {Yoldas},
  {{\v{Z}}erjal}, {Ziaeepour}, {Zorec}, {Zschocke}, {Zucker}, {Zurbach}, \&
  {Zwitter}}]{2018A&A...616A...1G}
{Gaia Collaboration}, {Brown}, A.~G.~A., {Vallenari}, A., {et~al.} 2018, \aap,
  616, A1

\bibitem[{{Gaia Collaboration} {et~al.}(2016{\natexlab{a}}){Gaia
  Collaboration}, {Brown}, {Vallenari}, {Prusti}, {de Bruijne}, {Mignard},
  {Drimmel}, {Babusiaux}, {Bailer-Jones}, {Bastian}, \&
  et~al.}]{2016A&A...595A...2G}
{Gaia Collaboration}, {Brown}, A.~G.~A., {Vallenari}, A., {et~al.}
  2016{\natexlab{a}}, \aap, 595, A2

\bibitem[{{Gaia Collaboration} {et~al.}(2021{\natexlab{b}}){Gaia
  Collaboration}, {Klioner}, \& {et al.}}]{EDR3-DPACP-133}
{Gaia Collaboration}, {Klioner}, S., \& {et al.} 2021{\natexlab{b}}, \aap\ in
  prep.

\bibitem[{{Gaia Collaboration} {et~al.}(2021{\natexlab{c}}){Gaia
  Collaboration}, {Klioner}, {Mignard}, {Lindegren}, {Bastian}, {McMillan},
  {Hern{\'a}ndez}, {Hobbs}, {Ramos-Lerate}, {Biermann}, {Bombrun}, {de Torres},
  {Gerlach}, {Geyer}, {Hilger}, {Lammers}, {Steidelm{\"u}ller}, {Stephenson},
  {Brown}, {Vallenari}, {Prusti}, {de Bruijne}, {Babusiaux}, {Creevey},
  {Evans}, {Eyer}, {Hutton}, {Jansen}, {Jordi}, {Luri}, {Panem}, {Pourbaix},
  {Randich}, {Sartoretti}, {Soubiran}, {Walton}, {Arenou}, {Bailer-Jones},
  {Cropper}, {Drimmel}, {Katz}, {Lattanzi}, {van Leeuwen}, {Bakker},
  {Casta{\~n}eda}, {De Angeli}, {Ducourant}, {Fabricius}, {Fouesneau},
  {Fr{\'e}mat}, {Guerra}, {Guerrier}, {Guiraud}, {Jean-Antoine Piccolo},
  {Masana}, {Messineo}, {Mowlavi}, {Nicolas}, {Nienartowicz}, {Pailler},
  {Panuzzo}, {Riclet}, {Roux}, {Seabroke}, {Sordo}, {Tanga}, {Th{\'e}venin},
  {Gracia-Abril}, {Portell}, {Teyssier}, {Altmann}, {Andrae}, {Bellas-Velidis},
  {Benson}, {Berthier}, {Blomme}, {Brugaletta}, {Burgess}, {Busso}, {Carry},
  {Cellino}, {Cheek}, {Clementini}, {Damerdji}, {Davidson}, {Delchambre},
  {Dell'Oro}, {Fern{\'a}ndez-Hern{\'a}ndez}, {Galluccio}, {Garc{\'\i}a-Lario},
  {Garcia-Reinaldos}, {Gonz{\'a}lez-N{\'u}{\~n}ez}, {Gosset}, {Haigron},
  {Halbwachs}, {Hambly}, {Harrison}, {Hatzidimitriou}, {Heiter}, {Hestroffer},
  {Hodgkin}, {Holl}, {Jan{\ss}en}, {Jevardat de Fombelle}, {Jordan},
  {Krone-Martins}, {Lanzafame}, {L{\"o}ffler}, {Lorca}, {Manteiga}, {Marchal},
  {Marrese}, {Moitinho}, {Mora}, {Muinonen}, {Osborne}, {Pancino}, {Pauwels},
  {Recio-Blanco}, {Richards}, {Riello}, {Rimoldini}, {Robin}, {Roegiers},
  {Rybizki}, {Sarro}, {Siopis}, {Smith}, {Sozzetti}, {Ulla}, {Utrilla}, {van
  Leeuwen}, {van Reeven}, {Abbas}, {Abreu Aramburu}, {Accart}, {Aerts},
  {Aguado}, {Ajaj}, {Altavilla}, {{\'A}lvarez}, {{\'A}lvarez Cid-Fuentes},
  {Alves}, {Anderson}, {Anglada Varela}, {Antoja}, {Audard}, {Baines}, {Baker},
  {Balaguer-N{\'u}{\~n}ez}, {Balbinot}, {Balog}, {Barache}, {Barbato},
  {Barros}, {Barstow}, {Bartolom{\'e}}, {Bassilana}, {Bauchet},
  {Baudesson-Stella}, {Becciani}, {Bellazzini}, {Bernet}, {Bertone}, {Bianchi},
  {Blanco-Cuaresma}, {Boch}, {Bossini}, {Bouquillon}, {Bramante}, {Breedt},
  {Bressan}, {Brouillet}, {Bucciarelli}, {Burlacu}, {Busonero}, {Butkevich},
  {Buzzi}, {Caffau}, {Cancelliere}, {C{\'a}novas}, {Cantat-Gaudin}, {Carballo},
  {Carlucci}, {Carnerero}, {Carrasco}, {Casamiquela}, {Castellani},
  {Castro-Ginard}, {Castro Sampol}, {Chaoul}, {Charlot}, {Chemin}, {Chiavassa},
  {Comoretto}, {Cooper}, {Cornez}, {Cowell}, {Crifo}, {Crosta}, {Crowley},
  {Dafonte}, {Dapergolas}, {David}, {David}, {de Laverny}, {De Luise}, {De
  March}, {De Ridder}, {de Souza}, {de Teodoro}, {del Peloso}, {del Pozo},
  {Delgado}, {Delgado}, {Delisle}, {Di Matteo}, {Diakite}, {Diener},
  {Distefano}, {Dolding}, {Eappachen}, {Enke}, {Esquej}, {Fabre}, {Fabrizio},
  {Faigler}, {Fedorets}, {Fernique}, {Fienga}, {Figueras}, {Fouron},
  {Fragkoudi}, {Fraile}, {Franke}, {Gai}, {Garabato}, {Garcia-Gutierrez},
  {Garc{\'\i}a-Torres}, {Garofalo}, {Gavras}, {Giacobbe}, {Gilmore}, {Girona},
  {Giuffrida}, {Gomez}, {Gonzalez-Santamaria}, {Gonz{\'a}lez-Vidal}, {Granvik},
  {Guti{\'e}rrez-S{\'a}nchez}, {Guy}, {Hauser}, {Haywood}, {Helmi}, {Hidalgo},
  {H{\l}adczuk}, {Holland}, {Huckle}, {Jasniewicz}, {Jonker}, {Juaristi
  Campillo}, {Julbe}, {Karbevska}, {Kervella}, {Khanna}, {Kochoska},
  {Kordopatis}, {Korn}, {Kostrzewa-Rutkowska}, {Kruszy{\'n}ska}, {Lambert},
  {Lanza}, {Lasne}, {Le Campion}, {Le Fustec}, {Lebreton}, {Lebzelter},
  {Leccia}, {Leclerc}, {Lecoeur-Taibi}, {Liao}, {Licata}, {Lindstr{\o}m},
  {Lister}, {Livanou}, {Lobel}, {Madrero Pardo}, {Managau}, {Mann}, {Marchant},
  {Marconi}, {Marcos Santos}, {Marinoni}, {Marocco}, {Marshall}, {Martin Polo},
  {Mart{\'\i}n-Fleitas}, {Masip}, {Massari}, {Mastrobuono-Battisti}, {Mazeh},
  {Messina}, {Michalik}, {Millar}, {Mints}, {Molina}, {Molinaro}, {Moln{\'a}r},
  {Montegriffo}, {Mor}, {Morbidelli}, {Morel}, {Morris}, {Mulone}, {Munoz},
  {Muraveva}, {Murphy}, {Musella}, {Noval}, {Ord{\'e}novic}, {Orr{\`u}},
  {Osinde}, {Pagani}, {Pagano}, {Palaversa}, {Palicio}, {Panahi}, {Pawlak},
  {Pe{\~n}alosa Esteller}, {Penttil{\"a}}, {Piersimoni}, {Pineau}, {Plachy},
  {Plum}, {Poggio}, {Poretti}, {Poujoulet}, {Pr{\v{s}}a}, {Pulone}, {Racero},
  {Ragaini}, {Rainer}, {Raiteri}, {Rambaux}, {Ramos}, {Re Fiorentin}, {Regibo},
  {Reyl{\'e}}, {Ripepi}, {Riva}, {Rixon}, {Robichon}, {Robin}, {Roelens},
  {Rohrbasser}, {Romero-G{\'o}mez}, {Rowell}, {Royer}, {Rybicki}, {Sadowski},
  {Sagrist{\`a} Sell{\'e}s}, {Sahlmann}, {Salgado}, {Salguero}, {Samaras},
  {Sanchez Gimenez}, {Sanna}, {Santove{\~n}a}, {Sarasso}, {Schultheis},
  {Sciacca}, {Segol}, {Segovia}, {S{\'e}gransan}, {Semeux}, {Siddiqui},
  {Siebert}, {Siltala}, {Slezak}, {Smart}, {Solano}, {Solitro}, {Souami},
  {Souchay}, {Spagna}, {Spoto}, {Steele}, {S{\"u}veges}, {Szabados},
  {Szegedi-Elek}, {Taris}, {Tauran}, {Taylor}, {Teixeira}, {Thuillot},
  {Tonello}, {Torra}, {Torra}, {Turon}, {Unger}, {Vaillant}, {van Dillen},
  {Vanel}, {Vecchiato}, {Viala}, {Vicente}, {Voutsinas}, {Weiler}, {Wevers},
  {Wyrzykowski}, {Yoldas}, {Yvard}, {Zhao}, {Zorec}, {Zucker}, {Zurbach}, \&
  {Zwitter}}]{EDR3-DPACP-134}
{Gaia Collaboration}, {Klioner}, S.~A., {Mignard}, F., {et~al.}
  2021{\natexlab{c}}, \aap, 649, A9

\bibitem[{{Gaia Collaboration} {et~al.}(2021{\natexlab{d}}){Gaia
  Collaboration}, {Luri}, {Chemin}, {Clementini}, {Delgado}, {McMillan},
  {Romero-G{\'o}mez}, {Balbinot}, {Castro-Ginard}, {Mor}, {Ripepi}, {Sarro},
  {Cioni}, {Fabricius}, {Garofalo}, {Helmi}, {Muraveva}, {Brown}, {Vallenari},
  {Prusti}, {de Bruijne}, {Babusiaux}, {Biermann}, {Creevey}, {Evans}, {Eyer},
  {Hutton}, {Jansen}, {Jordi}, {Klioner}, {Lammers}, {Lindegren}, {Mignard},
  {Panem}, {Pourbaix}, {Randich}, {Sartoretti}, {Soubiran}, {Walton}, {Arenou},
  {Bailer-Jones}, {Bastian}, {Cropper}, {Drimmel}, {Katz}, {Lattanzi}, {van
  Leeuwen}, {Bakker}, {Casta{\~n}eda}, {De Angeli}, {Ducourant}, {Fouesneau},
  {Fr{\'e}mat}, {Guerra}, {Guerrier}, {Guiraud}, {Jean-Antoine Piccolo},
  {Masana}, {Messineo}, {Mowlavi}, {Nicolas}, {Nienartowicz}, {Pailler},
  {Panuzzo}, {Riclet}, {Roux}, {Seabroke}, {Sordo}, {Tanga}, {Th{\'e}venin},
  {Gracia-Abril}, {Portell}, {Teyssier}, {Altmann}, {Andrae}, {Bellas-Velidis},
  {Benson}, {Berthier}, {Blomme}, {Brugaletta}, {Burgess}, {Busso}, {Carry},
  {Cellino}, {Cheek}, {Damerdji}, {Davidson}, {Delchambre}, {Dell'Oro},
  {Fern{\'a}ndez-Hern{\'a}ndez}, {Galluccio}, {Garc{\'\i}a-Lario},
  {Garcia-Reinaldos}, {Gonz{\'a}lez-N{\'u}{\~n}ez}, {Gosset}, {Haigron},
  {Halbwachs}, {Hambly}, {Harrison}, {Hatzidimitriou}, {Heiter},
  {Hern{\'a}ndez}, {Hestroffer}, {Hodgkin}, {Holl}, {Jan{\ss}en}, {Jevardat de
  Fombelle}, {Jordan}, {Krone-Martins}, {Lanzafame}, {L{\"o}ffler}, {Lorca},
  {Manteiga}, {Marchal}, {Marrese}, {Moitinho}, {Mora}, {Muinonen}, {Osborne},
  {Pancino}, {Pauwels}, {Recio-Blanco}, {Richards}, {Riello}, {Rimoldini},
  {Robin}, {Roegiers}, {Rybizki}, {Siopis}, {Smith}, {Sozzetti}, {Ulla},
  {Utrilla}, {van Leeuwen}, {van Reeven}, {Abbas}, {Abreu Aramburu}, {Accart},
  {Aerts}, {Aguado}, {Ajaj}, {Altavilla}, {{\'A}lvarez}, {{\'A}lvarez
  Cid-Fuentes}, {Alves}, {Anderson}, {Anglada Varela}, {Antoja}, {Audard},
  {Baines}, {Baker}, {Balaguer-N{\'u}{\~n}ez}, {Balog}, {Barache}, {Barbato},
  {Barros}, {Barstow}, {Bartolom{\'e}}, {Bassilana}, {Bauchet},
  {Baudesson-Stella}, {Becciani}, {Bellazzini}, {Bernet}, {Bertone}, {Bianchi},
  {Blanco-Cuaresma}, {Boch}, {Bombrun}, {Bossini}, {Bouquillon}, {Bragaglia},
  {Bramante}, {Breedt}, {Bressan}, {Brouillet}, {Bucciarelli}, {Burlacu},
  {Busonero}, {Butkevich}, {Buzzi}, {Caffau}, {Cancelliere}, {C{\'a}novas},
  {Cantat-Gaudin}, {Carballo}, {Carlucci}, {Carnerero}, {Carrasco},
  {Casamiquela}, {Castellani}, {Castro Sampol}, {Chaoul}, {Charlot},
  {Chiavassa}, {Comoretto}, {Cooper}, {Cornez}, {Cowell}, {Crifo}, {Crosta},
  {Crowley}, {Dafonte}, {Dapergolas}, {David}, {David}, {de Laverny}, {De
  Luise}, {De March}, {De Ridder}, {de Souza}, {de Teodoro}, {de Torres}, {del
  Peloso}, {del Pozo}, {Delgado}, {Delisle}, {Di Matteo}, {Diakite}, {Diener},
  {Distefano}, {Dolding}, {Eappachen}, {Enke}, {Esquej}, {Fabre}, {Fabrizio},
  {Faigler}, {Fedorets}, {Fernique}, {Fienga}, {Figueras}, {Fouron},
  {Fragkoudi}, {Fraile}, {Franke}, {Gai}, {Garabato}, {Garcia-Gutierrez},
  {Garc{\'\i}a-Torres}, {Gavras}, {Gerlach}, {Geyer}, {Giacobbe}, {Gilmore},
  {Girona}, {Giuffrida}, {Gomez}, {Gonzalez-Santamaria}, {Gonz{\'a}lez-Vidal},
  {Granvik}, {Guti{\'e}rrez-S{\'a}nchez}, {Guy}, {Hauser}, {Haywood},
  {Hidalgo}, {Hilger}, {H{\l}adczuk}, {Hobbs}, {Holland}, {Huckle},
  {Jasniewicz}, {Jonker}, {Juaristi Campillo}, {Julbe}, {Karbevska},
  {Kervella}, {Khanna}, {Kochoska}, {Kontizas}, {Kordopatis}, {Korn},
  {Kostrzewa-Rutkowska}, {Kruszy{\'n}ska}, {Lambert}, {Lanza}, {Lasne}, {Le
  Campion}, {Le Fustec}, {Lebreton}, {Lebzelter}, {Leccia}, {Leclerc},
  {Lecoeur-Taibi}, {Liao}, {Licata}, {Lindstr{\o}m}, {Lister}, {Livanou},
  {Lobel}, {Madrero Pardo}, {Managau}, {Mann}, {Marchant}, {Marconi}, {Marcos
  Santos}, {Marinoni}, {Marocco}, {Marshall}, {Martin Polo},
  {Mart{\'\i}n-Fleitas}, {Masip}, {Massari}, {Mastrobuono-Battisti}, {Mazeh},
  {Messina}, {Michalik}, {Millar}, {Mints}, {Molina}, {Molinaro}, {Moln{\'a}r},
  {Montegriffo}, {Morbidelli}, {Morel}, {Morris}, {Mulone}, {Munoz}, {Murphy},
  {Musella}, {Noval}, {Ord{\'e}novic}, {Orr{\`u}}, {Osinde}, {Pagani},
  {Pagano}, {Palaversa}, {Palicio}, {Panahi}, {Pawlak}, {Pe{\~n}alosa
  Esteller}, {Penttil{\"a}}, {Piersimoni}, {Pineau}, {Plachy}, {Plum},
  {Poggio}, {Poretti}, {Poujoulet}, {Pr{\v{s}}a}, {Pulone}, {Racero},
  {Ragaini}, {Rainer}, {Raiteri}, {Rambaux}, {Ramos}, {Ramos-Lerate}, {Re
  Fiorentin}, {Regibo}, {Reyl{\'e}}, {Riva}, {Rixon}, {Robichon}, {Robin},
  {Roelens}, {Rohrbasser}, {Rowell}, {Royer}, {Rybicki}, {Sadowski},
  {Sagrist{\`a} Sell{\'e}s}, {Sahlmann}, {Salgado}, {Salguero}, {Samaras},
  {Gimenez}, {Sanna}, {Santove{\~n}a}, {Sarasso}, {Schultheis}, {Sciacca},
  {Segol}, {Segovia}, {S{\'e}gransan}, {Semeux}, {Siddiqui}, {Siebert},
  {Siltala}, {Slezak}, {Smart}, {Solano}, {Solitro}, {Souami}, {Souchay},
  {Spagna}, {Spoto}, {Steele}, {Steidelm{\"u}ller}, {Stephenson},
  {S{\"u}veges}, {Szabados}, {Szegedi-Elek}, {Taris}, {Tauran}, {Taylor},
  {Teixeira}, {Thuillot}, {Tonello}, {Torra}, {Torra}, {Turon}, {Unger},
  {Vaillant}, {van Dillen}, {Vanel}, {Vecchiato}, {Viala}, {Vicente},
  {Voutsinas}, {Weiler}, {Wevers}, {Wyrzykowski}, {Yoldas}, {Yvard}, {Zhao},
  {Zorec}, {Zucker}, {Zurbach}, \& {Zwitter}}]{EDR3-DPACP-109}
{Gaia Collaboration}, {Luri}, X., {Chemin}, L., {et~al.} 2021{\natexlab{d}},
  \aap, 649, A7

\bibitem[{{Gaia Collaboration} {et~al.}(2016{\natexlab{b}}){Gaia
  Collaboration}, {Prusti}, {de Bruijne}, {Brown}, {Vallenari}, {Babusiaux},
  {Bailer-Jones}, {Bastian}, {Biermann}, {Evans}, \&
  et~al.}]{2016A&A...595A...1G}
{Gaia Collaboration}, {Prusti}, T., {de Bruijne}, J.~H.~J., {et~al.}
  2016{\natexlab{b}}, \aap, 595, A1

\bibitem[{{Gaia Collaboration} {et~al.}(2021{\natexlab{e}}){Gaia
  Collaboration}, {Smart}, {Sarro}, {Rybizki}, {Reyl{\'e}}, {Robin}, {Hambly},
  {Abbas}, {Barstow}, {de Bruijne}, {Bucciarelli}, {Carrasco}, {Cooper},
  {Hodgkin}, {Masana}, {Michalik}, {Sahlmann}, {Sozzetti}, {Brown},
  {Vallenari}, {Prusti}, {Babusiaux}, {Biermann}, {Creevey}, {Evans}, {Eyer},
  {Hutton}, {Jansen}, {Jordi}, {Klioner}, {Lammers}, {Lindegren}, {Luri},
  {Mignard}, {Panem}, {Pourbaix}, {Randich}, {Sartoretti}, {Soubiran},
  {Walton}, {Arenou}, {Bailer-Jones}, {Bastian}, {Cropper}, {Drimmel}, {Katz},
  {Lattanzi}, {van Leeuwen}, {Bakker}, {Casta{\~n}eda}, {De Angeli},
  {Ducourant}, {Fabricius}, {Fouesneau}, {Fr{\'e}mat}, {Guerra}, {Guerrier},
  {Guiraud}, {Jean-Antoine Piccolo}, {Messineo}, {Mowlavi}, {Nicolas},
  {Nienartowicz}, {Pailler}, {Panuzzo}, {Riclet}, {Roux}, {Seabroke}, {Sordo},
  {Tanga}, {Th{\'e}venin}, {Gracia-Abril}, {Portell}, {Teyssier}, {Altmann},
  {Andrae}, {Bellas-Velidis}, {Benson}, {Berthier}, {Blomme}, {Brugaletta},
  {Burgess}, {Busso}, {Carry}, {Cellino}, {Cheek}, {Clementini}, {Damerdji},
  {Davidson}, {Delchambre}, {Dell'Oro}, {Fern{\'a}ndez-Hern{\'a}ndez},
  {Galluccio}, {Garc{\'\i}a-Lario}, {Garcia-Reinaldos},
  {Gonz{\'a}lez-N{\'u}{\~n}ez}, {Gosset}, {Haigron}, {Halbwachs}, {Harrison},
  {Hatzidimitriou}, {Heiter}, {Hern{\'a}ndez}, {Hestroffer}, {Holl},
  {Jan{\ss}en}, {Jevardat de Fombelle}, {Jordan}, {Krone-Martins}, {Lanzafame},
  {L{\"o}ffler}, {Lorca}, {Manteiga}, {Marchal}, {Marrese}, {Moitinho}, {Mora},
  {Muinonen}, {Osborne}, {Pancino}, {Pauwels}, {Recio-Blanco}, {Richards},
  {Riello}, {Rimoldini}, {Roegiers}, {Siopis}, {Smith}, {Ulla}, {Utrilla}, {van
  Leeuwen}, {van Reeven}, {Abreu Aramburu}, {Accart}, {Aerts}, {Aguado},
  {Ajaj}, {Altavilla}, {{\'A}lvarez}, {{\'A}lvarez Cid-Fuentes}, {Alves},
  {Anderson}, {Anglada Varela}, {Antoja}, {Audard}, {Baines}, {Baker},
  {Balaguer-N{\'u}{\~n}ez}, {Balbinot}, {Balog}, {Barache}, {Barbato},
  {Barros}, {Bartolom{\'e}}, {Bassilana}, {Bauchet}, {Baudesson-Stella},
  {Becciani}, {Bellazzini}, {Bernet}, {Bertone}, {Bianchi}, {Blanco-Cuaresma},
  {Boch}, {Bombrun}, {Bossini}, {Bouquillon}, {Bragaglia}, {Bramante},
  {Breedt}, {Bressan}, {Brouillet}, {Burlacu}, {Busonero}, {Butkevich},
  {Buzzi}, {Caffau}, {Cancelliere}, {C{\'a}novas}, {Cantat-Gaudin}, {Carballo},
  {Carlucci}, {Carnerero}, {Casamiquela}, {Castellani}, {Castro-Ginard},
  {Castro Sampol}, {Chaoul}, {Charlot}, {Chemin}, {Chiavassa}, {Cioni},
  {Comoretto}, {Cornez}, {Cowell}, {Crifo}, {Crosta}, {Crowley}, {Dafonte},
  {Dapergolas}, {David}, {David}, {de Laverny}, {De Luise}, {De March}, {De
  Ridder}, {de Souza}, {de Teodoro}, {de Torres}, {del Peloso}, {del Pozo},
  {Delgado}, {Delgado}, {Delisle}, {Di Matteo}, {Diakite}, {Diener},
  {Distefano}, {Dolding}, {Eappachen}, {Edvardsson}, {Enke}, {Esquej}, {Fabre},
  {Fabrizio}, {Faigler}, {Fedorets}, {Fernique}, {Fienga}, {Figueras},
  {Fouron}, {Fragkoudi}, {Fraile}, {Franke}, {Gai}, {Garabato},
  {Garcia-Gutierrez}, {Garc{\'\i}a-Torres}, {Garofalo}, {Gavras}, {Gerlach},
  {Geyer}, {Giacobbe}, {Gilmore}, {Girona}, {Giuffrida}, {Gomel}, {Gomez},
  {Gonzalez-Santamaria}, {Gonz{\'a}lez-Vidal}, {Granvik},
  {Guti{\'e}rrez-S{\'a}nchez}, {Guy}, {Hauser}, {Haywood}, {Helmi}, {Hidalgo},
  {Hilger}, {H{\l}adczuk}, {Hobbs}, {Holland}, {Huckle}, {Jasniewicz},
  {Jonker}, {Juaristi Campillo}, {Julbe}, {Karbevska}, {Kervella}, {Khanna},
  {Kochoska}, {Kontizas}, {Kordopatis}, {Korn}, {Kostrzewa-Rutkowska},
  {Kruszy{\'n}ska}, {Lambert}, {Lanza}, {Lasne}, {Le Campion}, {Le Fustec},
  {Lebreton}, {Lebzelter}, {Leccia}, {Leclerc}, {Lecoeur-Taibi}, {Liao},
  {Licata}, {Lindstr{\o}m}, {Lister}, {Livanou}, {Lobel}, {Madrero Pardo},
  {Managau}, {Mann}, {Marchant}, {Marconi}, {Marcos Santos}, {Marinoni},
  {Marocco}, {Marshall}, {Martin Polo}, {Mart{\'\i}n-Fleitas}, {Masip},
  {Massari}, {Mastrobuono-Battisti}, {Mazeh}, {McMillan}, {Messina}, {Millar},
  {Mints}, {Molina}, {Molinaro}, {Moln{\'a}r}, {Montegriffo}, {Mor},
  {Morbidelli}, {Morel}, {Morris}, {Mulone}, {Munoz}, {Muraveva}, {Murphy},
  {Musella}, {Noval}, {Ord{\'e}novic}, {Orr{\`u}}, {Osinde}, {Pagani},
  {Pagano}, {Palaversa}, {Palicio}, {Panahi}, {Pawlak}, {Pe{\~n}alosa
  Esteller}, {Penttil{\"a}}, {Piersimoni}, {Pineau}, {Plachy}, {Plum},
  {Poggio}, {Poretti}, {Poujoulet}, {Pr{\v{s}}a}, {Pulone}, {Racero},
  {Ragaini}, {Rainer}, {Raiteri}, {Rambaux}, {Ramos}, {Ramos-Lerate}, {Re
  Fiorentin}, {Regibo}, {Ripepi}, {Riva}, {Rixon}, {Robichon}, {Robin},
  {Roelens}, {Rohrbasser}, {Romero-G{\'o}mez}, {Rowell}, {Royer}, {Rybicki},
  {Sadowski}, {Sagrist{\`a} Sell{\'e}s}, {Salgado}, {Salguero}, {Samaras},
  {Sanchez Gimenez}, {Sanna}, {Santove{\~n}a}, {Sarasso}, {Schultheis},
  {Sciacca}, {Segol}, {Segovia}, {S{\'e}gransan}, {Semeux}, {Shahaf},
  {Siddiqui}, {Siebert}, {Siltala}, {Slezak}, {Solano}, {Solitro}, {Souami},
  {Souchay}, {Spagna}, {Spoto}, {Steele}, {Steidelm{\"u}ller}, {Stephenson},
  {S{\"u}veges}, {Szabados}, {Szegedi-Elek}, {Taris}, {Tauran}, {Taylor},
  {Teixeira}, {Thuillot}, {Tonello}, {Torra}, {Torra}, {Turon}, {Unger},
  {Vaillant}, {van Dillen}, {Vanel}, {Vecchiato}, {Viala}, {Vicente},
  {Voutsinas}, {Weiler}, {Wevers}, {Wyrzykowski}, {Yoldas}, {Yvard}, {Zhao},
  {Zorec}, {Zucker}, {Zurbach}, \& {Zwitter}}]{EDR3-DPACP-81}
{Gaia Collaboration}, {Smart}, R.~L., {Sarro}, L.~M., {et~al.}
  2021{\natexlab{e}}, \aap, 649, A6

\bibitem[{{Hambly} {et~al.}(2018){Hambly}, {Cropper}, {Boudreault}, {Crowley},
  {Kohley}, {de Bruijne}, {Dolding}, {Fabricius}, {Seabroke}, {Davidson},
  {Rowell}, {Collins}, {Cross}, {Mart{\'\i}n-Fleitas}, {Baker}, {Smith},
  {Sartoretti}, {Marchal}, {Katz}, {De Angeli}, {Busso}, {Riello}, {Allende
  Prieto}, {Els}, {Corcione}, {Masana}, {Luri}, {Chassat}, {Fusero},
  {Pasquier}, {V{\'e}tel}, {Sarri}, \& {Gare}}]{2018A&A...616A..15H}
{Hambly}, N.~C., {Cropper}, M., {Boudreault}, S., {et~al.} 2018, \aap, 616, A15

\bibitem[{Harris {et~al.}(2020)Harris, Millman, van~der Walt, Gommers,
  Virtanen, Cournapeau, Wieser, Taylor, Berg, Smith, Kern, Picus, Hoyer, van
  Kerkwijk, Brett, Haldane, Fernández~del Río, Wiebe, Peterson,
  Gérard-Marchant, Sheppard, Reddy, Weckesser, Abbasi, Gohlke, \&
  Oliphant}]{numpy:2020}
Harris, C.~R., Millman, K.~J., van~der Walt, S.~J., {et~al.} 2020, Nature, 585,
  357–362

\bibitem[{{Henden} {et~al.}(2015){Henden}, {Levine}, {Terrell}, \&
  {Welch}}]{2015AAS...22533616H}
{Henden}, A.~A., {Levine}, S., {Terrell}, D., \& {Welch}, D.~L. 2015, in
  American Astronomical Society Meeting Abstracts, Vol. 225, American
  Astronomical Society Meeting Abstracts \#225, 336.16

\bibitem[{{Henden} {et~al.}(2016){Henden}, {Templeton}, {Terrell}, {Smith},
  {Levine}, \& {Welch}}]{apass9}
{Henden}, A.~A., {Templeton}, M., {Terrell}, D., {et~al.} 2016, VizieR Online
  Data Catalogue, 2336

\bibitem[{{H{\o}g} {et~al.}(2000){H{\o}g}, {Fabricius}, {Makarov}, {Urban},
  {Corbin}, {Wycoff}, {Bastian}, {Schwekendiek}, \&
  {Wicenec}}]{2000A&A...355L..27H}
{H{\o}g}, E., {Fabricius}, C., {Makarov}, V.~V., {et~al.} 2000, \aap, 355, L27

\bibitem[{{Hunter}(2007)}]{matplotlib:2007}
{Hunter}, J.~D. 2007, Computing in Science and Engineering, 9, 90

\bibitem[{{Katz} {et~al.}(2019){Katz}, {Sartoretti}, {Cropper}, {Panuzzo},
  {Seabroke}, {Viala}, {Benson}, {Blomme}, {Jasniewicz}, {Jean-Antoine},
  {Huckle}, {Smith}, {Baker}, {Crifo}, {Damerdji}, {David}, {Dolding},
  {Fr{\'e}mat}, {Gosset}, {Guerrier}, {Guy}, {Haigron}, {Jan{\ss}en},
  {Marchal}, {Plum}, {Soubiran}, {Th{\'e}venin}, {Ajaj}, {Allende Prieto},
  {Babusiaux}, {Boudreault}, {Chemin}, {Delle Luche}, {Fabre}, {Gueguen},
  {Hambly}, {Lasne}, {Meynadier}, {Pailler}, {Panem}, {Royer}, {Tauran},
  {Zurbach}, {Zwitter}, {Arenou}, {Bossini}, {Gerssen}, {G{\'o}mez},
  {Lemaitre}, {Leclerc}, {Morel}, {Munari}, {Turon}, {Vallenari}, \&
  {{\v{Z}}erjal}}]{2019A&A...622A.205K}
{Katz}, D., {Sartoretti}, P., {Cropper}, M., {et~al.} 2019, \aap, 622, A205

\bibitem[{{Kunder} {et~al.}(2017){Kunder}, {Kordopatis}, {Steinmetz},
  {Zwitter}, {McMillan}, {Casagrande}, {Enke}, {Wojno}, {Valentini},
  {Chiappini}, {Matijevi{\v c}}, {Siviero}, {de Laverny}, {Recio-Blanco},
  {Bijaoui}, {Wyse}, {Binney}, {Grebel}, {Helmi}, {Jofre}, {Antoja}, {Gilmore},
  {Siebert}, {Famaey}, {Bienaym{\'e}}, {Gibson}, {Freeman}, {Navarro},
  {Munari}, {Seabroke}, {Anguiano}, {{\v Z}erjal}, {Minchev}, {Reid},
  {Bland-Hawthorn}, {Kos}, {Sharma}, {Watson}, {Parker}, {Scholz}, {Burton},
  {Cass}, {Hartley}, {Fiegert}, {Stupar}, {Ritter}, {Hawkins}, {Gerhard},
  {Chaplin}, {Davies}, {Elsworth}, {Lund}, {Miglio}, \&
  {Mosser}}]{2017AJ....153...75K}
{Kunder}, A., {Kordopatis}, G., {Steinmetz}, M., {et~al.} 2017, \aj, 153, 75

\bibitem[{{Lasker} {et~al.}(2008){Lasker}, {Lattanzi}, {McLean}, {Bucciarelli},
  {Drimmel}, {Garcia}, {Greene}, {Guglielmetti}, {Hanley}, {Hawkins},
  {Laidler}, {Loomis}, {Meakes}, {Mignani}, {Morbidelli}, {Morrison},
  {Pannunzio}, {Rosenberg}, {Sarasso}, {Smart}, {Spagna}, {Sturch},
  {Volpicelli}, {White}, {Wolfe}, \& {Zacchei}}]{2008AJ....136..735L}
{Lasker}, B.~M., {Lattanzi}, M.~G., {McLean}, B.~J., {et~al.} 2008, \aj, 136,
  735

\bibitem[{Lindegren(2018)}]{ruwe}
Lindegren, L. 2018, technical note GAIA-C3-TN-LU-LL-124

\bibitem[{{Lindegren}(2020{\natexlab{a}})}]{2020A&A...633A...1L}
{Lindegren}, L. 2020{\natexlab{a}}, \aap, 633, A1

\bibitem[{{Lindegren}(2020{\natexlab{b}})}]{2020A&A...637C...5L}
{Lindegren}, L. 2020{\natexlab{b}}, \aap, 637, C5

\bibitem[{{Lindegren} {et~al.}(2021{\natexlab{a}}){Lindegren}, {Bastian},
  {Biermann}, {Bombrun}, {de Torres}, {Gerlach}, {Geyer}, {Hern{\'a}ndez},
  {Hilger}, {Hobbs}, {Klioner}, {Lammers}, {McMillan}, {Ramos-Lerate},
  {Steidelm{\"u}ller}, {Stephenson}, \& {van Leeuwen}}]{EDR3-DPACP-132}
{Lindegren}, L., {Bastian}, U., {Biermann}, M., {et~al.} 2021{\natexlab{a}},
  \aap, 649, A4

\bibitem[{{Lindegren} {et~al.}(2018){Lindegren}, {Hern{\'a}ndez}, {Bombrun},
  {Klioner}, {Bastian}, {Ramos-Lerate}, {de Torres}, {Steidelm{\"u}ller},
  {Stephenson}, {Hobbs}, {Lammers}, {Biermann}, {Geyer}, {Hilger}, {Michalik},
  {Stampa}, {McMillan}, {Casta{\~n}eda}, {Clotet}, {Comoretto}, {Davidson},
  {Fabricius}, {Gracia}, {Hambly}, {Hutton}, {Mora}, {Portell}, {van Leeuwen},
  {Abbas}, {Abreu}, {Altmann}, {Andrei}, {Anglada}, {Balaguer-N{\'u}{\~n}ez},
  {Barache}, {Becciani}, {Bertone}, {Bianchi}, {Bouquillon}, {Bourda},
  {Br{\"u}semeister}, {Bucciarelli}, {Busonero}, {Buzzi}, {Cancelliere},
  {Carlucci}, {Charlot}, {Cheek}, {Crosta}, {Crowley}, {de Bruijne}, {de
  Felice}, {Drimmel}, {Esquej}, {Fienga}, {Fraile}, {Gai}, {Garralda},
  {Gonz{\'a}lez-Vidal}, {Guerra}, {Hauser}, {Hofmann}, {Holl}, {Jordan},
  {Lattanzi}, {Lenhardt}, {Liao}, {Licata}, {Lister}, {L{\"o}ffler},
  {Marchant}, {Martin-Fleitas}, {Messineo}, {Mignard}, {Morbidelli}, {Poggio},
  {Riva}, {Rowell}, {Salguero}, {Sarasso}, {Sciacca}, {Siddiqui}, {Smart},
  {Spagna}, {Steele}, {Taris}, {Torra}, {van Elteren}, {van Reeven}, \&
  {Vecchiato}}]{2018A&A...616A...2L}
{Lindegren}, L., {Hern{\'a}ndez}, J., {Bombrun}, A., {et~al.} 2018, \aap, 616,
  A2

\bibitem[{{Lindegren} {et~al.}(2021{\natexlab{b}}){Lindegren}, {Klioner},
  {Hern{\'a}ndez}, {Bombrun}, {Ramos-Lerate}, {Steidelm{\"u}ller}, {Bastian},
  {Biermann}, {de Torres}, {Gerlach}, {Geyer}, {Hilger}, {Hobbs}, {Lammers},
  {McMillan}, {Stephenson}, {Casta{\~n}eda}, {Davidson}, {Fabricius},
  {Gracia-Abril}, {Portell}, {Rowell}, {Teyssier}, {Torra}, {Bartolom{\'e}},
  {Clotet}, {Garralda}, {Gonz{\'a}lez-Vidal}, {Torra}, {Abbas}, {Altmann},
  {Anglada Varela}, {Balaguer-N{\'u}{\~n}ez}, {Balog}, {Barache}, {Becciani},
  {Bernet}, {Bertone}, {Bianchi}, {Bouquillon}, {Brown}, {Bucciarelli},
  {Busonero}, {Butkevich}, {Buzzi}, {Cancelliere}, {Carlucci}, {Charlot},
  {Cioni}, {Crosta}, {Crowley}, {del Peloso}, {del Pozo}, {Drimmel}, {Esquej},
  {Fienga}, {Fraile}, {Gai}, {Garcia-Reinaldos}, {Guerra}, {Hambly}, {Hauser},
  {Jan{\ss}en}, {Jordan}, {Kostrzewa-Rutkowska}, {Lattanzi}, {Liao}, {Licata},
  {Lister}, {L{\"o}ffler}, {Marchant}, {Masip}, {Mignard}, {Mints}, {Molina},
  {Mora}, {Morbidelli}, {Murphy}, {Pagani}, {Panuzzo}, {Pe{\~n}alosa Esteller},
  {Poggio}, {Re Fiorentin}, {Riva}, {Sagrist{\`a} Sell{\'e}s}, {Sanchez
  Gimenez}, {Sarasso}, {Sciacca}, {Siddiqui}, {Smart}, {Souami}, {Spagna},
  {Steele}, {Taris}, {Utrilla}, {van Reeven}, \& {Vecchiato}}]{EDR3-DPACP-128}
{Lindegren}, L., {Klioner}, S.~A., {Hern{\'a}ndez}, J., {et~al.}
  2021{\natexlab{b}}, \aap, 649, A2

\bibitem[{{Lindegren} {et~al.}(2012){Lindegren}, {Lammers}, {Hobbs},
  {O'Mullane}, {Bastian}, \& {Hern{\'a}ndez}}]{Lindegren2012}
{Lindegren}, L., {Lammers}, U., {Hobbs}, D., {et~al.} 2012, \aap, 538, A78

\bibitem[{{Luri} {et~al.}(2018){Luri}, {Brown}, {Sarro}, {Arenou},
  {Bailer-Jones}, {Castro-Ginard}, {de Bruijne}, {Prusti}, {Babusiaux}, \&
  {Delgado}}]{2018A&A...616A...9L}
{Luri}, X., {Brown}, A.~G.~A., {Sarro}, L.~M., {et~al.} 2018, \aap, 616, A9

\bibitem[{{Luri} {et~al.}(2014){Luri}, {Palmer}, {Arenou}, {Masana}, {de
  Bruijne}, {Antiche}, {Babusiaux}, {Borrachero}, {Sartoretti}, {Julbe},
  {Isasi}, {Martinez}, {Robin}, {Reyl\'e}, {Jordi}, \& {Carrasco}}]{gog}
{Luri}, X., {Palmer}, M., {Arenou}, F., {et~al.} 2014, A\&A, 566, A119

\bibitem[{{Ma{\'\i}z Apell{\'a}niz} \& {Weiler}(2018)}]{2018A&A...619A.180M}
{Ma{\'\i}z Apell{\'a}niz}, J. \& {Weiler}, M. 2018, \aap, 619, A180

\bibitem[{{Marrese} {et~al.}(2019){Marrese}, {Marinoni}, {Fabrizio}, \&
  {Altavilla}}]{2019A&A...621A.144M}
{Marrese}, P.~M., {Marinoni}, S., {Fabrizio}, M., \& {Altavilla}, G. 2019,
  \aap, 621, A144

\bibitem[{{Marrese} {et~al.}(2021){Marrese}, {Marinoni}, {Fabrizio}, \&
  {Altavilla}}]{EDR3-DPACP-129}
{Marrese}, P.~M., {Marinoni}, S., {Fabrizio}, M., \& {Altavilla}, G. 2021,
  \aap\ in prep.

\bibitem[{{Marrese} {et~al.}(2017){Marrese}, {Marinoni}, {Fabrizio}, \&
  {Giuffrida}}]{2017A&A...607A.105M}
{Marrese}, P.~M., {Marinoni}, S., {Fabrizio}, M., \& {Giuffrida}, G. 2017,
  \aap, 607, A105

\bibitem[{{Michalik} {et~al.}(2015){Michalik}, {Lindegren}, {Hobbs}, \&
  {Butkevich}}]{2015A&A...583A..68M}
{Michalik}, D., {Lindegren}, L., {Hobbs}, D., \& {Butkevich}, A.~G. 2015, \aap,
  583, A68

\bibitem[{{Montegriffo et al.}(2021)}]{EDR3-DPACP-120}
{Montegriffo et al.} 2021, \aap\ in prep.

\bibitem[{{Onken} {et~al.}(2019){Onken}, {Wolf}, {Bessell}, {Chang}, {Da
  Costa}, {Luvaul}, {Mackey}, {Schmidt}, \& {Shao}}]{2019PASA...36...33O}
{Onken}, C.~A., {Wolf}, C., {Bessell}, M.~S., {et~al.} 2019, \pasa, 36, e033

\bibitem[{{Pancino} {et~al.}(2012){Pancino}, {Altavilla}, {Marinoni},
  {Cocozza}, {Carrasco}, {Bellazzini}, {Bragaglia}, {Federici}, {Rossetti},
  {Cacciari}, {Balaguer N{\'u}{\~n}ez}, {Castro}, {Figueras}, {Fusi Pecci},
  {Galleti}, {Gebran}, {Jordi}, {Lardo}, {Masana}, {Mongui{\'o}},
  {Montegriffo}, {Ragaini}, {Schuster}, {Trager}, {Vilardell}, \&
  {Voss}}]{2012MNRAS.426.1767P}
{Pancino}, E., {Altavilla}, G., {Marinoni}, S., {et~al.} 2012, \mnras, 426,
  1767

\bibitem[{P\'erez \& Granger(2007)}]{ipython:2007}
P\'erez, F. \& Granger, B.~E. 2007, Computing in Science and Engineering, 9, 21

\bibitem[{{Riello} {et~al.}(2021){Riello}, {De Angeli}, {Evans}, {Montegriffo},
  {Carrasco}, {Busso}, {Palaversa}, {Burgess}, {Diener}, {Davidson}, {Rowell},
  {Fabricius}, {Jordi}, {Bellazzini}, {Pancino}, {Harrison}, {Cacciari}, {van
  Leeuwen}, {Hambly}, {Hodgkin}, {Osborne}, {Altavilla}, {Barstow}, {Brown},
  {Castellani}, {Cowell}, {De Luise}, {Gilmore}, {Giuffrida}, {Hidalgo},
  {Holland}, {Marinoni}, {Pagani}, {Piersimoni}, {Pulone}, {Ragaini}, {Rainer},
  {Richards}, {Sanna}, {Walton}, {Weiler}, \& {Yoldas}}]{EDR3-DPACP-117}
{Riello}, M., {De Angeli}, F., {Evans}, D.~W., {et~al.} 2021, \aap, 649, A3

\bibitem[{{Robin} {et~al.}(2012){Robin}, {Luri}, {Reyl{\'e}}, {Isasi}, {Grux},
  {Blanco-Cuaresma}, {Arenou}, {Babusiaux}, {Belcheva}, {Drimmel}, {Jordi},
  {Krone-Martins}, {Masana}, {Mauduit}, {Mignard}, {Mowlavi},
  {Rocca-Volmerange}, {Sartoretti}, {Slezak}, \&
  {Sozzetti}}]{2012A&A...543A.100R}
{Robin}, A.~C., {Luri}, X., {Reyl{\'e}}, C., {et~al.} 2012, \aap, 543, A100

\bibitem[{{Rowell} {et~al.}(2021){Rowell}, {Davidson}, {Lindegren}, {van
  Leeuwen}, {Casta{\~n}eda}, {Fabricius}, {Bastian}, {Hambly}, {Hern{\'a}ndez},
  {Bombrun}, {Evans}, {De Angeli}, {Riello}, {Busonero}, {Crowley}, {Mora},
  {Lammers}, {Gracia}, {Portell}, {Biermann}, \& {Brown}}]{EDR3-DPACP-73}
{Rowell}, N., {Davidson}, M., {Lindegren}, L., {et~al.} 2021, \aap, 649, A11

\bibitem[{{Sartoretti} {et~al.}(2018){Sartoretti}, {Katz}, {Cropper},
  {Panuzzo}, {Seabroke}, {Viala}, {Benson}, {Blomme}, {Jasniewicz},
  {Jean-Antoine}, {Huckle}, {Smith}, {Baker}, {Crifo}, {Damerdji}, {David},
  {Dolding}, {Fr{\'e}mat}, {Gosset}, {Guerrier}, {Guy}, {Haigron},
  {Jan{\ss}en}, {Marchal}, {Plum}, {Soubiran}, {Th{\'e}venin}, {Ajaj}, {Allende
  Prieto}, {Babusiaux}, {Boudreault}, {Chemin}, {Delle Luche}, {Fabre},
  {Gueguen}, {Hambly}, {Lasne}, {Meynadier}, {Pailler}, {Panem}, {Riclet},
  {Royer}, {Tauran}, {Zurbach}, {Zwitter}, {Arenou}, {Gomez}, {Lemaitre},
  {Leclerc}, {Morel}, {Munari}, {Turon}, \&
  {{\v{Z}}erjal}}]{2018A&A...616A...6S}
{Sartoretti}, P., {Katz}, D., {Cropper}, M., {et~al.} 2018, \aap, 616, A6

\bibitem[{{Seabroke} {et~al.}(2021){Seabroke}, {Fabricius}, {Teyssier},
  {Sartoretti}, {Katz}, \& {et al.}}]{EDR3-DPACP-121}
{Seabroke}, G., {Fabricius}, C., {Teyssier}, D., {et~al.} 2021, submitted to
  \aap

\bibitem[{{Skrutskie} {et~al.}(2006){Skrutskie}, {Cutri}, {Stiening},
  {Weinberg}, {Schneider}, {Carpenter}, {Beichman}, {Capps}, {Chester},
  {Elias}, {Huchra}, {Liebert}, {Lonsdale}, {Monet}, {Price}, {Seitzer},
  {Jarrett}, {Kirkpatrick}, {Gizis}, {Howard}, {Evans}, {Fowler}, {Fullmer},
  {Hurt}, {Light}, {Kopan}, {Marsh}, {McCallon}, {Tam}, {Van Dyk}, \&
  {Wheelock}}]{2006AJ....131.1163S}
{Skrutskie}, M.~F., {Cutri}, R.~M., {Stiening}, R., {et~al.} 2006, \aj, 131,
  1163

\bibitem[{{Smart} \& {Nicastro}(2014)}]{2014A&A...570A..87S}
{Smart}, R.~L. \& {Nicastro}, L. 2014, \aap, 570, A87

\bibitem[{{Taylor}(2005)}]{topcat:2005}
{Taylor}, M.~B. 2005, in Astronomical Society of the Pacific Conference Series,
  Vol. 347, Astronomical Data Analysis Software and Systems XIV, ed.
  P.~{Shopbell}, M.~{Britton}, \& R.~{Ebert}, 29

\bibitem[{{Torra} {et~al.}(2021){Torra}, {Casta{\~n}eda}, {Fabricius},
  {Lindegren}, {Clotet}, {Gonz{\'a}lez-Vidal}, {Bartolom{\'e}}, {Bastian},
  {Bernet}, {Biermann}, {Garralda}, {G{\'u}rpide}, {Lammers}, {Portell}, \&
  {Torra}}]{EDR3-DPACP-124}
{Torra}, F., {Casta{\~n}eda}, J., {Fabricius}, C., {et~al.} 2021, \aap, 649,
  A10

\bibitem[{van Leeuwen(2007)}]{book:newhip}
van Leeuwen, F. 2007, {H}ipparcos, the {N}ew {R}eduction of the {R}aw {D}ata,
  {A}strophysics and {S}pace {S}cience {L}ibrary. {V}ol. 350 edn. (Springer)

\bibitem[{{Virtanen} {et~al.}(2020){Virtanen}, {Gommers}, {Oliphant},
  {Haberland}, {Reddy}, {Cournapeau}, {Burovski}, {Peterson}, {Weckesser},
  {Bright}, {van der Walt}, {Brett}, {Wilson}, {Jarrod Millman}, {Mayorov},
  {Nelson}, {Jones}, {Kern}, {Larson}, {Carey}, {Polat}, {Feng}, {Moore}, {Vand
  erPlas}, {Laxalde}, {Perktold}, {Cimrman}, {Henriksen}, {Quintero}, {Harris},
  {Archibald}, {Ribeiro}, {Pedregosa}, {van Mulbregt}, \&
  {Contributors}}]{scipy:2020}
{Virtanen}, P., {Gommers}, R., {Oliphant}, T.~E., {et~al.} 2020, Nature
  Methods, 17, 261

\bibitem[{{Wright} {et~al.}(2010){Wright}, {Eisenhardt}, {Mainzer}, {Ressler},
  {Cutri}, {Jarrett}, {Kirkpatrick}, {Padgett}, {McMillan}, {Skrutskie},
  {Stanford}, {Cohen}, {Walker}, {Mather}, {Leisawitz}, {Gautier}, {McLean},
  {Benford}, {Lonsdale}, {Blain}, {Mendez}, {Irace}, {Duval}, {Liu}, {Royer},
  {Heinrichsen}, {Howard}, {Shannon}, {Kendall}, {Walsh}, {Larsen}, {Cardon},
  {Schick}, {Schwalm}, {Abid}, {Fabinsky}, {Naes}, \&
  {Tsai}}]{2010AJ....140.1868W}
{Wright}, E.~L., {Eisenhardt}, P.~R.~M., {Mainzer}, A.~K., {et~al.} 2010, \aj,
  140, 1868

\bibitem[{{Zacharias} {et~al.}(2015){Zacharias}, {Finch}, {Subasavage},
  {Bredthauer}, {Crockett}, {Divittorio}, {Ferguson}, {Harris}, {Harris},
  {Henden}, {Kilian}, {Munn}, {Rafferty}, {Rhodes}, {Schultheiss}, {Tilleman},
  \& {Wieder}}]{2015AJ....150..101Z}
{Zacharias}, N., {Finch}, C., {Subasavage}, J., {et~al.} 2015, \aj, 150, 101

\end{thebibliography}

%
%

\begin{appendix}

    \section{$G$-band corrections for sources with 2-parameter or 6-parameter astrometric solutions}
    \label{app:gcorrection}

    Figure \ref{tab:gbandcorrs} shows how to formulate an ADQL query, to be executed in the \edr{3} archive, that
    contains an on-the-fly calculation of the corrected $G$-band fluxes or magnitudes. These queries are
    somewhat complex and create a performance overhead. Hence downloading the requisite \edr{3} fields and calculating
    the corrections a posteriori may be more efficient. Example Python code to do this is included in
    Fig.~\ref{tab:gcorrpython}. The Python code is also available as a Jupyter notebook at the following link:
    \url{https://github.com/agabrown/gaiaedr3-6p-gband-correction}.

    \begin{figure*}
        \caption{Example queries that can be submitted to the {\gaia} archive in the Astronomical Data
        Query Language to retrieve corrected $G$-band photometry.\label{tab:gbandcorrs}}

        \medskip\noindent
        Query that includes a calculation of the $G$-band flux correction. The condition `\texttt{bp\_rp > -20}' ensures
        that no correction is attempted in case the \bpminrp\ colour is not available (`\texttt{bp\_rp is not null}'
        does not work). The condition on \texttt{random\_index} is included to retrieve example data for a random sample
        of sources.\\
        \begin{lstlisting}[language=SQL]
select source_id, astrometric_params_solved, bp_rp, phot_g_mean_mag, phot_g_mean_flux,
if_then_else(
  bp_rp > -20,
  case_condition(
    phot_g_mean_flux * (1.00525 -0.02323*greatest(0.25, least(bp_rp, 3)) 
            +0.01740*power(greatest(0.25, least(bp_rp, 3)),2) 
            -0.00253*power(greatest(0.25, least(bp_rp, 3)),3)),
    astrometric_params_solved = 31,
    phot_g_mean_flux,
    phot_g_mean_mag < 13,
    phot_g_mean_flux,
    phot_g_mean_mag < 16,
    phot_g_mean_flux * (1.00876 -0.02540*greatest(0.25, least(bp_rp, 3)) 
                    +0.01747*power(greatest(0.25, least(bp_rp, 3)),2) 
                    -0.00277*power(greatest(0.25, least(bp_rp, 3)),3))
  ),
  phot_g_mean_flux
) as phot_g_mean_flux_corr
from gaiaedr3.gaia_source
where random_index between 1000000 and 1999999
        \end{lstlisting}

        Query that includes a calculation of the $G$-band magnitude correction. We note the type-cast
        `\texttt{to\_real()}' of the return value of the conditional part of the query.\\
        \begin{lstlisting}[language=SQL]
select source_id, astrometric_params_solved, bp_rp, phot_g_mean_mag, phot_g_mean_flux,
if_then_else(
  bp_rp > -20,
  to_real(case_condition(
    phot_g_mean_mag - 2.5*log10( (1.00525 -0.02323*greatest(0.25, least(bp_rp, 3)) 
                                  +0.01740*power(greatest(0.25, least(bp_rp, 3)),2) 
                                  -0.00253*power(greatest(0.25, least(bp_rp, 3)),3)) ),
    astrometric_params_solved = 31,
    phot_g_mean_mag,
    phot_g_mean_mag < 13,
    phot_g_mean_mag,
    phot_g_mean_mag < 16,
    phot_g_mean_mag - 2.5*log10( (1.00876 -0.02540*greatest(0.25, least(bp_rp, 3)) 
                                  +0.01747*power(greatest(0.25, least(bp_rp, 3)),2) 
                                  -0.00277*power(greatest(0.25, least(bp_rp, 3)),3)) )
  )),
  phot_g_mean_mag
) as phot_g_mean_mag_corr
from gaiaedr3.gaia_source
where random_index between 5000000 and 5999999
        \end{lstlisting}
    \end{figure*}

    \begin{figure*}
        \caption{Python code for calculating the corrections to the $G$-band photometry for sources with 2-parameter or
        6-parameter astrometric solutions.\label{tab:gcorrpython}}
        \begin{lstlisting}[language=Python]
import numpy as np

def correct_gband(bp_rp, astrometric_params_solved, phot_g_mean_mag, phot_g_mean_flux):
    """
    Correct the G-band fluxes and magnitudes for the input list of Gaia EDR3 data.
    
    Parameters
    ----------
    
    bp_rp: float, array_like
        The (BP-RP) colour listed in the Gaia EDR3 archive.
    astrometric_params_solved: int, array_like
        The astrometric solution type listed in the Gaia EDR3 archive.
    phot_g_mean_mag: float, array_like
        The G-band magnitude as listed in the Gaia EDR3 archive.
    phot_g_mean_flux: float, array_like
        The G-band flux as listed in the Gaia EDR3 archive.
        
    Returns
    -------
    
    The corrected G-band magnitudes and fluxes. The corrections are only applied to
    sources with a 2-parameter or 6-parameter astrometric solution fainter than G=13,
    for which a (BP-RP) colour is available.
    
    Example
    
    gmag_corr, gflux_corr = correct_gband(bp_rp, astrometric_params_solved, 
                                          phot_g_mean_mag, phot_g_mean_flux)
    """
    if np.isscalar(bp_rp) or np.isscalar(astrometric_params_solved) or \
                    np.isscalar(phot_g_mean_mag) or np.isscalar(phot_g_mean_flux):
        bp_rp = np.float64(bp_rp)
        astrometric_params_solved = np.int64(astrometric_params_solved)
        phot_g_mean_mag = np.float64(phot_g_mean_mag)
        phot_g_mean_flux = np.float64(phot_g_mean_flux)
    
    if not (bp_rp.shape == astrometric_params_solved.shape \
                        == phot_g_mean_mag.shape == phot_g_mean_flux.shape):
        raise ValueError('Function parameters must be of the same shape!')
    
    do_not_correct = np.isnan(bp_rp) | (phot_g_mean_mag<13) | \
                                       (astrometric_params_solved == 31)
    bright_correct = np.logical_not(do_not_correct) & (phot_g_mean_mag>=13) & \
                                   (phot_g_mean_mag<=16)
    faint_correct = np.logical_not(do_not_correct) & (phot_g_mean_mag>16)
    bp_rp_c = np.clip(bp_rp, 0.25, 3.0)
    
    correction_factor = np.ones_like(phot_g_mean_mag)
    correction_factor[faint_correct] = 1.00525 - 0.02323*bp_rp_c[faint_correct] + \
        0.01740*np.power(bp_rp_c[faint_correct],2) - \
        0.00253*np.power(bp_rp_c[faint_correct],3)
    correction_factor[bright_correct] = 1.00876 - 0.02540*bp_rp_c[bright_correct] + \
        0.01747*np.power(bp_rp_c[bright_correct],2) - \
        0.00277*np.power(bp_rp_c[bright_correct],3)
    
    gmag_corrected = phot_g_mean_mag - 2.5*np.log10(correction_factor)
    gflux_corrected = phot_g_mean_flux * correction_factor
    
    return gmag_corrected, gflux_corrected
        \end{lstlisting}
    \end{figure*}

    \section{Calculating the corrected flux excess factor}
    \label{app:excessfact}

    Figure \ref{tab:fluxexcesscorr} shows how to formulate an ADQL query, to be executed in the \edr{3} archive, that
    contains an on-the-fly calculation of the corrected flux excess factor. This query is somewhat complex and incurs
    a performance overhead. Hence downloading the requisite \edr{3} fields and calculating the corrections a posteriori
    may be more efficient. Example Python code to do this is included in Fig.~\ref{tab:fluxexcesscorrpython}.
    The Python code is also available as a Jupyter notebook at the following link:
    \url{https://github.com/agabrown/gaiaedr3-flux-excess-correction}.

    \begin{figure*}
        \caption{Example query that can be submitted to the {\gaia} archive in the Astronomical Data Query Language to
        retrieve the corrected flux excess factor presented in \cite{EDR3-DPACP-117}.\label{tab:fluxexcesscorr}}

        \medskip\noindent
        Query that includes a calculation of the correction of the flux excess factor. The condition `\texttt{bp\_rp >
        -20}' ensures that no correction is attempted in case the \bpminrp\ colour is not available (`\texttt{bp\_rp is
    not null}' does not work). We note the type-cast `\texttt{to\_real()}' of the return value of the conditional part
    of the query. The condition on \texttt{random\_index} is included to retrieve example data for a random sample of
    sources.
    \begin{lstlisting}[language=SQL]
select source_id, bp_rp, phot_bp_rp_excess_factor,
if_then_else(
    bp_rp > -20,
    to_real(case_condition(
        phot_bp_rp_excess_factor - (1.162004 + 0.011464*bp_rp 
                                    + 0.049255*power(bp_rp,2) 
                                    - 0.005879*power(bp_rp,3)),
        bp_rp < 0.5,
        phot_bp_rp_excess_factor - (1.154360 + 0.033772*bp_rp 
                                             + 0.032277*power(bp_rp,2)),
        bp_rp >= 4.0,
        phot_bp_rp_excess_factor - (1.057572 + 0.140537*bp_rp)
    )),
    phot_bp_rp_excess_factor
) as phot_bp_rp_excess_factor_corr
from gaiaedr3.gaia_source
where random_index between 1000000 and 1999999
    \end{lstlisting}
    \end{figure*}

    \begin{figure*}
        \caption{Python code for calculating the corrected flux excess factor presented in
        \cite{EDR3-DPACP-117}.\label{tab:fluxexcesscorrpython}}

        \begin{lstlisting}[language=Python]
import numpy as np

def correct_flux_excess_factor(bp_rp, phot_bp_rp_excess_factor):
    """
    Calculate the corrected flux excess factor for the input Gaia EDR3 data.
    
    Parameters
    ----------
    
    bp_rp: float, array_like
        The (BP-RP) colour listed in the Gaia EDR3 archive.
    phot_bp_rp_flux_excess_factor: float, array_like
        The flux excess factor listed in the Gaia EDR3 archive.
        
    Returns
    -------
    
    The corrected value for the flux excess factor, which is zero for "normal" stars.
    
    Example
    -------
    
    phot_bp_rp_excess_factor_corr = correct_flux_excess_factor(bp_rp, 
                                                phot_bp_rp_flux_excess_factor)
    """

    if np.isscalar(bp_rp) or np.isscalar(phot_bp_rp_excess_factor):
        bp_rp = np.float64(bp_rp)
        phot_bp_rp_excess_factor = np.float64(phot_bp_rp_excess_factor)
    
    if bp_rp.shape != phot_bp_rp_excess_factor.shape:
        raise ValueError('Function parameters must be of the same shape!')
    
    do_not_correct = np.isnan(bp_rp)
    bluerange = np.logical_not(do_not_correct) & (bp_rp < 0.5)
    greenrange = np.logical_not(do_not_correct) & (bp_rp >= 0.5) & (bp_rp < 4.0)
    redrange = np.logical_not(do_not_correct) & (bp_rp > 4.0)
    
    correction = np.zeros_like(bp_rp)
    correction[bluerange] = 1.154360 + 0.033772*bp_rp[bluerange] + 
                                       0.032277*np.power(bp_rp[bluerange],2)
    correction[greenrange] = 1.162004 + 0.011464*bp_rp[greenrange] + \
                                        0.049255*np.power(bp_rp[greenrange],2) \
                                      - 0.005879*np.power(bp_rp[greenrange],3)
    correction[redrange] = 1.057572 + 0.140537*bp_rp[redrange]
    
    return phot_bp_rp_excess_factor - correction
    \end{lstlisting}
    \end{figure*}

    \end{appendix}

\end{document}